\documentclass[
superscriptaddress,
 amsmath,amssymb,  
 aps,prx,twocolumn,longbibliography
]{revtex4-2}

\usepackage[utf8]{inputenc}
\usepackage[T1]{fontenc}
\usepackage{lineno}
\usepackage{graphicx}
\usepackage{upgreek}

\usepackage{xcolor}
\usepackage[table]{xcolor}

\usepackage{svg}
\usepackage{textcomp}
\usepackage{booktabs}
\usepackage{hyperref}
\usepackage{cleveref}
\usepackage{float}
\usepackage{braket}
\usepackage[title,titletoc]{appendix}
\usepackage{siunitx}
\preprint{APS/123-QED}
\usepackage{todonotes}

\usepackage{algorithm}
\usepackage{algpseudocode}
\hypersetup{colorlinks=true,
linkcolor=blue,
citecolor=blue,
urlcolor=blue,
filecolor=blue
}

\usepackage[normalem]{ulem}
\newcommand{\delete}[1]{} 
\newcommand{\add}[1]{\textcolor{black}{#1}}

\newcommand{\EVS}{\ensuremath{E_\mathrm{VS}}}

\begin{document}
\title{Mapping g-factors and complex intervalley coupling in Si/SiGe by conveyor-mode shuttling}

\author{Mats Volmer}

\affiliation{JARA-FIT Institute for Quantum Information, Forschungszentrum J\"ulich GmbH and RWTH Aachen University, Aachen, Germany}
\author{Tom Struck}
\affiliation{JARA-FIT Institute for Quantum Information, Forschungszentrum J\"ulich GmbH and RWTH Aachen University, Aachen, Germany}
\author{Arnau Sala}
\affiliation{JARA-FIT Institute for Quantum Information, Forschungszentrum J\"ulich GmbH and RWTH Aachen University, Aachen, Germany}
\author{Jhih-Sian Tu}
\author{Stefan Trellenkamp}
\affiliation{Helmholtz Nano Facility (HNF), Forschungszentrum J\"ulich, J\"ulich, Germany}
\author{Davide Degli Esposti}
\author{Giordano Scappucci}
\affiliation{QuTech and Kavli Institute of Nanoscience, Delft University \\ of Technology, Lorentzweg 1, 2628 CJ Delft, The Netherlands}
\author{Hendrik Bluhm}
\affiliation{JARA-FIT Institute for Quantum Information, Forschungszentrum J\"ulich GmbH and RWTH Aachen University, Aachen, Germany}
\author{{\L}ukasz Cywi{\'n}ski}
\affiliation{Institute of Physics, Polish Academy of Sciences, Warsaw, Poland}
\author{Lars R. Schreiber}
\email{lars.schreiber@physik.rwth-aachen.de}
\affiliation{JARA-FIT Institute for Quantum Information, Forschungszentrum J\"ulich GmbH and RWTH Aachen University, Aachen, Germany}

\begin{abstract}
As silicon spin qubit chips are increasing in qubit number and area, methods for the screening of qubit related material parameters become vital. Here we demonstrate the two-dimensional mapping of small variations of the electron g-factor of QDs formed in planar Si/SiGe quantum wells with precision better than 10$^{-3}$ and with nanometer lateral resolution. 
We scan the electron g-factor across a $40\,\mathrm{nm}\times400\,\mathrm{nm}$ area and observe two $g$-factors per QD site which obey a striking symmetry and bimodal distribution across the area. These two $g$-factors relate to valley states of the electron in the QD in agreement with a recent theoretical model. Using conveyor-belt shuttling of entangled electron spin pairs, complementary to the mapping of the local valley-splitting, we map the g-factor. We compare g-factor and valley splitting maps measured on the same device, and extract the complex intervalley coupling parameter along the shuttle trajectories applying a theoretical model of $g$-factor dependence on intervalley coupling. These maps will allow unprecedented insights into the spin-valley dynamics during qubit manipulation, readout and shuttling and serve as a benchmark for the engineering of Si/SiGe heterostructures for large-scale quantum chips.
\end{abstract}

\flushbottom
\maketitle
\section{Introduction}
For electron spin qubits in Si/SiGe heterostructures, all the main building blocks of gate-based quantum computing~\cite{DiVincenzo00} have been demonstrated with sufficient fidelity required for fault-tolerant quantum computing  ~\cite{Yoneda2018,watson18,Struck2020,Noiri2022,Mills2022,Xue2022,philips22, Stano22}. The fabrication of the quantum chips is compatible with processes in silicon foundries~\cite{neyens24,Huckemann25}. With spin-qubit shuttling~\cite{Seidler22,Kuenne23,Xue23,Struck23,desmet24,matsumoto25,Beer25} allowing for the co-integration of cryo-electronics operating the quantum chips \cite{Vandersypen17, Boter22, Zhao25}, chip architectures with high qubit count~\cite{Boter22,Kuenne23, Ginzel24} are suggested. Since these proposed chips contain millions of quantum dots (QDs), and cover a larger area, a remaining challenge is the precise characterization and control of the spin-qubit related material parameters across the quantum chip~\cite{Langrock23,philips22,Losert23}.

\add{In Si-based quantum dots an electron can occupy one of two low-energy valley states. The difference of energies of these two states, the valley splitting ($\EVS$), is a key parameter to be considered in the context of operation of spin qubits. There is a consensus in the community that maximizing the value of $\EVS$ \emph{in all the quantum dots in the operated device}, or at least knowing where the regions of low $\EVS$ are, is critical for scalability of Si-based spin qubits \cite{Zwanenburg13,Vandersypen17,Langrock23,Kuenne23}}. 

When $\EVS$ is comparable to Zeeman splitting, or to thermal excitation energy, both initialization, manipulation, and readout of spin qubits can lead to unwanted transitions between valley states. Dependence of the electron spin $g$-factor \cite{Kawakami2014,Veldhorst_PRB15,Ferdous2018} and two-qubit exchange interaction \cite{Tariq_NPJQI22} on valley state leads then to a plethora of errors in coherent control of qubits. It is also now recognized \cite{Langrock23,Losert24,Volmer25} that electron shuttling through regions with low $\EVS$ can negatively affect the coherence of the shuttled qubits, due to increased probability of inter-valley excitation, and the valley dependence of the electron $g$-factor.

\add{Another key parameter of systems containing multiple spin qubits is the QD  position-dependent $g$-factor of the electron. Apart from its above-mentioned dependence on the valley state, the $g$-factor in Si-based QDs is known to depend on the lateral position of the QD \cite{volmer24}. These lateral $g$-factor variations are relevant for the coupling of charge noise to the spin-qubit, for spin dephasing during shuttling, and for driving multiple qubits at the same frequency~\cite{Langrock23,delima25_2,Veldhorst2017,Patomaeki24}. }

\add{The third important parameter is the so-called intervalley coupling $\Delta \! \equiv \! |\Delta|e^{i\phi}$, which is a complex quantity the phase of which, $\phi$, determines the composition of the valley eigenstates \cite{Friesen2007,Zwanenburg13,Losert23}, and amplitude of which determines the valley splitting, as $\EVS \! =\! 2|\Delta|$. The presence of nonzero valley phase difference between two tunnel-coupled QDs leads to finite inter-valley tunnel couplings, the presence of which complicates the initialization of charge-separated singlet states \cite{Cywinski26, King26} and the physics of readout via Pauli Spin Blockade \cite{Tagliaferri_PRB18,Cywinski26}. The presence of finite intra- and inter-valley tunnel couplings leads also to valley-dependence of exchange interaction of two spins in adjacent QDs \cite{Tariq_NPJQI22}, and it complicates the flopping-mode 
spin resonance control \cite{Teske_PRB23,Losert_arXiv25}.
Finally, while the shuttling through regions of low $\EVS$ is expected to lead to an enhancement of probability of valley excitation, the knowledge of the spatial dependence of the valley phase is necessary for making precise predictions for the valley excitation probability \cite{Langrock23,delima25_2,Volmer25}. It is thus a map of the complex $\Delta$ that is the key to understand exchange interaction, spin manipulation \cite{Teske_PRB23,Losert24,Losert_arXiv25} by electric dipole spin resonance, or to apply mitigation strategies for qubit shuttling ~\cite{Losert24,Pazhedath24, Nemeth24}.
}

\add{There are ongoing experimental and theoretical efforts aimed at obtaining a large \emph{and spatially uniform} value of $\EVS$ \cite{McJunkin21,Wuetz2022,Esposti23,Losert23,Klos2024, thayil25, thayil2025arXiv,Cvitkovich_arXiv26, rahlff_arXiv2026}. The current theoretical models for the valley coupling in Si/SiGe QDs focus on the influence of alloy disorder at realistically diffuse Si/SiGe interfaces  \cite{Wuetz2022,Losert23} and particular spatial distributions of Ge in the Si quantum well \cite{woods24,Cvitkovich_arXiv26}. A comparison of statistical properties of calculated and measured landscapes of $\EVS(x,y)$ over a large area of a Si/SiGe quantum well would allow for testing the correctness of these theoretical models applied to particular samples. Measured $\EVS$ maps \cite{volmer24} have so far shown qualitative agreement with theoretical predictions of alloy-disorder based theory, However, as argued recently in \cite{Woods25}, mapping of the valley phase (in addition to only $\EVS$) would allow for a more robust comparison of measured and calculated landscapes.  
A theoretical model connecting the spatial dependence of $g$-factors for the two valley states to the spatial dependence of the valley phase was put forth in \cite{Woods24_2,Woods25}. Reconstructed maps of $\EVS(x,y)$ and of valley-dependent $g$-factors can thus be used to verify that model, and in case of confirming its validity, as a viable way to reconstruct the $\Delta(x,y)$ landscape. The detailed knowledge of $\EVS$, electron $g$-factor and valley coupling phase will then provide crucial feedback to these recent theories, opening way to establishment of reliable models of variability of valley couplings and $g$-factors, allowing then for systematic optimization of the growth and strain-engineering of the Si/SiGe material \cite{Wuetz2022, marcogliese25}. Let us also remind that apart from this possibility of achieving full understanding on the relationship between microscopic features of the heterostructure and statistics of $\Delta$ and $g$-factors, having a map of these quantities in a given structure would allow the optimization of qubit control, operation, and shuttling.
}

To understand the full qubit dynamics related to valley, e.g. valley excitations during spin-qubit shuttling and manipulation, knowledge about the laterally varying intervalley coupling $\Delta$ is required~\cite{Langrock23,delima25_2,Volmer25}.
This complex value is related to the valley splitting and the valley-dependent electron g-factor~\cite{Woods25}. 
A map of the complex $\Delta$ is the key to understand exchange-interaction, spin manipulation \cite{Losert24} by electric dipole spin resonance, or to apply mitigation strategies for qubit shuttling~\cite{Losert24,Pazhedath24, Nemeth24}. 
The detailed knowledge of $\EVS$, electron g-factor and $\Delta$ provides valuable feedback to operate qubit chips, and to systematically optimize the growth and strain-engineering of the Si/SiGe material \cite{Wuetz2022, marcogliese25}.

Measurement techniques for measuring $\EVS$ have been progressing from single QD site~\cite{Borselli11,Dodson22, McJunkin21}, via multiple sites~\cite{Hollmann20,Chen21, Wuetz2022, Marcks25}  to full maps with  high spatial resolution obtained with conveyor-mode spin shuttling~\cite{volmer24,Volmer25}. 
In this work, we introduce the nanometer-resolution two-dimensional mapping of the local valley-dependent electron $g$-factors of a QD formed in a Si/SiGe heterostructure. Our method is based on the separation of an  entangled two-spin state by conveyor-belt spin shuttling~\cite{Struck23,desmet24, volmer24}. The spin singlet-triplet oscillations of a static reference QD and a mobile QD  probe the $g$-factor difference $\Delta g$ of the mobile QD shuttled to the position of interest compared to the reference QD. We measure a two-dimensional map of $\Delta g$-variations spanning a total area of $\SI{40}{\nano\meter}\times \SI{400}{\nano\meter}$. The map reveals a striking symmetry with respect to the valley states that is in good statistical agreement with theoretical expectations \cite{Woods24_2,Woods25}. We reconstruct lateral traces of the complex valley coupling $\Delta$ ~\cite{Woods25} by  combining the $g$-factor map with the corresponding two-dimensional valley splitting map measured on the same device. \add{In this way we show how one can obtain detailed knowledge of $\EVS$, the electron $g$-factor, and the alley coupling phase that will then provide valuable feedback to the operation of qubit chips, and to the systematical optimization of the growth and strain-engineering of the Si/SiGe material \cite{Wuetz2022, marcogliese25}.}

\section{Device and Method}  \label{sec:device}
\begin{figure*}
    \centering
    \includegraphics[width=\linewidth]{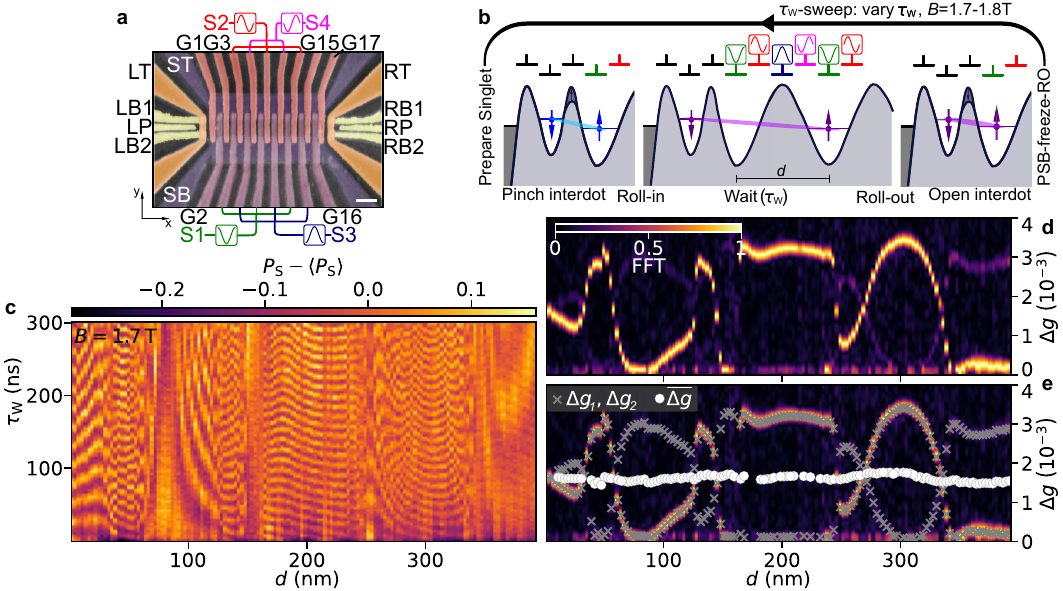}
    \caption{Sample and experimental method. (a) Scanning electron micrograph of the device used in this experiment. The scale bar corresponds to \SI{200}{\nano \meter}. (b) Schematic illustrating the experiments execution. (c) $\tau_\text{W}$-sweep: Raw data of a measurement where we shuttle the electron to position d and vary the wait time $\tau_\text{W}$ while keeping the magnetic field constant ($B = \SI{1.7}{\tesla}$). (d) Extracted $g$-factor variation by Fourier transformation of (c).(e) Version of (d) marked with found peaks and their average.}
    \label{fig1}
\end{figure*}
The shuttle device (Fig.~\ref{fig1}a), referred to as QuBus, is identical to the one used in Ref.~\cite{Volmer25} and described in detail in the Methods section. Two single electron transistors (SETs) located at the ends of a one-dimensional electron channel (1DEC) can be used for charge readout. The 1DEC along which we shuttle is formed in the QW by two \SI{1.2}{\micro \meter} long screening gates (SB, ST; purple).  On top 17 clavier gates (G1-G17) generate the periodic shuttle potential. Clavier gates are connected into four gate sets (S1-S4; dark magenta and bright red) ~\cite{Seidler22,Xue23,Struck23,volmer24}. \add{Further details on device design and fabrication as well as the experimental setup can be found in Appendix~\ref{app:Device_design_and_fabrication} and \ref{app:Experimental_setup}.}

\begin{figure*}
    \centering
    \includegraphics[width=\linewidth]{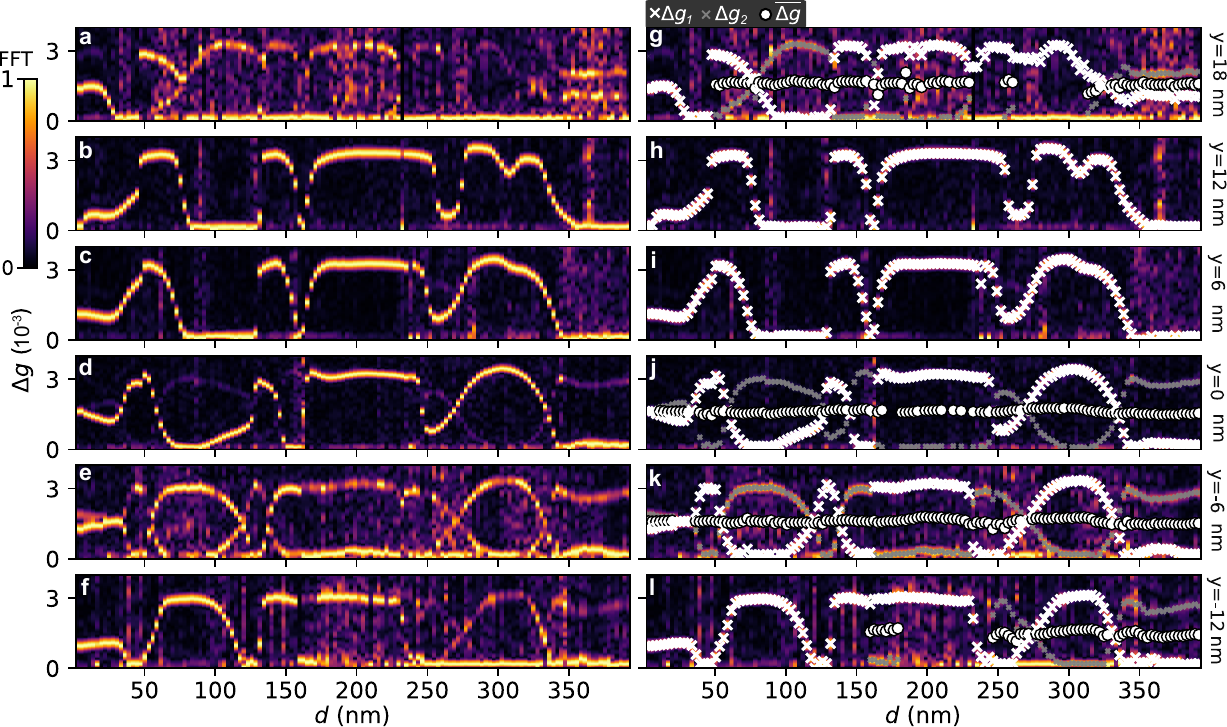}
    \caption{Extraction of the $\Delta g$ values. (a-f) Fourier transformation of the raw data from $y=\SI{18}{\nano \meter}$ to $y=\SI{-12}{\nano \meter}$ to extract the local $\Delta g$. (g-l) Fourier transformations with one to two marked peaks (white/grey crosses; white cross represents the dominant component) and their respective averages (white circles; only if two peaks are identified).}
    \label{fig2}
\end{figure*}

We map the electron $g$-factor by loading a spin singlet (S) state from the left SET in the first quantum dot (QD) of the shuttle (Fig.~\ref{fig1}b). Then we separate the S in two steps: First step is fast tunnelling of one electron into the second QD by a 1.2 ns detuning ramp (limited by the \SI{300}{\mega \hertz} bandwidth of our waveform generators). Second, after pinching the tunnel barrier within this double quantum dot (DQD), we shuttle the right electron in conveyor-mode with a velocity of \SI{5.6}{\meter \per \second} up to a distance $d$ determined by stopping the sinusoidal control signal $V_{\mathrm{S},i}(t)$ applied to the four shuttle gate sets S1-S4 \cite{Struck23}:
\begin{equation} \label{eq:signal}
    V_{\mathrm{S},i}(t) = A_i \cos\!\left( 2\pi f t - (i-1)\frac{\pi}{2} \right) + B_i ,
\end{equation}
where $f$ is the drive frequency, $A_i$ are the drive amplitudes ($\SI{150}{\milli\volt}$ for S1 and S3; $\SI{180}{\milli\volt}$ for S2 and S4), and $B_i$ are static offsets ($\SI{0.7}{\volt}$ for S1 and S3). At separation distance $d$, we wait for a time $\tau_\mathrm{W}$ in order to accumulate a singlet-triplet (ST) oscillation phase $\varphi_\text{W}$:
\begin{equation}\label{eq:phase}
    \varphi_\text{W}(B, \tau_\text{W})=\frac{\mu_B B}{\hbar} \cdot \Delta g(d) \cdot \tau_\text{W}, 
\end{equation}
where $B$, $\mu_B$ and $\hbar$ are the global inplane magnetic field, the Bohr magneton and reduced Planck's constant, respectively,  and $\Delta g(d)$ is the difference of the electron $g$-factors between the inert electron and the shuttled electron positioned at distance $d$ (thus $\mu_B \Delta g B$ is the difference in Zeeman energies between the two QDs). Then, we shuttle the electron back, and we reverse the separation process and open the interdot barrier of the DQD to detect the spin state of the entangled electrons in the S-T basis by Pauli-spin-blockade (PSB). Finally, we freeze the PSB charge state by raising the DQD tunnel barrier, and read out (RO) the charge state binning current levels across the left SET. We enhance the PSB contrast by operating the DQD in the (4,0)-(3,1) charge regime, effectively enhancing the singlet-triplet splitting. Thus, we load four electrons in total and three electrons remain in the inert outermost left QD, where two of the three electrons occupy one of the valley eigenstates forming a spin singlet, and thus can be neglected \cite{volmer24}.
The pulse sequence consists of up to 200 points of $\tau_\text{w}$, each repeated 800-times for each $d$ changed in the outermost loop. This provides us the singlet-return-probability $P_\textrm{S}(\tau_\text{w},d)$, from which we calculate the scanline $g(d)$, defined as one scanline. To measure a two-dimensional map of $\Delta g(d,y)$, we displace the 1DEC in $y$-direction by changing the voltages applied to the screening gates $V_\text{ST}$ and  $V_\text{SP}$ in opposite direction:
\begin{equation} \label{eq:displace}
    V_\text{ST, SB}=\SI{100}{\milli \volt}\pm \frac{\SI{50}{\milli \volt}}{\SI{6}{\nano\meter}} \cdot y, 
\end{equation}
where the $+/-$ refers to ST/SB gate, respectively. The factor in Eq. \ref{eq:displace} is experimentally determined from a localization scheme of various QD positions using the capacitance coupling of the QD to its surrounding gates in a shuttle device with similar gate geometry (cf. Supplements of Ref. \cite{volmer24}). The accuracy of setting spatial coordinates $(d,y)$ is mainly limited by electrostatic noise. As we observe reproducibility of the mapping within a few nanometers (cf. Ref. \cite{Volmer25}), the displacement due  to noise is small or at least static. Using an entangled state of the two spins to map the difference $\Delta g(d)$ with respect to the spin in a static QD yields a high accuracy of $g$-factor measurement that is below $10^{-3}$, and which  is ultimately limited by the dephasing time of the entangled spins. 

\section{Results} \label{sec:results}

\subsection{Measurement of the $\Delta g$ map}  \label{sec:measurements}

In order to extract $\Delta g(d)$ from a $P_\textrm{S}(d)$ measurement, we vary $\tau_\text{W}$ in Eq.~\ref{eq:phase}, while keeping the other parameters fixed. An exemplary measurement is shown in Fig.~\ref{fig1}c. As expected from Eq.~\ref{eq:phase}, we observe ST-oscillations as a function $\tau_\text{W}$ (Fig.~\ref{fig1}c). The frequency of these oscillations varies for each $d$, which confirms that $\Delta g$ is actually $d$-dependent. We choose $B = \SI{1.7}{\tesla}-\SI{1.8}{\tesla}$ (\SI{1.7}{\tesla} for $y=\SI{0}{\nano \meter}$, \SI{1.8}{\tesla} otherwise) to stay well above any of the spin-valley resonances that could cause decoherence during shuttling (cf. Ref. \cite{Volmer25}). Then, we transform the raw data column-wise by FFT and calculate $\Delta g(d)$ (Fig. \ref{fig1}d). Note that it is the modulus of $\Delta g(d)$ that is in fact measured, but for a clearer notation we define now $\Delta g(d)$ as positive. An alternative measurement scheme for which $B$ is varied while $\tau_\text{W}$ is kept fixed is discussed in Appendix~\ref{app:Magnetic_field_vs_wait_time}.

Interestingly, we observe the presence of two frequency components of $P_\textrm{S}(d)$ for most $d$. We use a peak-finder to mark them by grey crosses in Fig. \ref{fig1}e, and label them as $\Delta g_k(d)$, with $k\! =\! 1$ corresponding to the peak with larger FFT amplitude, i.e.~the ``dominant'' frequency. For more details on the peak identification, see Appendix~\ref{app:Extraction_of_the_dominant_component}. The symmetry of $\Delta g_1(d)$ and $\Delta g_2(d)$ is remarkable, \add{with the two frequencies appearing to be mirror images of each other with respect to $\Delta g \! \approx \! 1.5\cdot 10^{-3}$ line.} For each $d$, we therefore calculate the average $\overline{\Delta g}(d)=[\Delta g_1(d)+\Delta g_2(d)]/2$, marked by circles. Strikingly, this average $\overline{\Delta g}(d)$ is approximately $d$-independent. 

Next we check these remarkable observations by adding five more scanlines recorded at different $y$-positions in a range from  $y=$\SI{-12}{\nano\meter} to \SI{18}{\nano\meter} by varying the DC-voltages $V_\text{ST}$ and  $V_\text{SP}$ applied to the screening gates according to Eq. \ref{eq:displace} for each scanline. Applying FFT to the $P_\textrm{S}(d, \tau_\text{W})$  for each $d$ shown in Fig. \ref{fig2} a-f, we extract $\Delta g_{1(2)}(d,y)$, and calculate their average $\overline{\Delta g}(d,y)$, if two $\Delta g(d,y)$ components are visible. For the scanline at $y=\SI{-6}{\nano\meter}$, we clearly observe two frequency components while for scanlines at $y=\SI{18}{\nano\meter}$ and $y=\SI{-12}{\nano\meter}$ the second component is visible only at some $d$. Still, the calculated average $\overline{\Delta g}(d,y)$ is nearly constant as we have observed for the $y=\SI{0}{\nano\meter}$. The FFT intensity of the second component varies compared to the dominant one. For other values of $y$ only one component is visible.

\begin{figure*}
    \centering
    \includegraphics[width=\linewidth]{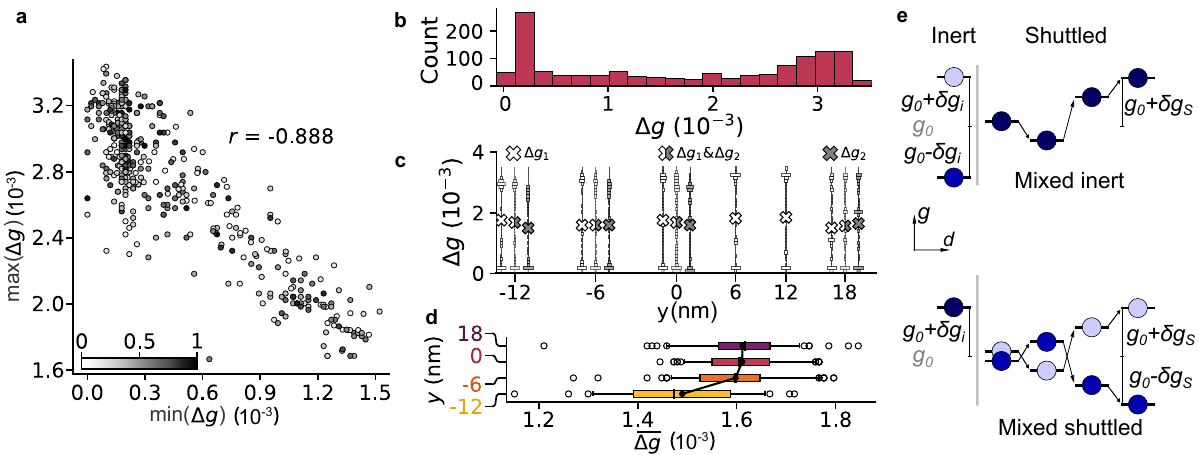}
    \caption{ Statistical analysis of the $\Delta g$ measurement. (a) Self correlation of the data points with two $\Delta g$-factor peaks. The circles are shaded by the relative strength of the secondary peak. The Pearson $r$ value is displayed. (b) Histogram of the extracted $\Delta g$ values. \add{(c) Statistics of the dominant ($\Delta g_1$) and subdominant component ($\Delta g_2$). The vertical histograms represent the distribution and average (cross) of $\Delta_1$ (open cross), $\Delta_2$ (grey cross) and both components (half-open cross) , for each scanline parametrized by $y$.} (d) Boxplots for the $\left< \overline{\Delta g}(d,y)\right>_d$ averaged over all $d$ for each scanline. The box spans from the first quartile to third quartile, the median is marked by a vertical line and the averages are marked by black points. The whiskers indicate the 5\% and 95\% mark with outliers indicated by circles. (e) Schematic of the origin of the two components $\Delta g_{1(2)}$ and $\Delta g_l$ and their symmetry. In the upper scenario only the shuttled electron is in a mixed valley state, and in the lower scenario only the inert (static) electrons are in a mixed valley state. The respective valley-dependent $g$-factors differ by $2 \delta g_s$ and $2 \delta g_i$, respectively.}
    \label{fig3}
\end{figure*}

To confirm the anti-correlation of the $\Delta g_1(d,y)$ and $\Delta g_2(d,y)$, we use all the picked values from Fig. \ref{fig2}g-l and plot all the pairs of values  in the point cloud in Fig. \ref{fig3}a, in which the $x$ coordinate corresponds to the smaller $\Delta g$ value in a given pair.
Indeed, in each pair the presence of a relatively large $\Delta g(d,y)$ corresponds to a relatively small second value of $\Delta g(d,y)$, and vice versa, with Pearson factor -0.9. The outliers from this anti-correlations dominantly exhibit a weak intensity of one component, which implies a relatively large error in the peak picking. 

Next, we investigate the distribution of all the $\Delta g(d,y)$ values from Fig. \ref{fig2}g-l in a histogram (Fig.~\ref{fig3}b).
Strikingly, we find a bimodal distribution with some values in the middle, while others are close to zero. This reflects the fact that the $\Delta g(d,y)$ values exhibit dominantly two values that differ by approximately $3\cdot 10^{-3}$. \add{We observe this bimodal distribution for each scanline $y$ and each component $\Delta g_i$ (Fig. \ref{fig3}c).} Finally, we analyse the distribution of $\overline{\Delta g}(d,y)$ for each scanline. Here we need to focus on the scanlines that exhibit two frequency components. The corresponding boxplot (Fig. \ref{fig3}d) reveals that the $d$-averaged $\left< \overline{\Delta g}(y)\right>_d$ is approximately equal for all the scanlines within $10^{-4}$ of the electron g-factor despite a slight trend. This is not surprising, as we expect that the g-factor map is a property of the Si/SiGe heterostructure, so it should not be influenced by shuttle directions (d, y coordinates), and hence averaging of $\Delta g(d,y)$ should be independent of direction in the limit of an infinite map. More than 50\% of the $\overline{\Delta g}$ values are within a range of $0.2 \cdot 10^{-3}$ around their mean value $\left< \overline{\Delta g}(y)\right>_d$.

\add{A summary of the above observations of spatial dependence of $\Delta g_{1,2}(d,y)$ is the following. The average of the two frequencies seen in the signal, $\overline{\Delta g}(d,y)$, is practically independent of $d$ and $y$ (``spatial uniformity of $\overline{\Delta g}$''). The probability density function of measured values of $\Delta g_{1,2}$ is bimodal, with maxima at values of $\Delta g$ close to the maximal and minimal measured values (``bimodality''). Furthermore, the data suggests that spatial averages of $\Delta g_1$ and $\Delta g_2$ are very close to $\overline{\Delta g}$, see Fig.~\ref{fig3}c for averages of $\Delta g_i$ over $d$, and compare with statistics of $\overline{\Delta g}$ shown in Fig.~\ref{fig3}d. This means that we have $\Delta g_{1,2}(d,y) \! \approx \! \overline{\Delta g} \pm f(d,y)$, where spatial average of function $f(d,y)$ is close to zero (precisely, it is $\ll \! \overline{\Delta g}$). 
We will refer to this last property as the ``mirror symmetry of deviations from $\overline{\Delta g}(d,y)$''.}

\subsection{Model for a $\Delta g$ map}  \label{sec:model}
Now we discuss the implications of these rich observations of the mapped $\Delta g_{1(2)}(d,y)$ and $\overline{\Delta g}(d,y)$ for symmetries and spatial dependence of valley-dependent $g$-factors the measured QDs.
The key for understanding lies in a valley occupation dependence of the $P_\textrm{S}(d,y)$ signal.  
\add{Presence of more than one frequency of this signal was reported previously \cite{Struck23}, and recently we have given a detailed account of this effect, showing that is stems from creation of statistical mixture of distinct valley occupation patterns in the two QDs \cite{Cywinski26}. This can happen during the initialization of the spatially separated singlet state from $(4,0)$ charge configuration \cite{Cywinski26}. The $(3,1)$ singlets created after charge separation, $\ket{S_{v_i,v_s}}$, are labeled by $v_{i(s)}$ valley states that are singly-occupied in the inert (shuttled) QD.
The creation of a statistical mixture of valley occupations in the shuttled dot can also occur during shuttling, due to a nonadiabatic passage through a region of low valley splitting and a strong spatial dependence of the phase of the intervalley coupling matrix element \cite{Langrock23,Volmer25}. Consequently, distinct $\Delta g(d)$ traces correspond to distinct pairs of $(v_i,v_s)$ indices.}

A fully general expression for the valley-dependent $g$-factor in inert/shuttled ($D\!=\! i/s$) QD is $g_{D,v}(\mathbf{r}_D) = g_0 + \delta g_{D,v_D}(\mathbf{r}_D)$, where $v_D=-$ ($+$) for ground (excited) valley eigenstate, \add{and $\mathbf{r}_D$ is the position of the inert/shuttled $D$.}
\add{The measured frequencies in $P_\textrm{S}(d)$ are proportional to the {\it modulus} of the $g$-factor difference between electron in valley $v_i$ in the inert QD and in valley $v_s$ in the shuttled QD: 
\begin{align}
\Delta g (\mathbf{r}_i,\mathbf{r}_s) &  = | \delta g_{i,v_i}(\mathbf{r}_i) - \delta g_{s,v_s}(\mathbf{r}_s)| \,\, , \\
& = \xi_i(\mathbf{r}_i) |\delta g_{i,v_i}(\mathbf{r}_i)| + \xi_s(\mathbf{r}_s) |\delta g_{s,v_s}(\mathbf{r}_s)| \,\, ,
\end{align}
where $\xi_D \! =\! \pm 1$, and we have $\xi_i\! =\! \xi_s\! =\! 1$ when the signs of $\delta g_{i,v_i}$ and $\delta g_{s,v_s}$ are opposite, and $\xi_i \! = \! \mathrm{sgn}(\delta g_{i,v_i} - \delta g_{s,v_s})$, $\xi_s \! =\! -\xi_i$, when these signs are the same.}

\add{ The $d$-dependence of $\Delta g$ in Fig.~\ref{fig2} follows from $d$-dependence of $\delta g_{s,v_s}(d,y)$. When considering two (note that up to four could be possible in theory) simultaneously visible $\Delta g$ traces, corresponding to $(v_i,v_s)$ and $(v'_i,v'_s)$ valley occupations, and  $(\xi_i,\xi_s)$ and $(\xi'_i,\xi'_s)$ indices, a necessary condition for the spatial uniformity of $\overline{\Delta g}$ is to have $\xi_s \! =\! -\xi_s'$ and $|\delta g_{s,v_s}|\! =\! |\delta g_{s,v'_s}|.$ This condition is sketched in Fig.~\ref{fig3}e for the shuttled QD: In the upper part, we have $v_s \! =\! v'_s$, while in the lower part two valleys are occupied in the shuttled dot, but $\delta g_{s,v_s}$ have opposite signs but equal moduli.}

\add{The first scenario in which this holds is that of having statistical mixture of valley occupations only in the inert QD: $v_s\! =\! v_{s'}$ and $v_i\! \neq \! v'_i$, while  $\xi_s \! =\! -\xi_s'$ at all $d$ at which the two frequencies are visible. We have then $\overline{\Delta g} \! = (\Delta g_1+ \Delta g_2)/2\! =\! (\xi_i |\delta g_{i,v_i}| + \xi'_i|\delta g_{i,v'_i}|)/2$. Furthermore, in order to recover the ``mirror symmetry of deviations from $\overline{\Delta g}$'' we need $\xi_i \! =\! \xi'_i$ and $\delta g_{i,+} \! = \! -\delta g_{i,-} \! \equiv \! -\delta g_i$, which results in $\overline{\Delta g} \! =\! |\delta g_i|$, and $\Delta g_{1(2)} \! = \! \overline{\Delta g_i} \pm |\delta g_{s,v_s}(d)|$, where $\delta g_{s,v_s}(d)$ is now a random function with zero spatial average. The necessary values of all the $\xi$ coefficients are obtained if we assume that $|\delta g_i| \! > \! |\delta g_s(d)|$. The valley occupation pattern for both dots in this scenario is illustrated in the top part of Fig.~\ref{fig3}e. }

\begin{figure*}
        \includegraphics[width=\textwidth]{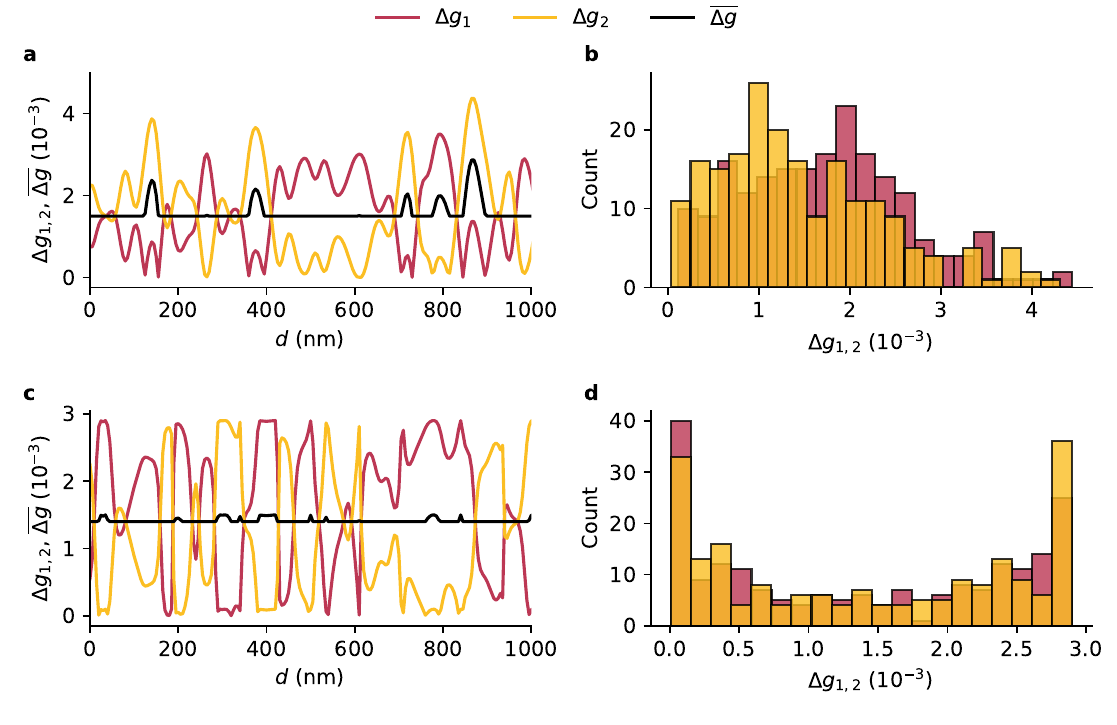}
    \caption{\add{Simulations of spatially dependent $\Delta g$ in the scenario from the upper part of Fig.~\ref{fig3}e (mixed valley occupation in the inert QD). In the first row we assume that $\delta g_s(d)$ in the shuttled QD is normally distributed with zero mean and a standard deviation $1.5 \cdot 10^{-3}$, with a Gaussian autocorrelation function having autocorrelation length of $20$ nm. The values of $\delta g_i$ are taken to be $\pm 1.5\cdot 10^{-3}$. This choice results in approximate spatial uniformity of $\overline{\Delta g}(d)$ and mirror symmetry of deviations of $\Delta g_{1,2}$ from this value, both visible in (a). In (b) we show histograms of values of $\Delta g_1$ and $\Delta g_2$ taken from 200 points across $10$ microns. In the second row we use the model of a valley-dependent $g$-factor disorder based on theory of random valley coupling $\Delta(d) \! =\! \frac{1}{2}\EVS(d)e^{i\phi(d)}$ from \cite{Losert23}, and of dependence of $g(d)$ on $\Delta(d)$ from \cite{Woods24_2,Woods25}: $g(d) = g_0 + \delta g_{\mathrm{max}} \cos[\phi(d)]$. We use $\delta g_{\mathrm{max}}\! =\! 1.5\cdot 10^{-3}$ and  assume $\delta g_i \! = \pm 1.4\cdot 10^{-3}$, i.e.~$|\delta g_{i}|$ is slightly smaller than $\delta g_{\mathrm{max}}$, resulting in appearance of deviations of $\overline{Delta g}(d)$ from the constant value given by $|\delta g_i|$. Note that within this model the assumption that for most $d$ we have $|\delta g_i| \! \geq \! |\delta g_s(d)|$ is much more natural than in the model used in panels a) and b) in which $\delta g$ is normally distributed. In the simulation of random $\Delta(d)$ trajectory we assume that the average $\Delta$ is zero, and the standard deviation of $\EVS$ is $50$ $\mu$eV. In (c) we see $\Delta g_{1,2}$ again approximately showing spatial uniformity of $\overline{\Delta g}(d)$ and mirror symmetry of deviations of $\Delta g_{1,2}$, but with presence of characteristic plateaus $\Delta g_{1,2}$ at the limits of its variability range. In (d) we see a bimodal distribution of $\Delta g_{1,2}$ values.}
    }
    \label{fig4}
\end{figure*}

\add{Let us assume for a moment  $\delta g_s(d)$ to be a Gaussian random field with zero mean. In principle, such a scenario could be generated if the g-factor variation is dominated by strong variations of the QD confinement during shuttling due to electrostatic disorder and a $g$-factor dependence on the QD confinement, as observed for materials with high spin-orbit interaction. In order to stay close to our observations, we take a standard deviation of $\delta g_s(d)$ given by $1.5\cdot 10^{-3}$, and correlation length of the order of lateral QD size. Then we simulate the $\Delta g_{1(2)}$ trajectories shown in Fig.~\ref{fig4}a, with the resulting histogram of values of $\Delta g$ (taken at positions separated by $\Delta d \! = \! 50$\,nm across $10$\,$\mu$m range of $d$) shown in Fig.~\ref{fig4}b. The observed bimodal structure of distribution of $\Delta g$ is clearly absent in this simulation. Furthermore, in order to obtain this result we need to assume that $|\delta g_i|$ are taken from the tails of the considered distribution of $\delta g$. In order to improve the agreement with observations, we need to assume a bimodal shape of the distribution of $\delta g$, with maxima of this distribution being close to the extreme positive and negative values that $\delta g$ can possibly take. This will not only lead to a bimodal distribution of $\Delta g$, but it will also make the necessary tuning of $\delta g_i$ value more natural, as the probability of having ``outlier'' values for $\delta g_i$ with respect to values of $\delta g_{s,v_s}(d)$  in the considered range is much higher than in the case of normal distribution of $\delta g$.}

\add{
However, this scenario does not explain the persistence of spatial uniformity of $\overline{\Delta g}$ in measurements in which the second $\Delta g$ component appears at finite $d$, see e.g.~Fig.~\ref{fig2}a at $d \! \approx \! 50$ nm, or when the dominant frequency switches between two values as in Fig.~\ref{fig2}f at $d \! \approx \! 160$ nm
and $\approx \! 180$ nm. These effects at finite $d$ happen due to creation of finite occupations of two valleys in the shuttled QD, or a change in the ratio of two pre-existing occupations in it. This leads us to consideration of the second scenario in which the spatial uniformity of $\overline{\Delta g}$ is guaranteed: that of $\delta g_{s,-}(d) \! =\! - \delta g_{s,+}(d) \! \equiv \! -\delta g_s(d)$, while $\xi_s\! =\! -\xi'_s$. Consistency with the first scenario leads us to assume $\delta g_{i,-} \! = \! -\delta g_{i,+} \! \equiv \! \delta g_i$. Note that after requesting that $\delta g_{s,-}\!= \! -\delta g_{s,+}$ this is a rather natural assumption, as we do not expect the inert and shuttled QDs to be qualitatively different. Furthermore, we expect some variability of parameters of the shuttled QD as function of $d$ due to electrostatic disorder \cite{Langrock23}, so a global assumption of $\delta g_{s,-}\!= \! -\delta g_{s,+}$ implies that this property should be rather robust, and should apply as well to the inert QD. With these assumptions we see that the case of $v_i \! \neq v_i'$ for $(v_i,v_s)$ and $(v'_i,v'_s)$ valley occupations is ruled out, as it leads to $\Delta g_{1} \! =\! \Delta g_2 \! = \! |\delta g_i - \delta g_s|$. We are left only with the case in which for the two visible $\Delta g_{1,2}$ the valley state in the inert QD is well defined, and we have a finite probability of occupation of the two valley states in the shuttled QD. This valley occupation pattern is illustrated in the lower part of Fig.~\ref{fig3}e. 
We obtain again that when $|\delta g_i| \! > \! |\delta g_s(d)|$, we have $\overline{\Delta g} \! =\! |\delta g_i|$, and $\Delta g_{1(2)} \! =\!  \overline{\Delta g} \pm |\delta g_s(d)|$. We thus  end up in the same situation as in the first scenario: in order to explain the bimodality of $\Delta g$ distribution we need a bimodal distribution of $\delta g_{i/s}$ for both QDs, independent of their placement. Such a distribution will also make the fulfillment of $|\delta g_i| \! > \! |\delta g_s(d)|$ more probable, when considering a finite range of $d$. 
}

\add{We have thus arrived at the conclusion that the necessary condition to explain all of the observations is the following form of the valley dependence of the $g$-factor for QDs in the investigated Si/SiGe heterostructure:
\begin{equation}
   g_{D,\pm} \! =\! g_0 \pm \delta g_D \,\, , \label{eq:gpm} 
\end{equation}
and we need the probability density function of $\delta g_D$ to have zero mean, and a bimodal structure, in order to explain the observation of the bimodal distribution of the measured values of $\Delta g$. }

\add{
After spending some time deducing the properties of valley dependence of $g$-factors  from the experimental data, let us turn to theoretical models available in literature. The valley dependence of electron $g$-factor of the form given in Eq.~(\ref{eq:gpm}) was obtained in \cite{Ruskov_PRB18} for SiMOS QDs with large $\EVS$ and in presence of strong electric fields in the growth direction, and, very recently, for Si/SiGe QDs under generic assumptions \cite{Woods24_2,Woods25}. Crucially, according to the latter theoretical works we have 
\begin{equation}
g_{\pm} = g_0 \pm \delta g_{\mathrm{max}} \cos(\phi) \,\, , \label{eq:g}
\end{equation}
in which $\delta g_{\mathrm{max}}$ depends on direction of $B$ field and QD geometry, while $\phi$ is the phase of the valley coupling matrix element $\Delta\! \equiv \! |\Delta|e^{i\phi}$.  For valley coupling in the ``disordered'' regime \cite{Losert23}, which was identified as holding in the measured samples \cite{Volmer25}, the distribution of $\phi$ is flat, indeed leading to a bimodal distribution of $c\! \equiv \! \cos\phi$ given by $p(c) \! =\! 1/(\pi\sqrt{1-c^2})$. In Fig.~\ref{fig4}c we show simulated spatial dependence of $\Delta g(d)$ within this model, and the resulting bimodal histogram of values of $\Delta g$ is shown in Fig.~\ref{fig4}d.}

\begin{figure*}[t]
    \centering
    \includegraphics[width=\linewidth]{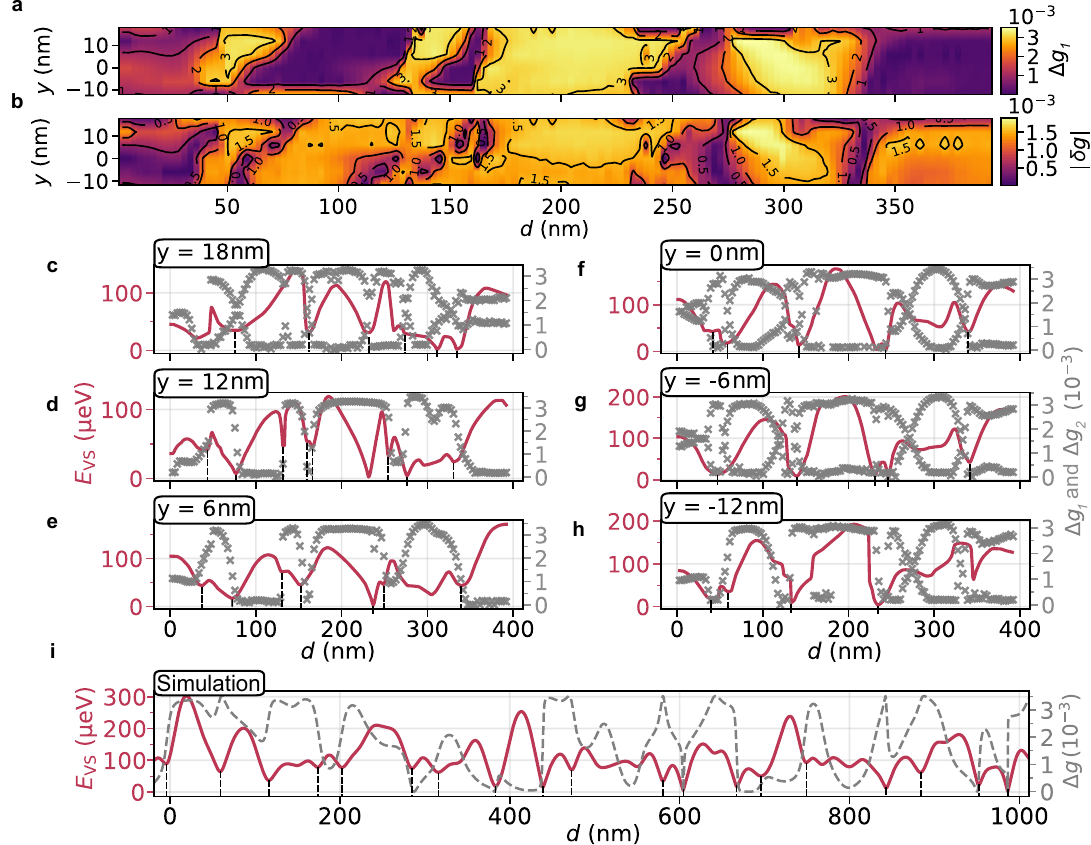}
    \caption{Comparison of a g-factor map to its corresponding valley splitting map. (a) Map of the absolute value of $|\delta g_s|$ as a function of $d$ and $y$. (b) Map of the dominant component $\Delta g_1$ as a function of $d$ and $y$. (c-h) Direct comparison of valley splitting scanlines from~\cite{Volmer25} against the extracted $\Delta g$-factors. (i) Simulation of a valley splitting and a $\Delta g$ trace.}
    \label{fig5}
\end{figure*}

\subsection{Juxtaposition of the $\Delta g$ map against the $E_\text{VS}$ map} \label{sec:juxtaposition}
\add{After arguing for applicability of the $g$-factor model from \cite{Woods24_2,Woods25} to the observations in the investigated Si/SiGe structure, we can use this model to translate the observations of $\Delta g_{1,2}(d,y)$ into maps of $g$-factor of the shuttled QD.}
In Fig.~\ref{fig5} we show the dominant $\Delta g_1(d,y)$ component (Fig. \ref{fig5}a), and the valley-induced difference from the mean value, $|\delta g(d,y)|\!= \! |\Delta g_1(d,y)- \overline{\Delta g}(d,y)|=|\Delta g_2(d,y)- \overline{\Delta g}(d,y)|$ (Fig. \ref{fig5}b). Both maps are interpolated from a $(d,y)$ mesh with a resolution smaller than the correlation length. A clear identification of the dominant of the two components is difficult at a few positions of the map, and can lead to sudden switches between low and high value in the map. These jumps are related to a change of the valley state occupation during shuttling across a certain region. Therefore, we manually adapted the choice of the dominant component in a few spots (see Appendix~\ref{app:Extraction_of_the_dominant_component}, white crosses in Fig.~\ref{fig2}g-l). The $|\delta g(d,y)|$ representation (Fig. \ref{fig5}b) does not suffer from this problem, as it is independent of the identification of the dominant component. However, it requires $\overline{\Delta g}(d,y)$, which we either calculate in the presence of two components at the same $(d,y)$, or extrapolate from the averaged $\overline{\Delta g}(d,y)$ using the fact that $\overline{\Delta g}(d,y)$ is nearly constant (Fig. \ref{fig3}c). The two maps reveal interesting similarities: There are patches of about constant $\Delta g_1(d,y)$ and $|\delta g(d,y)|$ interrupted by transition zones. The sizes of the $\Delta g_1(d,y)$ patches sometimes exceed the QD size, and $\Delta g_1(d,y)$ within them has alternating low and high values, reflecting the bimodal distribution of the $g$-factor values.

We compare this result with the corresponding valley splitting ($\EVS$) map measured in the exact same device, and in the same cool-down cycle \cite{Volmer25}. In Figs.~\ref{fig5}c-h, we compare $\EVS{}(d,y)$ with $\Delta g_{1(2)}(d,y)$ for each of the six scanlines. The $\EVS$ values are Rician distributed with QD size as correlation length, as we analysed previously \cite{Volmer25}. We find a remarkable connection between the $\Delta g_{1(2)}(d,y)$ components and the $\EVS(d,y)$: The transition zones between the extremal $\Delta g(d,y)$ values (e.g. vertical black dashed lines in Figs.~\ref{fig5}c-h) are related to deep $\EVS(d,y)$ minima, with some exceptions, e.g.~at $(d,y)=(130,0)$\,\SI{}{\nano\meter}. Reversely, not every $\EVS(d,y)$ minimum leads to a transition, see e.g.~ $(d,y)=(232,12)$\,$\SI{}{\nano\meter}$.

To understand this observation, we recall that $\EVS$ for a QD located at $d$ is given by $\EVS(d) \! =\! 2|\Delta(d)|$, where $\Delta(d)$ is the inter-valley coupling matrix element from above. According to the theoretical model \cite{Wuetz2022,Losert23} of spatial randomness of $\Delta(d)$ due to atomic disorder at Si/SiGe interface, real and imaginary parts of $\Delta(d)$ are independent Gaussian random fields with autocorrelation length given by the linear QD size. Measurements of statistics of $\EVS(d)$ in the samples investigated here have shown \cite{Volmer25} that the valley splitting is in the ``disordered'' regime \cite{Losert23}, in which the mean values of both  real and imaginary parts of $\Delta(d)$ are close to zero. Our simulations of trajectories of $\Delta(d)$ based on this model reveal that deep local minima of $\EVS(d)$ at $d\! =\! d_m$ are associated with $\Delta(d) \! \equiv \! |\Delta(d)|e^{i\phi(d)}$  trajectories ($d\in [d_i,d_f]$, where $d_i$ and $d_f$ are the initial and final points of the relevant trajectory located on the two sides of $\EVS(d)$ minimum) passing near the origin of the complex plane, with the minimal distance to the origin attained at $d\! =\! d_m$.  
The change of the valley phase along such a trajectory, $\delta\phi \equiv |\phi(d_i)-\phi(d_f)|$, is typically $>1$. When $\delta \phi \! \approx \! \pi$, upon passage through such a $\EVS(d)$ minimum, the $g$-factor given by Eq.~(\ref{eq:g}) will experience a rapid change by $\approx 2|\delta g_{\mathrm{max}}|$, provided that $\cos(\phi(d_i)) \! \approx \pm 1$. Hence, in the above model of dependence of $g$-factor on $\Delta(d)$, the presence of a deep $\EVS(d)$ minimum at $d_m$  implies with sizable probability the presence of a maximum-amplitude $g$-factor change in the vicinity of $d_m$, see Fig.~\ref{fig5}i as an illustrative simulation with deep local minima marked of $\EVS$ by black dashed vertical lines. This behaviour is similar to the one observed in Fig.~\ref{fig5}c-h. 

\subsection{Reconstruction of complex intervalley coupling} \label{sec:reconstruction}

\begin{figure*}[t]
    \centering
    \includegraphics[width=\linewidth]{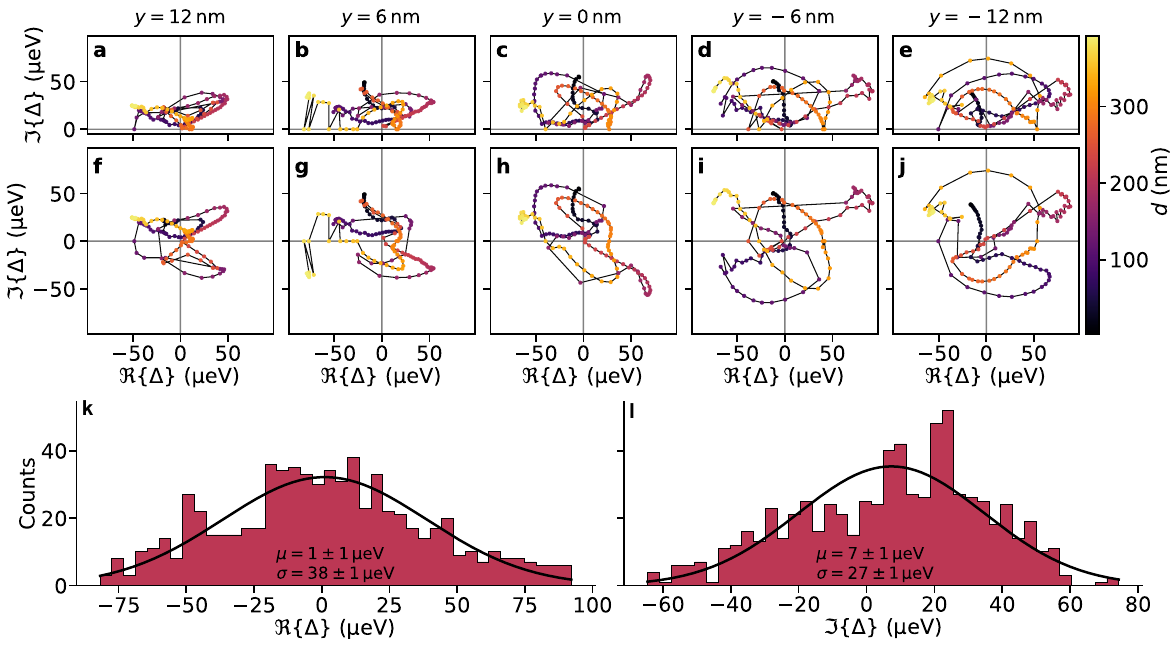}
    \caption{Reconstruction of the intervalley coupling $\Delta$. (a-e) Extracted intervalley coupling from the $g$-factor and $E_\text{VS}$ measurements from $y=\SI{12}{\nano \meter}$ to $y=\SI{-12}{\nano \meter}$. (f-g) Corresponding reconstructions of the full trace of the intervalley coupling. (i-j) Histogramm of the real and imaginary part of the all the intervalley couplings $\Delta$, which we calculated together with a least-square fit Gaussian function (black solid line). The mean $\mu$ and the variance $\sigma$ of the fits are given.}
    \label{fig6}
\end{figure*}

\add{The theoretical model of valley-dependent $g$-factor controlled by phase $\phi$ of the inter-valley coupling, $g_{\pm}(d,y) = g_0 \pm \delta g_{\mathrm{max}} \cos[\phi(d,y)]$, described in \cite{Woods24_2,Woods25}, explains in a natural way (i.e.~with no fine-tuning of parameters) the following properties of the measured $\Delta g_{1,2}(d,y)$ components: (1) the spatial uniformity of $\overline{\Delta g}$; (2) both the bimodality of distribution of $\Delta g_{1,2}$, and the fact that the two peaks of this distribution are at $\Delta g \! \approx \! \pm \delta g_{\mathrm{max}}$; (3) mirror symmetry of deviations from $\overline{\Delta g}$, and (4) correlation between positions of minima of $\EVS$ and maximal-amplitude changes of $\Delta g$. 
Note that (2) requires $\Delta$ values to have a cylindrically symmetric distribution around $0$, which means that intervalley coupling (and thus the valley splittings) in our sample is in the ``disordered'' regime \cite{Losert23}. While our fits of measured $\EVS$ distribution of values to the Rice distribution are consistent with this regime, the observation of bimodal distribution of $\Delta g$ stregthens this assignment. }

\add{Taking into account this level of agreement between theory and measurements concerning simultaneously all these nontrivial features, we take now this model as the correct description of physics of $g$-factor in our structure. Using the above-discussed results of $\delta g(d,y)$ to reconstruct the phase of $\Delta(d,y)$, and $\EVS(d,y) \! =\! 2|\Delta(d,y)|$ to obtain the modulus of this coupling, we can now set out to reconstruct the spatial dependencece of the complex valley coupling matrix element $\Delta\! \equiv \! |\Delta|e^{i\phi}$ with nanometer resolution.}
We demonstrate this ability for the scanlines for which we can identify one dominant component ($y=\SI{-12}{\nano \meter}$ to $y=\SI{12}{\nano \meter}$, see Fig.~\ref{fig2}) and renormalize each of them from -1 to 1. Taking the inverse cosine, we extract $\phi$ within a one $\pi$ range and compute $\Delta=\EVS e^{i\phi(d)}/2 $ (Fig.~\ref{fig6}a-e). \add{Inverting the $\cos(\phi)$ yields two possible solutions for $\phi$ and thus $\Delta$, mirrored with respect to the real axis. Through continuity analysis, we can make an educated guess, which of the two solutions is the correct one, based on the extracted $\Delta$ values before and after each point. Our reconstruction algorithm is described in detail in Appendix~\ref{app:Explanation_of_the_reconstruction_algorithm}.} 

We observe mostly smooth variations across the complex plane. The trace at $y=\SI{-6}{\nano \meter}$ exhibits one large jump, corresponding to a jump in the state population from the $\Delta g$ measurements in Fig.~\ref{fig2}k. The neighbouring $\Delta g$ in Fig.~\ref{fig2} measurements suggest that this fast change in $\Delta g$ is reasonable. Moreover, we observe that the reconstructed $\phi$ tend to cluster for values close, but not equal,  to $0$ and $\pi$. This can be explained by deviations from the theoretically predicted distribution of $g$-factor values which is bimodal with sharp cutoffs at the end. In reality, these cutoffs are smoothed out by crosstalk of the shuttle pulse on the static electron $g$-factor and other sources of noise affecting the identified value of $\Delta g$. Transforming this smoothed-out distribution of $\Delta g$ by the inverse cosine, we get accumulation of $\phi$ values depending on the exact parameters of the smoothing. \add{In Appendix~\ref{app:valley} we further discuss this effect and use simulations to visualize it.}

In Fig.~\ref{fig5}c-h, we observe some positions at which the trace seems to be reflected from the real axis. This is due to the fact that the valley phase is distributed on $[0,2\pi)$ but the inverse cosine only projects on $[0,\pi]$. Therefore, we reconstruct the intervalley coupling traces across the full complex plane  by applying a cost function based reconstruction algorithm (Fig.~\ref{fig6}f-j). More details on the reconstruction algorithm can be found in Appendix~\ref{app:Explanation_of_the_reconstruction_algorithm}. For each trace, the occupation of quadrants is close to being balanced. The relative rate of change of $\Delta$ can be estimated by the density of the dots (QD distance corresponds to \SI{2}{\nano \meter} shuttle distance). For every trace, we observe dots close to the origin due to the fact that $\EVS$ drops close to zero for every trace. We rarely find spots where the valley phase has a large gradient  while the $\EVS$ is large (most visible in Fig.~\ref{fig6}d,i). This implies that valley excitation due to shuttling \cite{Volmer25} at high $\EVS$ is rare. The histograms of the real and imaginary part of the intervalley coupling $\Delta$ show distributions centered around zero. This implies that $\Delta$ is dominated by alloy disorder and far away from a regime of a deterministically large $\EVS$ (Fig.~\ref{fig6}k,l). This is consistent with the mesured Rayleigh distribution of valley splittings ~\cite{Volmer25}.

\section{Discussion}
In this work, we have mapped the g-factor of a QD placed in an area of \SI{40}{\nano \meter} by \SI{400}{\nano \meter} with respect to a reference $g$-factor in a static QD. In the experiment we have measured the singlet-triplet oscillations of a spin-entangled electron spin pair that was spatially separated  by conveyor-mode shuttling. We have observed appearance of two oscillation frequencies that exhibit  a striking symmetry across the map, \add{and the distribution of which is visibly bimodal}.
 \add{Both of these features are following without any fine-tuning from recent}  theory~\cite{Woods24_2,Woods25}. 
 We compare the $g$-factor map with a map of valley splittings recorded with the same shuttle device. \add{The observed correlation between locations of valley splitting minima and large-amplitude changes of the $g$-factor, give further support to the model from \cite{Woods24_2,Woods25}.}
 \add{Using the valley-splitting map and this model that relates the electron $g$-factor with the phase of the intervalley coupling,} we reconstruct for the first time the complex intervalley coupling with high lateral resolution. 
\add{Knowledge of the spatial dependence of this coupling will aid in optimization of many features of spin qubits. In the context of qubit shuttling, it will allow for  precise calculation of the probability of valley excitations close to local minima of valley splitting, leading to a more precise analysis of their relevance than given in \cite{Volmer25}, where only the spatial dependence of $\EVS$ was available.}

\add{We conclude that our measurements of the statistics of $\EVS$ (which are consistent with Rice distribution characterized by zero average value of complex intervalley coupling), the symmetry of simultaneously visible $g$-factors corresponding to two valleys, and the bimodality of  distribution of these $g$-factors, give strong support to the recently elucidated theoretical picture of $\EVS$ and $g$-factor disorder in Si/SiGe structures \cite{Losert23,Woods24_2,Woods25}.}

Our new metrology provides valuable new insights in the theory of intervalley coupling, and allows for prediction of the valley dynamics of electrons in QD across the mapped area, which is important for high fidelity qubit manipulation. For scale-up of Si/SiGe quantum chips, a finite minimum of the valley splitting of at least a couple of tens of \si{\micro \electronvolt} and a relatively smooth valley phase is desirable across the whole area of the device. There are many proposals to engineer the Si/SiGe heterostructure towards this goal \cite{McJunkin21,Wuetz2022,Esposti23,Losert23,Klos2024, thayil25,thayil2025arXiv,marcogliese25,rahlff_arXiv2026,kanaar26}. Our mapping method will help to validate the progress, and reveal insights into the dominant enhancement mechanisms.

\section*{Data availability}
The data that support the findings of this study are available in the \href{https://doi.org/10.5281/zenodo.18792771}{Zenodo repository}.

\section*{Acknowledgements}
We acknowledge valuable discussions with Merritt P. Losert and Mark Friesen and the support of the Dresden High Magnetic Field Laboratory (HLD) at the Helmholtz-Zentrum Dresden - Rossendorf (HZDR), member of the European Magnetic Field Laboratory (EMFL). This work was funded by the German Research Foundation (DFG) within the project 289786932 (SCHR 1404/2-2) and under Germany's Excellence Strategy - Cluster of Excellence Matter and Light for Quantum Computing" (ML4Q) EXC 2004/2 - 390534769, and  by the European Union’s Horizon Research and Innovation Actions under Grant Agreement No. 101174557 (QLSI2).
The device fabrication has been done at HNF - Helmholtz Nano Facility, Research Center Juelich GmbH \cite{Albrecht17}. This research was sponsored in part by The Netherlands Ministry of Defence under Awards No. QuBits R23/009. The views, conclusions, and recommendations contained in this document are those of the authors and are not necessarily endorsed nor should they be interpreted as representing the official policies,
either expressed or implied, of The Netherlands Ministry of Defence. The Netherlands Ministry of Defence is authorized to reproduce and distribute reprints for Government purposes notwithstanding any copyright notation herein.

\section*{Author contributions}
M.V. set up and conducted the experiments assisted by T.S. Authors M.V.,   {\L}.C. and L.R.S. analyzed the data. The heterostructure was grown by D.D.E. and G.S. Device fabrication was done by J.T., and S.T and device pre-screening by A.S. Theory was developed by {\L}.C. with help of L.R.S. and M.V. The experiment was designed and supervised by L.R.S. L.R.S. and H.B. provided guidance to all authors. M.V., {\L}.C. and L.R.S. wrote the manuscript which was commented on by all other authors.

\section*{Competing interests}
H.B., L.R.S., T.S. and M.V. are co-inventors of patent applications that cover conveyor-mode shuttling and/or its applications. G.S. is founding advisor of Groove Quantum BV and
declares equity interests. L.R.S. and H.B. are founders and shareholders of ARQUE Systems GmbH. The other authors declare no competing interest.

\begin{appendix}
\section{Device design and fabrication}\label{app:Device_design_and_fabrication}

The device is fabricated on an isotopically enriched Si$_{0.7}$Ge$_{0.3}$/$^{28}$Si/Si$_{0.7}$Ge$_{0.3}$ heterostructure. Instead of employing an epitaxial silicon cap layer, the heterostructure is grown with an amorphous silicon-rich passivation layer. A \SI{5}{\nano\meter} thick $^{28}$Si quantum well is grown on top of a \SI{30}{\nano\meter} thick Si$_{0.7}$Ge$_{0.3}$ buffer. The small quantum well thickness is expected to increase the overlap of the electron wave function with the surrounding barriers considerably~\cite{paqueletwuetz23,Volmer25}. As the variation of the local $g$-factor mainly depends on the overlap of the wave function with the germanium atoms in the SiGe barrier, we therefore expect a large $g$-factor variation compared to works with thicker quantum wells~\cite{Struck23,volmer24}. This quantum well is grown on top of a \SI{2.4}{\micro\meter} thick Si$_{0.7}$Ge$_{0.3}$ buffer step, followed by a step step-graded Si$_{1-x}$Ge$_x$ buffer with three \SI{1}{\micro\meter} large steps, of which the lowest is pure silicon.

Ohmic contacts are fabricated by phosphorus ion implantation at the Helmholtz-Center Dresden-Rossendorf, followed by thermal activation at $730~^\circ\mathrm{C}$ for \SI{30}{\second}. The three metal gate layers are fabricated at the Helmholtz Nano Facility~\cite{Albrecht17} and are made of a Ti/Pt e-beam-evaporated stack~\cite{Volmer21} with a Ti thickness of \SI{5}{\nano\meter} and Pt thicknesses of \SI{15}{\nano\meter}, \SI{22}{\nano\meter}, and \SI{29}{\nano\meter} for L1, L2, and L3, respectively. The gates are patterned using \SI{100}{\kilo\electronvolt} electron-beam lithography and a metal lift-off process. They are electrically isolated from each other and from the SiGe heterostructure by three \SI{10}{\nano\meter} amorphous Al$_2$O$_3$ layers fabricated by atomic layer deposition, such that the gate stack is fully enclosed in amorphous Al$_2$O$_3$.

On top of this heterostructure, we define a gate-defined structure comprising three metal layers (L1, L2, and L3; see Fig.~\ref{fig1}a and Fig.~1b). Each charge sensor is formed by three finger gates (L1: LP, RP; L2: LB1, LB2, RB1, RB2; shown in yellow in Fig.~\ref{fig1}a) and a top gate (L3: LT, RT; orange) used to accumulate a single-electron transistor (SET) that serves as a charge sensor. The one-dimensional electron channel (1DEC) is defined between two screening gates (L1: SB, ST; purple) by 17 finger gates (L2: G2, S1, S3, G16; L3: G1, G3, S2, S4, G15, G17; dark magenta and bright red). The S1–S4 gates are cross-wired groups of two to three finger gates each, providing additional flexibility for shaping the confinement potential.

\section{Experimental setup}\label{app:Experimental_setup}
At room temperature, DC voltages are supplied by custom low‑noise digital‑to‑analog converters, and pulsed voltages on the RF lines are generated by Zürich Instruments HDAWG arbitrary‑waveform generators. For each high‑frequency gate, these voltage pulses are also summed onto the corresponding DC lines to suppress transients during long measurements. All gates are connected through low‑pass‑filtered loom lines (DC lines; cutoff \SI{10}{\kilo\hertz}). Signals on G2, G3, and S1--S4 are routed inside the refrigerator to the sample PCB via coaxial cables with a bandwidth of \SI{20}{\giga\hertz} (RF lines) and are combined with the DC lines using RC bias tees with a cutoff frequency of \SI{5}{\hertz}. The device is mounted on the mixing‑chamber stage of an Oxford Instruments dry dilution refrigerator at approximately \SI{60}{\milli\kelvin}, and a superconducting magnet provides a global in‑plane magnetic field along the 1DEC and parallel to the (110) crystallographic direction. The SET current is read out using Basel Precision Instruments transimpedance amplifiers and AlazarTech digitizer cards.

\section{Magnetic field vs. wait time sweep for $g$-factor extraction}\label{app:Magnetic_field_vs_wait_time}

\begin{figure*}
    \centering
    \includegraphics[width=\linewidth]{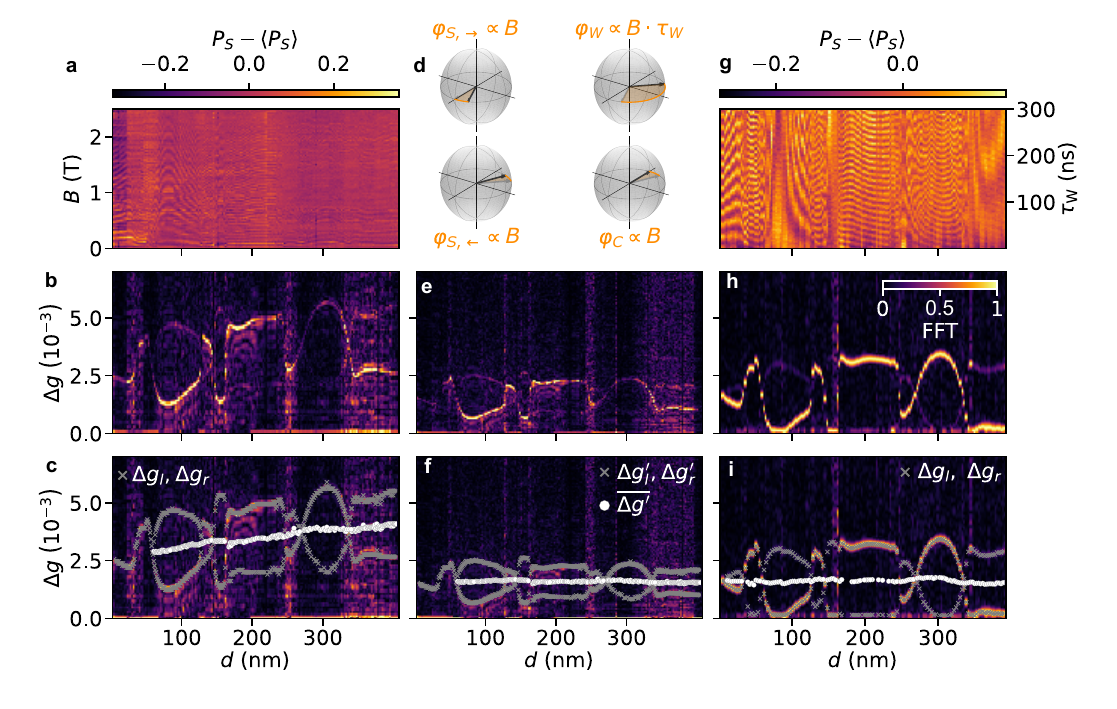}
    \caption{Method comparison for measuring the $g$-factor variation, differently evaluated. (a) Raw data of a measurement where we shuttle the electron to distance $d$ and vary the magnetic field while keeping $\tau_{W}=\SI{300}{\nano \second}$. (b) Extracted $g$-factor variation by Fourier transform of (a). (c) Version of (b) marked with found peaks and their average. (d) Bloch spheres illustrating parameter dependencies of collected phases during the entire coherent shuttling experiment.  (e) Extracted $g$-factor variation by Fourier transform of (a), using Eq.~\ref{eq:delg_magsweep}. (f) Version of (e) marked with found peaks and their average. (g) Raw data of a measurement where we shuttle the electron to distance $d$ and vary the wait time $\tau_{W}$ while keeping the magnetic field constant ($B=\SI{1.7}{\tesla}$). (h) Extracted $g$-factor variation by Fourier transform of (g). (i) Version of (h) marked with found peaks and their average.}
    \label{appfig1}
\end{figure*}

According to Eq.~2 from the main text,

\begin{equation}\label{eq:wait_phase}
    \varphi_\text{W}(B, \tau_\text{W})=\frac{\mu_B B}{\hbar} \cdot \Delta g(d) \cdot \tau_\text{W}, 
\end{equation}

\noindent the accumulated phase during the wait period is not only a function of the wait time $\tau_\text{W}$, but also a function of the magnetic field $B$. Thus, sweeping the magnetic field at fixed $\tau_\text{W}$ is another method to measure the g-factor. A representative measurement result for the same scanline $y=\SI{0}{\nano\meter}$ is shown in Fig. \ref{appfig1} for both dependencies $\tau_\text{W}$ and $B$.

As expected (Eq. \ref{eq:wait_phase}), we observe ST-oscillations either as a function of $B$ (Fig. \ref{appfig1}a) or $\tau_\text{W}$ (Fig. \ref{appfig1}b). The frequency of these oscillations varies for each distance $d$, which confirms that $\Delta g$ is actually $d$-dependent. We fix $\tau_\text{W}=\SI{300}{\nano \second}$ in Fig. \ref{appfig1}a well below the dephasing time of the ST-oscillations and yet long compared to the parasitic times required to separate the spin singlet. In Fig. \ref{appfig1}b, we choose $B = \SI{1.7}{\tesla}$ to stay well above any spin-valley resonance, which causes decoherence during shuttling (cf. Ref. \cite{Volmer25}). Then, we transform the raw data column-wise by FFT with normalisation 
\begin{align}
    \Delta g_B=&\frac{f_Bh}{\mu_B\tau_\text{W}},\\
    \Delta g_{\tau_\text{W}}=&\frac{f_{\tau_\text{W}}h}{\mu_BB}.
\end{align}
\noindent (Fig. \ref{appfig1}c,d). We mark peaks by grey crosses in Fig.~\ref{appfig1}e,f. If there are two peaks in the FFT, we indicate the average by a white circle. While the magnetic field sweep shows a trend tending to higher $g$-factor readings at larger distances, the wait time sweep keeps a constant average. There is more than one frequency peak per $d$ and the results for both extraction strategies are not equal despite some similarities. The discrepancy is related to the ST-oscillation phase $\varphi_{tot}$ accumulated during the total separation pulse. The total time the entangled S state is separated includes the charge operations within the DQD and the total shuttle time. The latter depends on $d$ and the corresponding additional phase depends on all the local $\Delta g(x)$ the electron crosses until reaching $d$. This parasitic phase accumulation is $\varphi_p(B,d)=\varphi_{tot}(B,d,\tau_\text{W})-\varphi_\text{W}(B,d,\tau_\text{W})$. Its unknown $B$-dependence complicates the interpretation of the FFT in Fig. \ref{appfig1}c. For each FFT along the $\tau_\text{W}$-axis (Fig. \ref{appfig1}d), this $\varphi_p(B,d)$ is constant and thus drops out. Ultimately, this uncompensated dependence results in a smoothing effect and explains the observed lower variation of $\Delta g(d)$ in Fig. \ref{appfig1}c,e compared to Fig. \ref{appfig1}d,f as well as the drift of the average and the offset. Sweeping $\tau_\mathrm{W}$ is thus clearly the better choice as the FFT reveals the pure $\Delta g(d)$. Furthermore, it requires no time-overhead for magnetic field sweeps and the $B$-field can be fixed at a sufficiently high value to avoid spin-valley resonances and thus decoherence of the entangled spin state during the shuttle process.

We examine the difference between the $B$- and $\tau_\text{W}$-sweep methods in detail and try to correct for the shortcomings of the $B$-sweep. We rigorously compute the extraction method that leads to the data as displayed in Fig.~\ref{appfig1}. During the actual experiment, in contrast to Eq.~\ref{eq:wait_phase}, the shuttled spin experiences a multitude of phases 
\begin{equation}\label{eq:all_phases}
\begin{aligned}
    \varphi(d,\tau_\mathrm{W}, B) &= \varphi_{S,\rightarrow}(B,d) 
     + \varphi_W(B,d,\tau_\mathrm{W})  \\ &\quad + \varphi_{S,\leftarrow}(B,d) + \varphi_C(B).
\end{aligned}
\end{equation}
In addition to the phase accumulated during waiting at distance $d$, the spin pair experiences a phase accumulation during shuttling in and out of the one-dimensional electron channel (1DEC) ($\varphi_{S,\rightarrow}$ and $\varphi_{S,\leftarrow}$) as well as another phase during waiting in the DQD system $\varphi_C$. None of these phases depend on $\tau_\mathrm{W}$, hence they are only a phase offset for the $\tau_\text{W}$-sweep from the main text. This phase offset disappears when applying the Fourier transform for those datasets. For the magnetic field sweep however, these phases change the obtainable data significantly. First, $\varphi_{S}=\varphi_{S,\rightarrow}+\varphi_{S,\leftarrow}$ represents the total phase accumulated during shuttling
\begin{equation}\label{eq:shuttle_phase}
    \varphi_S (B, d)=\frac{2\mu_B B}{h}\int_0^{\tau_S(d)} dt \Delta g(x(t)).
\end{equation}
Here, $\tau_S(d)$ relates to the one-way shuttle time it takes to reach distance $d$. Lastly, $\varphi_C$ represents the phase accumulated during waiting in the DQD
\begin{equation}\label{eq:constant_phase}
    \varphi_C(B)=\frac{\mu_B B \tau_C \Delta g_C}{h}.
\end{equation}
As all phases depend linearly on the magnetic field $B$, we can only determine the average over all measurement times. As
\begin{equation}\label{eq:fB_magsweep}
\begin{aligned}
    &f_B\cdot B=\varphi=\frac{\mu_B B}{h}\cdot\\
    &\cdot (\Delta g(d)\tau_\mathrm{W}+2\int_0^{\tau_S} dt \Delta g(x(t))+\Delta g_C\tau_C),
\end{aligned}
\end{equation}
we can compute only the time-wise mean of the $g$-factor variation. To indicate that this quantity required more overhead to compute, we describe it by $\Delta g'$
\begin{equation}\label{eq:delg_magsweep}
    \Delta g'(d)\approx \left<\Delta g\right>_t=\frac{h f_B}{\mu_B\cdot (\tau_C+2\tau_S(d)+\tau_\mathrm{W})}.
\end{equation}
\noindent This is the final formula applied in the evaluation of the measurement raw data with $\tau_\mathrm{W}=\SI{300}{\nano \second}$, $\tau_S(d)=\SI{200}{\nano \second}\cdot d/\SI{280}{\nano \meter}$ and $\tau_C=\SI{200}{\nano \second}$. That it is crucial to correct for the additional times here, can be easily seen by reevaluating the data. Note that there still is an offset as well as a reduced amplitude due to the parasitic phases within the $B$-sweep. The results of this correction can be seen in Fig.~\ref{appfig1}e,f. It is a replot of Fig.~\ref{appfig1}b,c exchanged using Eq.~\ref{eq:delg_magsweep}.

Here, we see that the adjustments in computing the $g$-factor difference for the $B$-sweep improved the correspondence to the $\tau_\text{W}$-sweep. The average is constant and at the same position. However, the strength of the relative variations is lowered significantly, as we effectively average over multiple phases which introduces a smoothing of the variations. This smoothing of the variations is increased with increased shuttle distance $d$, as the shuttle phase takes up an increasingly larger part of the experiment with increasing $d$ at constant velocity. We will analyze this next by looking at the ratio of variations between the grey crosses in Fig.~\ref{appfig1}e,f.

\begin{figure}
    \centering
    \includegraphics[width=\linewidth]{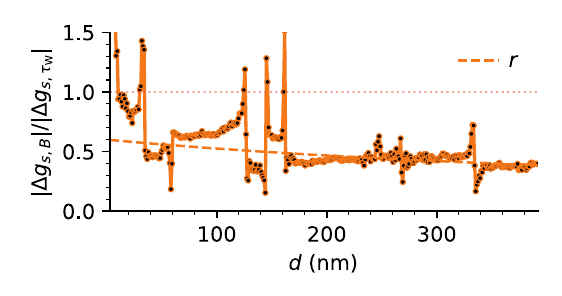}
    \caption{Ratios of $g$-factors between the $B$-sweep and the $\tau_\text{W}$-sweep. $|\Delta g_1(d)-\Delta g_2(d)|$ of the $B$-sweep is divided by the one of $\tau_\text{W}$-sweep. We also plot the model expectation in a red dashed line.}
    \label{appfig2}
\end{figure}

Next, we compare the ratio of $|\Delta g_1(d)- \Delta g_2(d)|$ extracted from both methods with the expected behavior. We divide the value extracted from the $B$-sweep by the value extracted by the $\tau_\text{W}$-sweep. In Fig.~\ref{appfig2}, the ratio between the two is plotted as a function of $d$. The ratio of 1 is added as a guide to the eye. Moreover, we added the function
\begin{equation}
    r=\frac{\tau_\text{W}}{\tau_C+\tau_\text{W}+2\tau_\text{S}}=\frac{300}{500+200\cdot d /\SI{280}{\nano \meter}},
\end{equation}
dashed in red in order to simulate the expected amplitude ratio due to the averaging in the $B$-sweep. Deviations are expected, as it only holds, when the approximation in Eq.~\ref{eq:delg_magsweep} is exactly equal. For this to hold, we need to be able to replace the integral over the shuttle time in Eq.~\ref{eq:fB_magsweep} by a multiplication of the average $g$-factor difference with the shuttle time. The longer we shuttle, the closer this integral is to the expectation value. Hence, we expect that the computed ratios converge to this added function at large $d$, which we also observe in Fig.~\ref{appfig2}.

\section{Extraction of the dominant component for the $g$-factor map}\label{app:Extraction_of_the_dominant_component}

\begin{figure*}
    \centering
    \includegraphics[width=\linewidth]{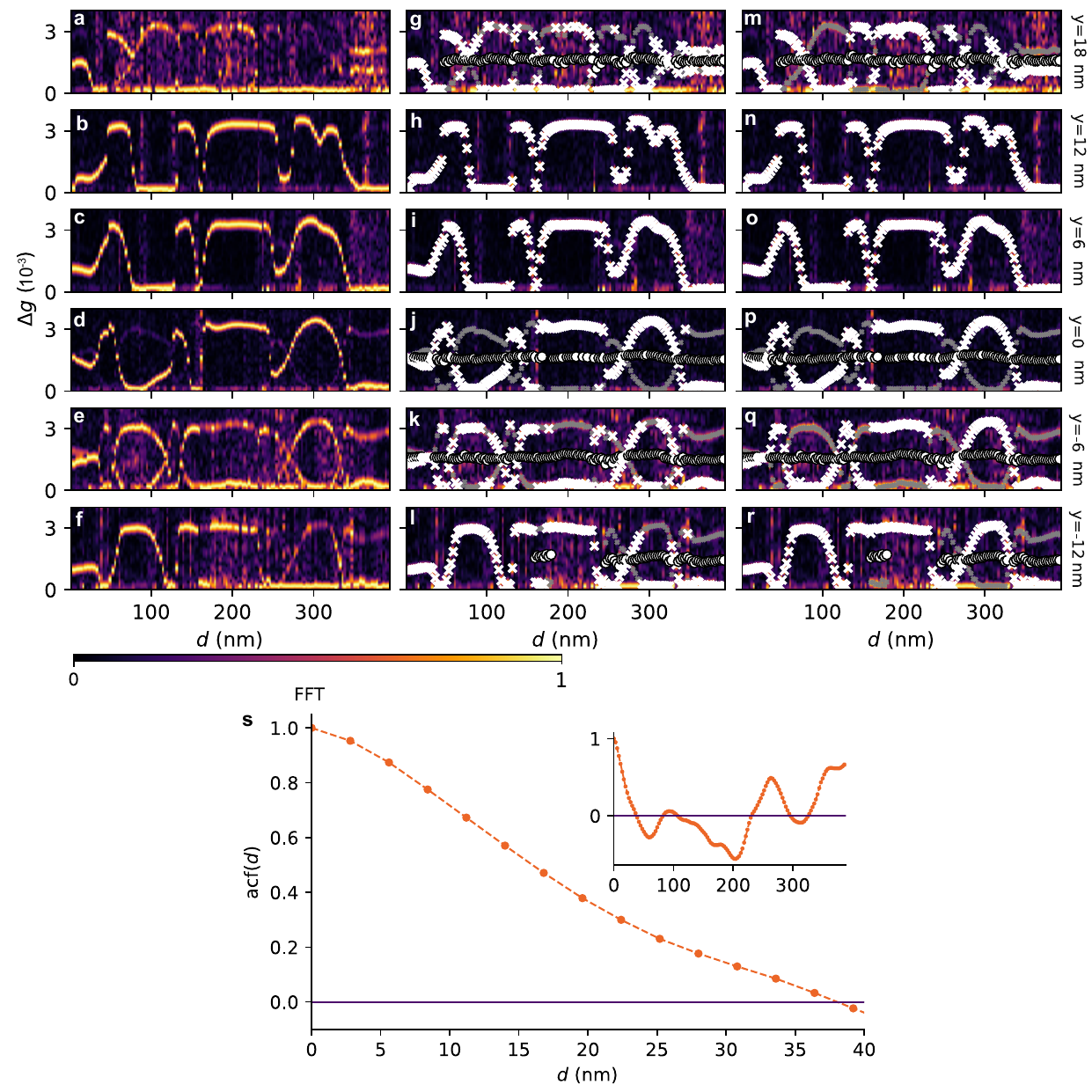}
    \caption{Extracting the dominant component of the $g$-factor map. (a-f) Raw data of the $g$-factor map. (g-l) Raw data with the two peaks marked. The dominant peak selected by signal strength is marked by a white cross while the subdominant component (if present) is marked with a grey cross. These points are used to generate the $g$-factor map. (m-r) Raw data with marked points from (g-l) post-processed and corrected by hand. This data is used to produce the $g$-factor map. (s) Autocorrelation function $acf(d)$ of the dominant component $\Delta g_1(d)$ as a function of shuttle distance, averaged over all traces with an inset displaying the zoom-out.}
    \label{appfig3}
\end{figure*}

In this section, we show how we extract and select the $g$-factor $\Delta g_1$ component that belongs to the dominating valley configuration. We start with the raw data as seen in Fig.~\ref{appfig3}a-f (see also Fig.~\ref{fig2}a-f). There, we firstly extract one or two peak frequencies by hand picking the range where we expect it. Thereafter a simple peak-finding algorithm takes care of finding the maximal point in the vicinity. Doing this yields one to two maxima that we firstly order by signal strength. The white crosses in Fig.~\ref{appfig3}g-l correspond to the peaks that range highest between the two. We observe that for the most part, this grants us a continuous property, as the initialized valley mixture is expected to be the same for one trace. For some traces however, the two participating valley configurations have close population probabilities (see for instance Fig.~\ref{appfig3}g,k), making it sometimes difficult to point out a dominant component by algorithmic selection alone. Thus, we post-correct the valley allocation by hand-selecting points that seem clearly switched in their valley allocation with respect to the rest of their trace (x-direction) as well as their neighbours in y-direction. The much more continuous result can be seen in Fig.~\ref{appfig3}m-r. For the most part, we achieve correspondence between neighbouring traces (y-direction) as well. There is one are that is difficult to correctly allocate at $x\in(\SI{40}{\nano \meter},\SI{130}{\nano \meter})$, $y \in (\SI{0}{\nano \meter},\SI{-12}{\nano \meter})$. There, at $y=\SI{-6}{\nano \meter}$ we find two similarly dominant components, with correlations in both directions in $y$. However, the valley populations between the $y>\SI{-6}{\nano \meter}$ seem to be flipped with respect to the valley populations at $y=\SI{-12}{\nano \meter}$. At $y=\SI{-12}{\nano \meter}$, the other component is not visible at the start, leaving us with a likely population flip here. Nevertheless, for the remaining part of the map, this allocation is a lot more simple and clear.

In Fig.~\ref{appfig3}s, we display the autocorrelation function of the final evaluation step as a function of shuttle distance. For the dominating component, the autocorrelation function $acf(d)$ shows a decay ending at zero at about $d=\SI{40}{\nano \meter}$. Beyond this length, it starts to fluctuate, due to the limited statistics in this dataset. The inset shows that beyond \SI{40}{\nano \meter}, the correlations become random fluctuations. 

\section{Explanation of the $\Delta$ reconstruction algorithm}\label{app:Explanation_of_the_reconstruction_algorithm}
 The computed solutions for main text Eq.~4 are either $\phi$ or $2\pi-\phi$. Therefore, for full reconstruction, we need to create an algorithm that decides which of the two solutions is the most fitting for each point in context of the other points. We create an algorithm that aims for the most continuous trace as the most physically plausible one. For this, we set up a cost function that is evaluated for both solutions at each step and picks the one with lower cost. The cost function is a sum of three components. The first component heavily penalizes angular deviations from the current trace's tangent vector. The second component is more costly, the more the step size deviates from the exponentially weighted recent steps. The third component is more costly, the further the next point is away.

We reconstruct a single complex path by choosing, at each step index \(j\), one of two mirror
candidates in the complex plane. Each candidate is described by a radius \(r_j\) and a phase \(\phi_j\).
The phase is obtained by mapping the locally measured \(\Delta g_j\) on that row monotonically onto
\([0,\pi]\) (small \(\Delta g\) near \(0\), large \(\Delta g\) near \(\pi\)); the two candidates are then
\[
z^{(a)}_j = r_j\,e^{i\phi_j},
\qquad
z^{(b)}_j = r_j\,e^{i(2\pi-\phi_j)}.
\]
The radius is taken from an interpolation of the valley–splitting map (to the more coarse grid of the $g$-factor measurements) at that step. In this way the geometry reflects the local energy scale when desired.

\medskip
To steer decisions, the algorithm maintains two short memories:
(1) a tangential vector \(d_j\) that indicates the recent direction of the path;
(2) a reference step length \(\ell_j\) that captures the typical size of recent moves.
Both memories are updated by an exponential moving average. The new value is the previous value plus a fraction \(\beta\) of the difference to the current measurement. For the heading we blend the previous direction with the current unit step and then rescale to length \(1\); for the step length we blend the previous reference with the current step length. A larger \(\beta\) reacts faster, while a smaller \(\beta\) retains more history. If needed, a small \(\varepsilon\!>\!0\) is used to avoid divisions by numbers that are too small.

\medskip
We define the step vector from the current point \(z_j\) to a candidate \(z_{j+1}\) as \(\mathbf{u} = z_{j+1} - z_j\), with magnitude \(n = |\mathbf{u}|\) and unit direction \(\hat{\mathbf{u}} = \mathbf{u}/n\).
The immediate cost \(c_{j+1}\) is a linear combination of three penalty terms,
\[
c_{j+1} \;=\;
w_{\text{ang}}\,a_{j+1} \;+\; \lambda_\ell\,\delta_{j+1} \;+\; \lambda_s\,s_{j+1},
\]
where the components penalize deviations from the running memory of the direction (\(\vec{d}_j\)) and characteristic step length (\(\ell_j\)):
\begin{align*}
    a_{j+1} &= 1 - \hat{\mathbf{u}} \cdot \vec{d}_j & \text{(alignment penalty)} \\
    \delta_{j+1} &= \left| \frac{n}{\ell_j} - 1 \right| & \text{(pace deviation)} \\
    s_{j+1} &= \frac{n}{\ell_j} & \text{(step magnitude)}
\end{align*}
After evaluating \(c_{j+1}\) and selecting the step, the memory states are updated via an exponential moving average (EMA) to guide future decisions:
\[
\vec{d}_{j+1} \propto (1-\beta_d)\vec{d}_j + \beta_d \hat{\mathbf{u}},
\qquad
\ell_{j+1} = (1-\beta_\ell)\ell_j + \beta_\ell n,
\]
where \(\vec{d}_{j+1}\) is re-normalized to unit length.

\medskip
Over the chosen path the algorithm minimizes the undiscounted total
\[
C \;=\; \sum_{j=0}^{N-2} c_{j+1}.
\]

\medskip
To circumvent local minima caused by measurement noise, the algorithm avoids irreversible "greedy" decisions by employing a nested beam-search strategy. Conceptually, this separates the algorithm's historical memory from its hypothetical future simulations. First, the algorithm maintains a main beam, which carries forward the top $k_{\text{main}}$ (denoted as $K$ in the Supplementary Videos) most promising historical trajectories up to the current spatial index. 

Second, to evaluate the true quality of a candidate step, the algorithm initiates a temporary lookahead simulation extending $h$ steps into the future (denoted as depth $D$ in the videos). Because the number of possible future branches grows exponentially, this simulation restricts its exploration using an internal lookahead beam, which tracks only the best $k_{\ell}$ (denoted as $k$) hypothetical future branches at each simulated step. 

For each child branch originating from a main survivor, we form discounted sums of the future costs evaluated during the lookahead:
\[
\gamma\,c_{j+2} \;+\; \gamma^2 c_{j+3} \;+\; \cdots \;+\; \gamma^{L} c_{j+1+L},
\]
where the discount factor $\gamma \in [0,1]$ controls how far the forecast tries to see: values near $1$ make the heuristic far-sighted and stabilizing, while smaller values make it short-sighted and reactive. The algorithm identifies the smallest such sum among the simulated futures to serve as the "forecast" penalty. Each candidate is then ranked by its ``current total cost so far'' $+$ ``forecast''. The best $k_{\text{main}}$ paths are kept to form the new main beam, and the procedure repeats. At the end, the survivor with the smallest undiscounted total $C$ is returned. For symmetry reasons, we couple memory and foresight by choosing a single number $\rho$ and setting $\beta_d = 1-\rho$ for the heading update and $\gamma=\rho$ for the forecast.

\medskip
In the algorithm \(w_{\text{ang}}\) enforces directional consistency; \(\lambda_\ell\) enforces pace consistency;
\(\lambda_s\) limits absolute step size. \(\beta_d\) (heading) and \(\beta_\ell\) (scale) set how
quickly the two memories adapt. \(\gamma\) expresses trust in the future during the forecast.
\(k_{\text{main}}\), \(k_{\ell}\), and \(h\) determine how much of the search space is explored
now, how broadly we peek ahead, and how far we look. Visual elements used in figures (fading trails,
normalized bars, and example future beams; see dataset) do not affect any of the decisions above; they only make
them inspectable. The concrete values of these parameters can be found in Tab.~\ref{tab:search_params}. Note that these parameters were chosen intuitively and we do not claim general correctness of the unfolded traces in the main text Fig.~5f-j.

\begin{table}[t]
    \centering
    \begin{tabular}{|c|c|c|}
        \hline
        Symbol        & Description   & Value \\
        \hline
        $w_{\text{ang}}$ & weight of turning penalty       & $1.0$  \\
        $\lambda_\ell$   & weight of pace deviation        & $0.4$  \\
        $\lambda_s$      & weight of absolute step size    & $2.0$  \\
        $\beta_d$        & EMA memory for direction        & $0.4$  \\
        $\beta_\ell$     & EMA memory for step scale       & $0.25$ \\
        $\gamma$         & discount for future costs       & $0.6$  \\
        $k_{\text{main}}$& main beam width (paths kept)    & $8$    \\
        $k_{\ell}$       & lookahead beam width            & $20$   \\
        $h$              & lookahead depth (steps ahead)   & $10$   \\
        \hline
    \end{tabular}
        \caption{Parameters of the cost and beam-search model and values used in this work.}
    \label{tab:search_params}
\end{table}

\section{Simulation of valley phase extraction with fluctuations}\label{app:Simulation_of_valley_phase_extraction}
\label{app:valley}
\begin{figure*}
    \centering
    \includegraphics[width=\linewidth]{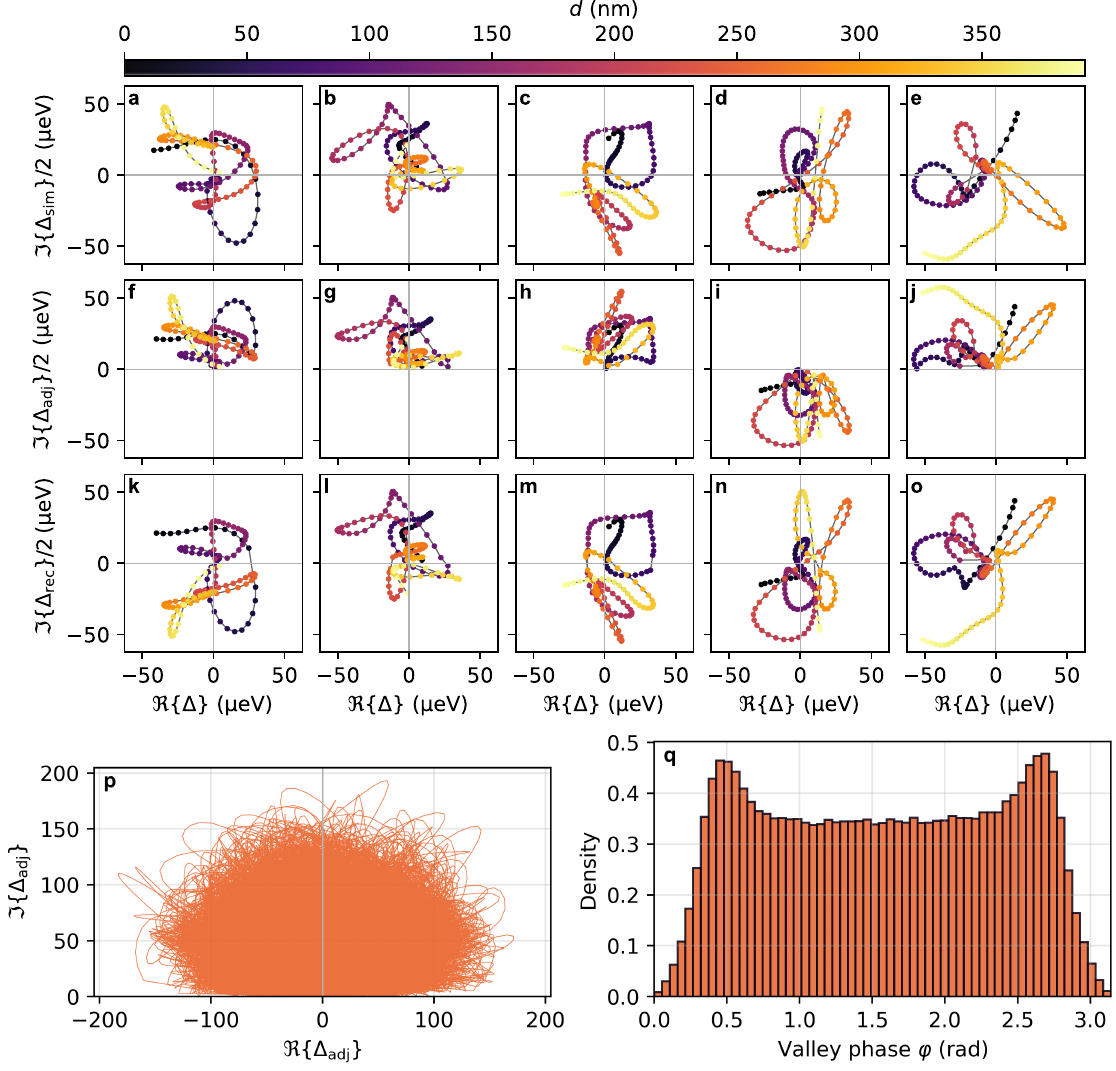}
    \caption{Simulation of the valley phase extraction within realistic measurement parameters. (a-e) Simulated traces of the intervalley coupling $\Delta_\text{sim}$. (f-j) Fluctuation adjusted traces of (a-e) evaluated by the $g$-factor measurements. (k-o) Reconstructed traces from (f-j) using the same reconstruction algorithm as in the main text. (p) Simulated and fluctuation adjusted trace of \SI{115}{\micro \meter} length of the intervalley coupling. (q) statistics of the valley phase in (p).}
    \label{appfig4}
\end{figure*}

In this section, we want to shed light on the key differences between the theoretic model relating the variations of the $g$-factor to the valley phase $\varphi$~\cite{woods24,Woods25}. For this, we first simulate some scanlines of $\Delta$ using statistical parameters of valley splitting disorder extracted from the experiment (Fig.~\ref{appfig4}a-e). Then, we compute the corresponding correction to the $g$-factor given by
\begin{equation}\label{eq:g-factors}
    \delta g =\delta g_{\mathrm{max}}\cos(\phi),
\end{equation}
and add Gaussian noise with a correlation length of \SI{100}{\nano \meter} and a width of 2.5\% on top. We do this to simulate correlated shifts in $g$-factor difference due to theoretically unaccounted imperfections like crosstalk of the shuttle pulse on the static electron. Thereafter, we compute the valley phase again and with that the adjusted intervalley coupling $\Delta_\text{adj}$ (see Fig.~\ref{appfig4}f-j). In a final step, we apply the reconstruction algorithm discussed in the main text and the previous section of the appendix to fully reconstruct the trace (Fig.~\ref{appfig4}k-o). It is evident that the reconstruction works well for the most part, only the final part of panels k and n is flipped to the wrong side. The deviations induced by the correlated noise can be seen in panels f-j or even more pronounced panels p and q, where we simulated statistics of a \SI{115}{\micro \meter} long trace of adjusted intervalley coupling and the corresponding statistics of the valley phase $\varphi$. As the valley phases without adjustment should be distributed uniformly, the corresponding $\delta g$ are distributed in a bimodal distribution after Eq.~(\ref{eq:g-factors}). The precise form of the distribution is
\begin{equation}
    f_G(x=\delta g/\delta g_{\mathrm{max}})=\frac{1}{\pi \sqrt{1-x^2}}, \quad x\in [-1,1].
\end{equation}
This is a bimodal distribution that has sharp cutoffs on both ends. Taking the cosine of values of only this precise distribution will give a uniform distribution of valley phases. As our measurements are subject to noise and fluctuations (a specific example would be crosstalk from the shuttle pulse shifting the $g$-factor of the static electron), the sharp peaks will be softened at the edges, leaving us with a non uniform valley phase distribution in Fig.~\ref{appfig4} q. We find that especially the edges of the distribution (meaning the $\Delta$ values close to the real axis) are affected by this: While only very few values remain close to the axis, we find accumulation angles that depend on the width of the Gaussian noise we assume. In experiments, this manifests itself by a funnel-like dataset (see panels i and j or Figure 5 from the main text). 
This does impact the performance of our reconstruction algorithm, as its main purpose is to detect and predict switching of the $\Delta$ scanline between the two half planes in the imaginary plane. This occurs when the scanline moves close to the real axis, which is also the region where the noise has the greatest impact. By this simulation, where we tried to match the experimental data as closely as possible, we find that the reconstruction algorithm matches the original data for approximately 90\% of the time.
\end{appendix}


\begin{thebibliography}{65}%
\makeatletter
\providecommand \@ifxundefined [1]{%
 \@ifx{#1\undefined}
}%
\providecommand \@ifnum [1]{%
 \ifnum #1\expandafter \@firstoftwo
 \else \expandafter \@secondoftwo
 \fi
}%
\providecommand \@ifx [1]{%
 \ifx #1\expandafter \@firstoftwo
 \else \expandafter \@secondoftwo
 \fi
}%
\providecommand \natexlab [1]{#1}%
\providecommand \enquote  [1]{``#1''}%
\providecommand \bibnamefont  [1]{#1}%
\providecommand \bibfnamefont [1]{#1}%
\providecommand \citenamefont [1]{#1}%
\providecommand \href@noop [0]{\@secondoftwo}%
\providecommand \href [0]{\begingroup \@sanitize@url \@href}%
\providecommand \@href[1]{\@@startlink{#1}\@@href}%
\providecommand \@@href[1]{\endgroup#1\@@endlink}%
\providecommand \@sanitize@url [0]{\catcode `\\12\catcode `\$12\catcode `\&12\catcode `\#12\catcode `\^12\catcode `\_12\catcode `\%12\relax}%
\providecommand \@@startlink[1]{}%
\providecommand \@@endlink[0]{}%
\providecommand \url  [0]{\begingroup\@sanitize@url \@url }%
\providecommand \@url [1]{\endgroup\@href {#1}{\urlprefix }}%
\providecommand \urlprefix  [0]{URL }%
\providecommand \Eprint [0]{\href }%
\providecommand \doibase [0]{https://doi.org/}%
\providecommand \selectlanguage [0]{\@gobble}%
\providecommand \bibinfo  [0]{\@secondoftwo}%
\providecommand \bibfield  [0]{\@secondoftwo}%
\providecommand \translation [1]{[#1]}%
\providecommand \BibitemOpen [0]{}%
\providecommand \bibitemStop [0]{}%
\providecommand \bibitemNoStop [0]{.\EOS\space}%
\providecommand \EOS [0]{\spacefactor3000\relax}%
\providecommand \BibitemShut  [1]{\csname bibitem#1\endcsname}%
\let\auto@bib@innerbib\@empty
\bibitem [{\citenamefont {DiVincenzo}(2000)}]{DiVincenzo00}%
  \BibitemOpen
  \bibfield  {author} {\bibinfo {author} {\bibfnamefont {D.~P.}\ \bibnamefont {DiVincenzo}},\ }\bibfield  {title} {\bibinfo {title} {The physical implementation of quantum computation},\ }\href {https://doi.org/https://doi.org/10.1002/1521-3978(200009)48:9/11<771::AID-PROP771>3.0.CO;2-E} {\bibfield  {journal} {\bibinfo  {journal} {Fortschr Phys.}\ }\textbf {\bibinfo {volume} {48}},\ \bibinfo {pages} {771} (\bibinfo {year} {2000})}\BibitemShut {NoStop}%
\bibitem [{\citenamefont {Yoneda}\ \emph {et~al.}(2018)\citenamefont {Yoneda}, \citenamefont {Takeda}, \citenamefont {Otsuka}, \citenamefont {Nakajima}, \citenamefont {Delbecq}, \citenamefont {Allison}, \citenamefont {Honda}, \citenamefont {Kodera}, \citenamefont {Oda}, \citenamefont {Hoshi}, \citenamefont {Usami}, \citenamefont {Itoh},\ and\ \citenamefont {Tarucha}}]{Yoneda2018}%
  \BibitemOpen
  \bibfield  {author} {\bibinfo {author} {\bibfnamefont {J.}~\bibnamefont {Yoneda}}, \bibinfo {author} {\bibfnamefont {K.}~\bibnamefont {Takeda}}, \bibinfo {author} {\bibfnamefont {T.}~\bibnamefont {Otsuka}}, \bibinfo {author} {\bibfnamefont {T.}~\bibnamefont {Nakajima}}, \bibinfo {author} {\bibfnamefont {M.~R.}\ \bibnamefont {Delbecq}}, \bibinfo {author} {\bibfnamefont {G.}~\bibnamefont {Allison}}, \bibinfo {author} {\bibfnamefont {T.}~\bibnamefont {Honda}}, \bibinfo {author} {\bibfnamefont {T.}~\bibnamefont {Kodera}}, \bibinfo {author} {\bibfnamefont {S.}~\bibnamefont {Oda}}, \bibinfo {author} {\bibfnamefont {Y.}~\bibnamefont {Hoshi}}, \bibinfo {author} {\bibfnamefont {N.}~\bibnamefont {Usami}}, \bibinfo {author} {\bibfnamefont {K.~M.}\ \bibnamefont {Itoh}},\ and\ \bibinfo {author} {\bibfnamefont {S.}~\bibnamefont {Tarucha}},\ }\bibfield  {title} {\bibinfo {title} {A quantum-dot spin qubit with coherence limited by charge noise and fidelity higher than 99.9\%},\ }\href
  {https://doi.org/10.1038/s41565-017-0014-x} {\bibfield  {journal} {\bibinfo  {journal} {Nat. Nanotechnol.}\ }\textbf {\bibinfo {volume} {13}},\ \bibinfo {pages} {102} (\bibinfo {year} {2018})}\BibitemShut {NoStop}%
\bibitem [{\citenamefont {Watson}\ \emph {et~al.}(2018)\citenamefont {Watson}, \citenamefont {Philips}, \citenamefont {Kawakami}, \citenamefont {Ward}, \citenamefont {Scarlino}, \citenamefont {Veldhorst}, \citenamefont {Savage}, \citenamefont {Lagally}, \citenamefont {Friesen}, \citenamefont {Coppersmith}, \citenamefont {Eriksson},\ and\ \citenamefont {Vandersypen}}]{watson18}%
  \BibitemOpen
  \bibfield  {author} {\bibinfo {author} {\bibfnamefont {T.~F.}\ \bibnamefont {Watson}}, \bibinfo {author} {\bibfnamefont {S.~G.~J.}\ \bibnamefont {Philips}}, \bibinfo {author} {\bibfnamefont {E.}~\bibnamefont {Kawakami}}, \bibinfo {author} {\bibfnamefont {D.~R.}\ \bibnamefont {Ward}}, \bibinfo {author} {\bibfnamefont {P.}~\bibnamefont {Scarlino}}, \bibinfo {author} {\bibfnamefont {M.}~\bibnamefont {Veldhorst}}, \bibinfo {author} {\bibfnamefont {D.~E.}\ \bibnamefont {Savage}}, \bibinfo {author} {\bibfnamefont {M.~G.}\ \bibnamefont {Lagally}}, \bibinfo {author} {\bibfnamefont {M.}~\bibnamefont {Friesen}}, \bibinfo {author} {\bibfnamefont {S.~N.}\ \bibnamefont {Coppersmith}}, \bibinfo {author} {\bibfnamefont {M.~A.}\ \bibnamefont {Eriksson}},\ and\ \bibinfo {author} {\bibfnamefont {L.~M.~K.}\ \bibnamefont {Vandersypen}},\ }\bibfield  {title} {\bibinfo {title} {A programmable two-qubit quantum processor in silicon},\ }\href {https://doi.org/10.1038/nature25766} {\bibfield  {journal} {\bibinfo  {journal}
  {Nature}\ }\textbf {\bibinfo {volume} {555}},\ \bibinfo {pages} {633} (\bibinfo {year} {2018})}\BibitemShut {NoStop}%
\bibitem [{\citenamefont {Struck}\ \emph {et~al.}(2020)\citenamefont {Struck}, \citenamefont {Hollmann}, \citenamefont {Schauer}, \citenamefont {Fedorets}, \citenamefont {Schmidbauer}, \citenamefont {Sawano}, \citenamefont {Riemann}, \citenamefont {Abrosimov}, \citenamefont {Cywi{\'n}ski}, \citenamefont {Bougeard},\ and\ \citenamefont {Schreiber}}]{Struck2020}%
  \BibitemOpen
  \bibfield  {author} {\bibinfo {author} {\bibfnamefont {T.}~\bibnamefont {Struck}}, \bibinfo {author} {\bibfnamefont {A.}~\bibnamefont {Hollmann}}, \bibinfo {author} {\bibfnamefont {F.}~\bibnamefont {Schauer}}, \bibinfo {author} {\bibfnamefont {O.}~\bibnamefont {Fedorets}}, \bibinfo {author} {\bibfnamefont {A.}~\bibnamefont {Schmidbauer}}, \bibinfo {author} {\bibfnamefont {K.}~\bibnamefont {Sawano}}, \bibinfo {author} {\bibfnamefont {H.}~\bibnamefont {Riemann}}, \bibinfo {author} {\bibfnamefont {N.~V.}\ \bibnamefont {Abrosimov}}, \bibinfo {author} {\bibfnamefont {{\L}.}~\bibnamefont {Cywi{\'n}ski}}, \bibinfo {author} {\bibfnamefont {D.}~\bibnamefont {Bougeard}},\ and\ \bibinfo {author} {\bibfnamefont {L.~R.}\ \bibnamefont {Schreiber}},\ }\bibfield  {title} {\bibinfo {title} {Low-frequency spin qubit energy splitting noise in highly purified $^{28}${Si/SiGe}},\ }\href {https://doi.org/10.1038/s41534-020-0276-2} {\bibfield  {journal} {\bibinfo  {journal} {npj Quantum Inf.}\ }\textbf {\bibinfo {volume} {6}},\
  \bibinfo {pages} {2056} (\bibinfo {year} {2020})}\BibitemShut {NoStop}%
\bibitem [{\citenamefont {Noiri}\ \emph {et~al.}(2022)\citenamefont {Noiri}, \citenamefont {Takeda}, \citenamefont {Nakajima}, \citenamefont {Kobayashi}, \citenamefont {Sammak}, \citenamefont {Scappucci},\ and\ \citenamefont {Tarucha}}]{Noiri2022}%
  \BibitemOpen
  \bibfield  {author} {\bibinfo {author} {\bibfnamefont {A.}~\bibnamefont {Noiri}}, \bibinfo {author} {\bibfnamefont {K.}~\bibnamefont {Takeda}}, \bibinfo {author} {\bibfnamefont {T.}~\bibnamefont {Nakajima}}, \bibinfo {author} {\bibfnamefont {T.}~\bibnamefont {Kobayashi}}, \bibinfo {author} {\bibfnamefont {A.}~\bibnamefont {Sammak}}, \bibinfo {author} {\bibfnamefont {G.}~\bibnamefont {Scappucci}},\ and\ \bibinfo {author} {\bibfnamefont {S.}~\bibnamefont {Tarucha}},\ }\bibfield  {title} {\bibinfo {title} {Fast universal quantum gate above the fault-tolerance threshold in silicon},\ }\href {https://doi.org/10.1038/s41586-021-04182-y} {\bibfield  {journal} {\bibinfo  {journal} {Nature}\ }\textbf {\bibinfo {volume} {601}},\ \bibinfo {pages} {338} (\bibinfo {year} {2022})}\BibitemShut {NoStop}%
\bibitem [{\citenamefont {Mills}\ \emph {et~al.}(2022)\citenamefont {Mills}, \citenamefont {Guinn}, \citenamefont {Gullans}, \citenamefont {Sigillito}, \citenamefont {Feldman}, \citenamefont {Nielsen},\ and\ \citenamefont {Petta}}]{Mills2022}%
  \BibitemOpen
  \bibfield  {author} {\bibinfo {author} {\bibfnamefont {A.~R.}\ \bibnamefont {Mills}}, \bibinfo {author} {\bibfnamefont {C.~R.}\ \bibnamefont {Guinn}}, \bibinfo {author} {\bibfnamefont {M.~J.}\ \bibnamefont {Gullans}}, \bibinfo {author} {\bibfnamefont {A.~J.}\ \bibnamefont {Sigillito}}, \bibinfo {author} {\bibfnamefont {M.~M.}\ \bibnamefont {Feldman}}, \bibinfo {author} {\bibfnamefont {E.}~\bibnamefont {Nielsen}},\ and\ \bibinfo {author} {\bibfnamefont {J.~R.}\ \bibnamefont {Petta}},\ }\bibfield  {title} {\bibinfo {title} {Two-qubit silicon quantum processor with operation fidelity exceeding 99\%},\ }\href {https://doi.org/10.1126/sciadv.abn5130} {\bibfield  {journal} {\bibinfo  {journal} {Sci. Adv.}\ }\textbf {\bibinfo {volume} {8}},\ \bibinfo {pages} {eabn5130} (\bibinfo {year} {2022})}\BibitemShut {NoStop}%
\bibitem [{\citenamefont {Xue}\ \emph {et~al.}(2022)\citenamefont {Xue}, \citenamefont {Russ}, \citenamefont {Samkharadze}, \citenamefont {Undseth}, \citenamefont {Sammak}, \citenamefont {Scappucci},\ and\ \citenamefont {Vandersypen}}]{Xue2022}%
  \BibitemOpen
  \bibfield  {author} {\bibinfo {author} {\bibfnamefont {X.}~\bibnamefont {Xue}}, \bibinfo {author} {\bibfnamefont {M.}~\bibnamefont {Russ}}, \bibinfo {author} {\bibfnamefont {N.}~\bibnamefont {Samkharadze}}, \bibinfo {author} {\bibfnamefont {B.}~\bibnamefont {Undseth}}, \bibinfo {author} {\bibfnamefont {A.}~\bibnamefont {Sammak}}, \bibinfo {author} {\bibfnamefont {G.}~\bibnamefont {Scappucci}},\ and\ \bibinfo {author} {\bibfnamefont {L.~M.~K.}\ \bibnamefont {Vandersypen}},\ }\bibfield  {title} {\bibinfo {title} {Quantum logic with spin qubits crossing the surface code threshold},\ }\href {https://doi.org/10.1038/s41586-021-04273-w} {\bibfield  {journal} {\bibinfo  {journal} {Nature}\ }\textbf {\bibinfo {volume} {601}},\ \bibinfo {pages} {343} (\bibinfo {year} {2022})}\BibitemShut {NoStop}%
\bibitem [{\citenamefont {Philips}\ \emph {et~al.}(2022)\citenamefont {Philips}, \citenamefont {M{\k{a}}dzik}, \citenamefont {Amitonov}, \citenamefont {de~Snoo}, \citenamefont {Russ}, \citenamefont {Kalhor}, \citenamefont {Volk}, \citenamefont {Lawrie}, \citenamefont {Brousse}, \citenamefont {Tryputen}, \citenamefont {Paquelet~Wuetz}, \citenamefont {Sammak}, \citenamefont {Veldhorst}, \citenamefont {Scappucci},\ and\ \citenamefont {Vandersypen}}]{philips22}%
  \BibitemOpen
  \bibfield  {author} {\bibinfo {author} {\bibfnamefont {S.~G.~J.}\ \bibnamefont {Philips}}, \bibinfo {author} {\bibfnamefont {M.~T.}\ \bibnamefont {M{\k{a}}dzik}}, \bibinfo {author} {\bibfnamefont {S.~V.}\ \bibnamefont {Amitonov}}, \bibinfo {author} {\bibfnamefont {S.~L.}\ \bibnamefont {de~Snoo}}, \bibinfo {author} {\bibfnamefont {M.}~\bibnamefont {Russ}}, \bibinfo {author} {\bibfnamefont {N.}~\bibnamefont {Kalhor}}, \bibinfo {author} {\bibfnamefont {C.}~\bibnamefont {Volk}}, \bibinfo {author} {\bibfnamefont {W.~I.~L.}\ \bibnamefont {Lawrie}}, \bibinfo {author} {\bibfnamefont {D.}~\bibnamefont {Brousse}}, \bibinfo {author} {\bibfnamefont {L.}~\bibnamefont {Tryputen}}, \bibinfo {author} {\bibfnamefont {B.}~\bibnamefont {Paquelet~Wuetz}}, \bibinfo {author} {\bibfnamefont {A.}~\bibnamefont {Sammak}}, \bibinfo {author} {\bibfnamefont {M.}~\bibnamefont {Veldhorst}}, \bibinfo {author} {\bibfnamefont {G.}~\bibnamefont {Scappucci}},\ and\ \bibinfo {author} {\bibfnamefont {L.~M.~K.}\ \bibnamefont {Vandersypen}},\
  }\bibfield  {title} {\bibinfo {title} {Universal control of a six-qubit quantum processor in silicon},\ }\href {https://doi.org/10.1038/s41586-022-05117-x} {\bibfield  {journal} {\bibinfo  {journal} {Nature}\ }\textbf {\bibinfo {volume} {609}},\ \bibinfo {pages} {919} (\bibinfo {year} {2022})}\BibitemShut {NoStop}%
\bibitem [{\citenamefont {Stano}\ and\ \citenamefont {Loss}(2022)}]{Stano22}%
  \BibitemOpen
  \bibfield  {author} {\bibinfo {author} {\bibfnamefont {P.}~\bibnamefont {Stano}}\ and\ \bibinfo {author} {\bibfnamefont {D.}~\bibnamefont {Loss}},\ }\bibfield  {title} {\bibinfo {title} {Review of performance metrics of spin qubits in gated semiconducting nanostructures},\ }\href {https://doi.org/10.1038/s42254-022-00484-w} {\bibfield  {journal} {\bibinfo  {journal} {Nat. Rev. Phys.}\ }\textbf {\bibinfo {volume} {4}},\ \bibinfo {pages} {672} (\bibinfo {year} {2022})}\BibitemShut {NoStop}%
\bibitem [{\citenamefont {Neyens}\ \emph {et~al.}(2024)\citenamefont {Neyens}, \citenamefont {Zietz}, \citenamefont {Watson}, \citenamefont {Luthi}, \citenamefont {Nethwewala}, \citenamefont {George}, \citenamefont {Henry}, \citenamefont {Islam}, \citenamefont {Wagner}, \citenamefont {Borjans}, \citenamefont {Connors}, \citenamefont {Corrigan}, \citenamefont {Curry}, \citenamefont {Keith}, \citenamefont {Kotlyar}, \citenamefont {Lampert}, \citenamefont {Mądzik}, \citenamefont {Millard}, \citenamefont {Mohiyaddin}, \citenamefont {Pellerano}, \citenamefont {Pillarisetty}, \citenamefont {Ramsey}, \citenamefont {Savytskyy}, \citenamefont {Schaal}, \citenamefont {Zheng}, \citenamefont {Ziegler}, \citenamefont {Bishop}, \citenamefont {Bojarski}, \citenamefont {Roberts},\ and\ \citenamefont {Clarke}}]{neyens24}%
  \BibitemOpen
  \bibfield  {author} {\bibinfo {author} {\bibfnamefont {S.}~\bibnamefont {Neyens}}, \bibinfo {author} {\bibfnamefont {O.~K.}\ \bibnamefont {Zietz}}, \bibinfo {author} {\bibfnamefont {T.~F.}\ \bibnamefont {Watson}}, \bibinfo {author} {\bibfnamefont {F.}~\bibnamefont {Luthi}}, \bibinfo {author} {\bibfnamefont {A.}~\bibnamefont {Nethwewala}}, \bibinfo {author} {\bibfnamefont {H.~C.}\ \bibnamefont {George}}, \bibinfo {author} {\bibfnamefont {E.}~\bibnamefont {Henry}}, \bibinfo {author} {\bibfnamefont {M.}~\bibnamefont {Islam}}, \bibinfo {author} {\bibfnamefont {A.~J.}\ \bibnamefont {Wagner}}, \bibinfo {author} {\bibfnamefont {F.}~\bibnamefont {Borjans}}, \bibinfo {author} {\bibfnamefont {E.~J.}\ \bibnamefont {Connors}}, \bibinfo {author} {\bibfnamefont {J.}~\bibnamefont {Corrigan}}, \bibinfo {author} {\bibfnamefont {M.~J.}\ \bibnamefont {Curry}}, \bibinfo {author} {\bibfnamefont {D.}~\bibnamefont {Keith}}, \bibinfo {author} {\bibfnamefont {R.}~\bibnamefont {Kotlyar}}, \bibinfo {author} {\bibfnamefont {L.~F.}\
  \bibnamefont {Lampert}}, \bibinfo {author} {\bibfnamefont {M.~T.}\ \bibnamefont {Mądzik}}, \bibinfo {author} {\bibfnamefont {K.}~\bibnamefont {Millard}}, \bibinfo {author} {\bibfnamefont {F.~A.}\ \bibnamefont {Mohiyaddin}}, \bibinfo {author} {\bibfnamefont {S.}~\bibnamefont {Pellerano}}, \bibinfo {author} {\bibfnamefont {R.}~\bibnamefont {Pillarisetty}}, \bibinfo {author} {\bibfnamefont {M.}~\bibnamefont {Ramsey}}, \bibinfo {author} {\bibfnamefont {R.}~\bibnamefont {Savytskyy}}, \bibinfo {author} {\bibfnamefont {S.}~\bibnamefont {Schaal}}, \bibinfo {author} {\bibfnamefont {G.}~\bibnamefont {Zheng}}, \bibinfo {author} {\bibfnamefont {J.}~\bibnamefont {Ziegler}}, \bibinfo {author} {\bibfnamefont {N.~C.}\ \bibnamefont {Bishop}}, \bibinfo {author} {\bibfnamefont {S.}~\bibnamefont {Bojarski}}, \bibinfo {author} {\bibfnamefont {J.}~\bibnamefont {Roberts}},\ and\ \bibinfo {author} {\bibfnamefont {J.~S.}\ \bibnamefont {Clarke}},\ }\bibfield  {title} {\bibinfo {title} {Probing single electrons across 300-mm spin
  qubit wafers},\ }\href {https://doi.org/10.1038/s41586-024-07275-6} {\bibfield  {journal} {\bibinfo  {journal} {Nature}\ }\textbf {\bibinfo {volume} {629}},\ \bibinfo {pages} {80} (\bibinfo {year} {2024})}\BibitemShut {NoStop}%
\bibitem [{\citenamefont {Huckemann}\ \emph {et~al.}(2025)\citenamefont {Huckemann}, \citenamefont {Muster}, \citenamefont {Langheinrich}, \citenamefont {Brackmann}, \citenamefont {Friedrich}, \citenamefont {Komerički}, \citenamefont {Diebel}, \citenamefont {Stieß}, \citenamefont {Bougeard}, \citenamefont {Yamamoto}, \citenamefont {Reichmann}, \citenamefont {Zoellner}, \citenamefont {Dahl}, \citenamefont {Schreiber},\ and\ \citenamefont {Bluhm}}]{Huckemann25}%
  \BibitemOpen
  \bibfield  {author} {\bibinfo {author} {\bibfnamefont {T.}~\bibnamefont {Huckemann}}, \bibinfo {author} {\bibfnamefont {P.}~\bibnamefont {Muster}}, \bibinfo {author} {\bibfnamefont {W.}~\bibnamefont {Langheinrich}}, \bibinfo {author} {\bibfnamefont {V.}~\bibnamefont {Brackmann}}, \bibinfo {author} {\bibfnamefont {M.}~\bibnamefont {Friedrich}}, \bibinfo {author} {\bibfnamefont {N.~D.}\ \bibnamefont {Komerički}}, \bibinfo {author} {\bibfnamefont {L.~K.}\ \bibnamefont {Diebel}}, \bibinfo {author} {\bibfnamefont {V.}~\bibnamefont {Stieß}}, \bibinfo {author} {\bibfnamefont {D.}~\bibnamefont {Bougeard}}, \bibinfo {author} {\bibfnamefont {Y.}~\bibnamefont {Yamamoto}}, \bibinfo {author} {\bibfnamefont {F.}~\bibnamefont {Reichmann}}, \bibinfo {author} {\bibfnamefont {M.~H.}\ \bibnamefont {Zoellner}}, \bibinfo {author} {\bibfnamefont {C.}~\bibnamefont {Dahl}}, \bibinfo {author} {\bibfnamefont {L.~R.}\ \bibnamefont {Schreiber}},\ and\ \bibinfo {author} {\bibfnamefont {H.}~\bibnamefont {Bluhm}},\ }\bibfield  {title}
  {\bibinfo {title} {Industrially fabricated single-electron quantum dots in {Si/SiGe} heterostructures},\ }\href {https://doi.org/https://ieeexplore.ieee.org/document/10937186} {\bibfield  {journal} {\bibinfo  {journal} {IEEE Electron Device Lett.}\ }\textbf {\bibinfo {volume} {46}},\ \bibinfo {pages} {868} (\bibinfo {year} {2025})}\BibitemShut {NoStop}%
\bibitem [{\citenamefont {Seidler}\ \emph {et~al.}(2022)\citenamefont {Seidler}, \citenamefont {Struck}, \citenamefont {Xue}, \citenamefont {Focke}, \citenamefont {Trellenkamp}, \citenamefont {Bluhm},\ and\ \citenamefont {Schreiber}}]{Seidler22}%
  \BibitemOpen
  \bibfield  {author} {\bibinfo {author} {\bibfnamefont {I.}~\bibnamefont {Seidler}}, \bibinfo {author} {\bibfnamefont {T.}~\bibnamefont {Struck}}, \bibinfo {author} {\bibfnamefont {R.}~\bibnamefont {Xue}}, \bibinfo {author} {\bibfnamefont {N.}~\bibnamefont {Focke}}, \bibinfo {author} {\bibfnamefont {S.}~\bibnamefont {Trellenkamp}}, \bibinfo {author} {\bibfnamefont {H.}~\bibnamefont {Bluhm}},\ and\ \bibinfo {author} {\bibfnamefont {L.~R.}\ \bibnamefont {Schreiber}},\ }\bibfield  {title} {\bibinfo {title} {{Conveyor-mode single-electron shuttling in {Si/SiGe} for a scalable quantum computing architecture}},\ }\href {https://doi.org/10.1038/s41534-022-00615-2} {\bibfield  {journal} {\bibinfo  {journal} {npj Quantum Inf.}\ }\textbf {\bibinfo {volume} {8}},\ \bibinfo {pages} {100} (\bibinfo {year} {2022})}\BibitemShut {NoStop}%
\bibitem [{\citenamefont {Künne}\ \emph {et~al.}(2024)\citenamefont {Künne}, \citenamefont {Willmes}, \citenamefont {Oberländer}, \citenamefont {Gorjaew}, \citenamefont {Teske}, \citenamefont {Bhardwaj}, \citenamefont {Beer}, \citenamefont {Kammerloher}, \citenamefont {Otten}, \citenamefont {Seidler}, \citenamefont {Xue}, \citenamefont {Schreiber},\ and\ \citenamefont {Bluhm}}]{Kuenne23}%
  \BibitemOpen
  \bibfield  {author} {\bibinfo {author} {\bibfnamefont {M.}~\bibnamefont {Künne}}, \bibinfo {author} {\bibfnamefont {A.}~\bibnamefont {Willmes}}, \bibinfo {author} {\bibfnamefont {M.}~\bibnamefont {Oberländer}}, \bibinfo {author} {\bibfnamefont {C.}~\bibnamefont {Gorjaew}}, \bibinfo {author} {\bibfnamefont {J.~D.}\ \bibnamefont {Teske}}, \bibinfo {author} {\bibfnamefont {H.}~\bibnamefont {Bhardwaj}}, \bibinfo {author} {\bibfnamefont {M.}~\bibnamefont {Beer}}, \bibinfo {author} {\bibfnamefont {E.}~\bibnamefont {Kammerloher}}, \bibinfo {author} {\bibfnamefont {R.}~\bibnamefont {Otten}}, \bibinfo {author} {\bibfnamefont {I.}~\bibnamefont {Seidler}}, \bibinfo {author} {\bibfnamefont {R.}~\bibnamefont {Xue}}, \bibinfo {author} {\bibfnamefont {L.~R.}\ \bibnamefont {Schreiber}},\ and\ \bibinfo {author} {\bibfnamefont {H.}~\bibnamefont {Bluhm}},\ }\bibfield  {title} {\bibinfo {title} {The {SpinBus} architecture for scaling spin qubits with electron shuttling},\ }\href {https://doi.org/10.1038/s41467-024-49182-4}
  {\bibfield  {journal} {\bibinfo  {journal} {Nat. Commun.}\ }\textbf {\bibinfo {volume} {15}},\ \bibinfo {pages} {4977} (\bibinfo {year} {2024})}\BibitemShut {NoStop}%
\bibitem [{\citenamefont {Xue}\ \emph {et~al.}(2024)\citenamefont {Xue}, \citenamefont {Beer}, \citenamefont {Seidler}, \citenamefont {Humpohl}, \citenamefont {Tu}, \citenamefont {Trellenkamp}, \citenamefont {Struck}, \citenamefont {Bluhm},\ and\ \citenamefont {Schreiber}}]{Xue23}%
  \BibitemOpen
  \bibfield  {author} {\bibinfo {author} {\bibfnamefont {R.}~\bibnamefont {Xue}}, \bibinfo {author} {\bibfnamefont {M.}~\bibnamefont {Beer}}, \bibinfo {author} {\bibfnamefont {I.}~\bibnamefont {Seidler}}, \bibinfo {author} {\bibfnamefont {S.}~\bibnamefont {Humpohl}}, \bibinfo {author} {\bibfnamefont {J.-S.}\ \bibnamefont {Tu}}, \bibinfo {author} {\bibfnamefont {S.}~\bibnamefont {Trellenkamp}}, \bibinfo {author} {\bibfnamefont {T.}~\bibnamefont {Struck}}, \bibinfo {author} {\bibfnamefont {H.}~\bibnamefont {Bluhm}},\ and\ \bibinfo {author} {\bibfnamefont {L.~R.}\ \bibnamefont {Schreiber}},\ }\bibfield  {title} {\bibinfo {title} {Si/{SiGe} {QuBus} for single electron information-processing devices with memory and micron-scale connectivity function},\ }\href {https://doi.org/10.1038/s41467-024-46519-x} {\bibfield  {journal} {\bibinfo  {journal} {Nat. Commun.}\ }\textbf {\bibinfo {volume} {15}},\ \bibinfo {pages} {2296} (\bibinfo {year} {2024})}\BibitemShut {NoStop}%
\bibitem [{\citenamefont {Struck}\ \emph {et~al.}(2024)\citenamefont {Struck}, \citenamefont {Volmer}, \citenamefont {Visser}, \citenamefont {Offermann}, \citenamefont {Xue}, \citenamefont {Tu}, \citenamefont {Trellenkamp}, \citenamefont {Cywi{\'n}ski}, \citenamefont {Bluhm},\ and\ \citenamefont {Schreiber}}]{Struck23}%
  \BibitemOpen
  \bibfield  {author} {\bibinfo {author} {\bibfnamefont {T.}~\bibnamefont {Struck}}, \bibinfo {author} {\bibfnamefont {M.}~\bibnamefont {Volmer}}, \bibinfo {author} {\bibfnamefont {L.}~\bibnamefont {Visser}}, \bibinfo {author} {\bibfnamefont {T.}~\bibnamefont {Offermann}}, \bibinfo {author} {\bibfnamefont {R.}~\bibnamefont {Xue}}, \bibinfo {author} {\bibfnamefont {J.-S.}\ \bibnamefont {Tu}}, \bibinfo {author} {\bibfnamefont {S.}~\bibnamefont {Trellenkamp}}, \bibinfo {author} {\bibfnamefont {{\L}.}~\bibnamefont {Cywi{\'n}ski}}, \bibinfo {author} {\bibfnamefont {H.}~\bibnamefont {Bluhm}},\ and\ \bibinfo {author} {\bibfnamefont {L.~R.}\ \bibnamefont {Schreiber}},\ }\bibfield  {title} {\bibinfo {title} {Spin-{EPR}-pair separation by conveyor-mode single electron shuttling in {Si}/{SiGe}},\ }\href {https://doi.org/10.1038/s41467-024-45583-7} {\bibfield  {journal} {\bibinfo  {journal} {Nat. Commun.}\ }\textbf {\bibinfo {volume} {15}},\ \bibinfo {pages} {1325} (\bibinfo {year} {2024})}\BibitemShut {NoStop}%
\bibitem [{\citenamefont {De~Smet}\ \emph {et~al.}(2025)\citenamefont {De~Smet}, \citenamefont {Matsumoto}, \citenamefont {Zwerver}, \citenamefont {Tryputen}, \citenamefont {de~Snoo}, \citenamefont {Amitonov}, \citenamefont {Katiraee-Far}, \citenamefont {Sammak}, \citenamefont {Samkharadze}, \citenamefont {G{\"u}l}, \citenamefont {Wasserman}, \citenamefont {Greplov{\'a}}, \citenamefont {Rimbach-Russ}, \citenamefont {Scappucci},\ and\ \citenamefont {Vandersypen}}]{desmet24}%
  \BibitemOpen
  \bibfield  {author} {\bibinfo {author} {\bibfnamefont {M.}~\bibnamefont {De~Smet}}, \bibinfo {author} {\bibfnamefont {Y.}~\bibnamefont {Matsumoto}}, \bibinfo {author} {\bibfnamefont {A.-M.~J.}\ \bibnamefont {Zwerver}}, \bibinfo {author} {\bibfnamefont {L.}~\bibnamefont {Tryputen}}, \bibinfo {author} {\bibfnamefont {S.~L.}\ \bibnamefont {de~Snoo}}, \bibinfo {author} {\bibfnamefont {S.~V.}\ \bibnamefont {Amitonov}}, \bibinfo {author} {\bibfnamefont {S.~R.}\ \bibnamefont {Katiraee-Far}}, \bibinfo {author} {\bibfnamefont {A.}~\bibnamefont {Sammak}}, \bibinfo {author} {\bibfnamefont {N.}~\bibnamefont {Samkharadze}}, \bibinfo {author} {\bibfnamefont {{\"O}.}~\bibnamefont {G{\"u}l}}, \bibinfo {author} {\bibfnamefont {R.~N.~M.}\ \bibnamefont {Wasserman}}, \bibinfo {author} {\bibfnamefont {E.}~\bibnamefont {Greplov{\'a}}}, \bibinfo {author} {\bibfnamefont {M.}~\bibnamefont {Rimbach-Russ}}, \bibinfo {author} {\bibfnamefont {G.}~\bibnamefont {Scappucci}},\ and\ \bibinfo {author} {\bibfnamefont {L.~M.~K.}\ \bibnamefont
  {Vandersypen}},\ }\bibfield  {title} {\bibinfo {title} {High-fidelity single-spin shuttling in silicon},\ }\href {https://doi.org/https://doi.org/10.1038/s41565-025-01920-5} {\bibfield  {journal} {\bibinfo  {journal} {Nat. Nanotechnol.}\ }\textbf {\bibinfo {volume} {20}},\ \bibinfo {pages} {866} (\bibinfo {year} {2025})}\BibitemShut {NoStop}%
\bibitem [{\citenamefont {Matsumoto}\ \emph {et~al.}(2026)\citenamefont {Matsumoto}, \citenamefont {Smet}, \citenamefont {Tryputen}, \citenamefont {de~Snoo}, \citenamefont {Amitonov}, \citenamefont {Sammak}, \citenamefont {Rimbach-Russ}, \citenamefont {Scappucci},\ and\ \citenamefont {Vandersypen}}]{matsumoto25}%
  \BibitemOpen
  \bibfield  {author} {\bibinfo {author} {\bibfnamefont {Y.}~\bibnamefont {Matsumoto}}, \bibinfo {author} {\bibfnamefont {M.~D.}\ \bibnamefont {Smet}}, \bibinfo {author} {\bibfnamefont {L.}~\bibnamefont {Tryputen}}, \bibinfo {author} {\bibfnamefont {S.~L.}\ \bibnamefont {de~Snoo}}, \bibinfo {author} {\bibfnamefont {S.~V.}\ \bibnamefont {Amitonov}}, \bibinfo {author} {\bibfnamefont {A.}~\bibnamefont {Sammak}}, \bibinfo {author} {\bibfnamefont {M.}~\bibnamefont {Rimbach-Russ}}, \bibinfo {author} {\bibfnamefont {G.}~\bibnamefont {Scappucci}},\ and\ \bibinfo {author} {\bibfnamefont {L.~M.~K.}\ \bibnamefont {Vandersypen}},\ }\bibfield  {title} {\bibinfo {title} {Two-qubit logic and teleportation with mobile spin qubits in silicon},\ }\href {https://doi.org/https://doi.org/10.1038/s41586-026-10423-9} {\bibfield  {journal} {\bibinfo  {journal} {Nature}\ }\textbf {\bibinfo {volume} {653}},\ \bibinfo {pages} {391–397} (\bibinfo {year} {2026})}\BibitemShut {NoStop}%
\bibitem [{\citenamefont {Beer}\ \emph {et~al.}()\citenamefont {Beer}, \citenamefont {Xue}, \citenamefont {Deda}, \citenamefont {Trellenkamp}, \citenamefont {Tu}, \citenamefont {Surrey}, \citenamefont {Seidler}, \citenamefont {Bluhm},\ and\ \citenamefont {Schreiber}}]{Beer25}%
  \BibitemOpen
  \bibfield  {author} {\bibinfo {author} {\bibfnamefont {M.}~\bibnamefont {Beer}}, \bibinfo {author} {\bibfnamefont {R.}~\bibnamefont {Xue}}, \bibinfo {author} {\bibfnamefont {L.}~\bibnamefont {Deda}}, \bibinfo {author} {\bibfnamefont {S.}~\bibnamefont {Trellenkamp}}, \bibinfo {author} {\bibfnamefont {J.-S.}\ \bibnamefont {Tu}}, \bibinfo {author} {\bibfnamefont {P.}~\bibnamefont {Surrey}}, \bibinfo {author} {\bibfnamefont {I.}~\bibnamefont {Seidler}}, \bibinfo {author} {\bibfnamefont {H.}~\bibnamefont {Bluhm}},\ and\ \bibinfo {author} {\bibfnamefont {L.~R.}\ \bibnamefont {Schreiber}},\ }\bibfield  {title} {\bibinfo {title} {Conveyor-mode electron shuttling through a {T}-junction in {Si/SiGe}},\ }\href {10.48550/arXiv.2601.03942} {\ }\Eprint {https://arxiv.org/abs/arXiv:2601.03942} {arXiv:2601.03942} \BibitemShut {NoStop}%
\bibitem [{\citenamefont {Vandersypen}\ \emph {et~al.}(2017)\citenamefont {Vandersypen}, \citenamefont {Bluhm}, \citenamefont {Clarke}, \citenamefont {Dzurak}, \citenamefont {Ishihara}, \citenamefont {Morello}, \citenamefont {Reilly}, \citenamefont {Schreiber},\ and\ \citenamefont {Veldhorst}}]{Vandersypen17}%
  \BibitemOpen
  \bibfield  {author} {\bibinfo {author} {\bibfnamefont {L.~M.~K.}\ \bibnamefont {Vandersypen}}, \bibinfo {author} {\bibfnamefont {H.}~\bibnamefont {Bluhm}}, \bibinfo {author} {\bibfnamefont {J.~S.}\ \bibnamefont {Clarke}}, \bibinfo {author} {\bibfnamefont {A.~S.}\ \bibnamefont {Dzurak}}, \bibinfo {author} {\bibfnamefont {R.}~\bibnamefont {Ishihara}}, \bibinfo {author} {\bibfnamefont {A.}~\bibnamefont {Morello}}, \bibinfo {author} {\bibfnamefont {D.~J.}\ \bibnamefont {Reilly}}, \bibinfo {author} {\bibfnamefont {L.~R.}\ \bibnamefont {Schreiber}},\ and\ \bibinfo {author} {\bibfnamefont {M.}~\bibnamefont {Veldhorst}},\ }\bibfield  {title} {\bibinfo {title} {Interfacing spin qubits in quantum dots and donors---hot, dense, and coherent},\ }\href {https://doi.org/10.1038/s41534-017-0038-y} {\bibfield  {journal} {\bibinfo  {journal} {npj Quantum Inf.}\ }\textbf {\bibinfo {volume} {3}},\ \bibinfo {pages} {34} (\bibinfo {year} {2017})}\BibitemShut {NoStop}%
\bibitem [{\citenamefont {Boter}\ \emph {et~al.}(2022)\citenamefont {Boter}, \citenamefont {Dehollain}, \citenamefont {van Dijk}, \citenamefont {Xu}, \citenamefont {Hensgens}, \citenamefont {Versluis}, \citenamefont {Naus}, \citenamefont {Clarke}, \citenamefont {Veldhorst}, \citenamefont {Sebastiano},\ and\ \citenamefont {Vandersypen}}]{Boter22}%
  \BibitemOpen
  \bibfield  {author} {\bibinfo {author} {\bibfnamefont {J.~M.}\ \bibnamefont {Boter}}, \bibinfo {author} {\bibfnamefont {J.~P.}\ \bibnamefont {Dehollain}}, \bibinfo {author} {\bibfnamefont {J.~P.}\ \bibnamefont {van Dijk}}, \bibinfo {author} {\bibfnamefont {Y.}~\bibnamefont {Xu}}, \bibinfo {author} {\bibfnamefont {T.}~\bibnamefont {Hensgens}}, \bibinfo {author} {\bibfnamefont {R.}~\bibnamefont {Versluis}}, \bibinfo {author} {\bibfnamefont {H.~W.}\ \bibnamefont {Naus}}, \bibinfo {author} {\bibfnamefont {J.~S.}\ \bibnamefont {Clarke}}, \bibinfo {author} {\bibfnamefont {M.}~\bibnamefont {Veldhorst}}, \bibinfo {author} {\bibfnamefont {F.}~\bibnamefont {Sebastiano}},\ and\ \bibinfo {author} {\bibfnamefont {L.~M.}\ \bibnamefont {Vandersypen}},\ }\bibfield  {title} {\bibinfo {title} {Spiderweb array: A sparse spin-qubit array},\ }\href {https://doi.org/10.1103/PhysRevApplied.18.024053} {\bibfield  {journal} {\bibinfo  {journal} {Phys. Rev. Appl.}\ }\textbf {\bibinfo {volume} {18}},\ \bibinfo {pages} {024053}
  (\bibinfo {year} {2022})}\BibitemShut {NoStop}%
\bibitem [{\citenamefont {Zhao}\ \emph {et~al.}(2025)\citenamefont {Zhao}, \citenamefont {Han}, \citenamefont {Han}, \citenamefont {Schreiber}, \citenamefont {Lee}, \citenamefont {Chiang}, \citenamefont {Radu}, \citenamefont {Enz}, \citenamefont {Grützmacher}, \citenamefont {Stampfer}, \citenamefont {Takagi},\ and\ \citenamefont {Knoch}}]{Zhao25}%
  \BibitemOpen
  \bibfield  {author} {\bibinfo {author} {\bibfnamefont {Q.-T.}\ \bibnamefont {Zhao}}, \bibinfo {author} {\bibfnamefont {Y.}~\bibnamefont {Han}}, \bibinfo {author} {\bibfnamefont {H.-C.}\ \bibnamefont {Han}}, \bibinfo {author} {\bibfnamefont {L.~R.}\ \bibnamefont {Schreiber}}, \bibinfo {author} {\bibfnamefont {T.-E.}\ \bibnamefont {Lee}}, \bibinfo {author} {\bibfnamefont {H.-L.}\ \bibnamefont {Chiang}}, \bibinfo {author} {\bibfnamefont {I.}~\bibnamefont {Radu}}, \bibinfo {author} {\bibfnamefont {C.}~\bibnamefont {Enz}}, \bibinfo {author} {\bibfnamefont {D.}~\bibnamefont {Grützmacher}}, \bibinfo {author} {\bibfnamefont {C.}~\bibnamefont {Stampfer}}, \bibinfo {author} {\bibfnamefont {S.}~\bibnamefont {Takagi}},\ and\ \bibinfo {author} {\bibfnamefont {J.}~\bibnamefont {Knoch}},\ }\bibfield  {title} {\bibinfo {title} {Ultra-low-power cryogenic complementary metal oxide semiconductor technology},\ }\href {https://doi.org/10.1038/s44287-025-00157-7} {\bibfield  {journal} {\bibinfo  {journal} {Nat. Rev. Electr.
  Eng.}\ }\textbf {\bibinfo {volume} {2}},\ \bibinfo {pages} {277} (\bibinfo {year} {2025})}\BibitemShut {NoStop}%
\bibitem [{\citenamefont {Ginzel}\ \emph {et~al.}(2024)\citenamefont {Ginzel}, \citenamefont {Fellner}, \citenamefont {Ertler}, \citenamefont {Schreiber}, \citenamefont {Bluhm},\ and\ \citenamefont {Lechner}}]{Ginzel24}%
  \BibitemOpen
  \bibfield  {author} {\bibinfo {author} {\bibfnamefont {F.}~\bibnamefont {Ginzel}}, \bibinfo {author} {\bibfnamefont {M.}~\bibnamefont {Fellner}}, \bibinfo {author} {\bibfnamefont {C.}~\bibnamefont {Ertler}}, \bibinfo {author} {\bibfnamefont {L.~R.}\ \bibnamefont {Schreiber}}, \bibinfo {author} {\bibfnamefont {H.}~\bibnamefont {Bluhm}},\ and\ \bibinfo {author} {\bibfnamefont {W.}~\bibnamefont {Lechner}},\ }\bibfield  {title} {\bibinfo {title} {Scalable parity architecture with a shuttling-based spin qubit processor},\ }\href {https://doi.org/10.1103/PhysRevB.110.075302} {\bibfield  {journal} {\bibinfo  {journal} {Phys. Rev. B}\ }\textbf {\bibinfo {volume} {110}},\ \bibinfo {pages} {075302} (\bibinfo {year} {2024})}\BibitemShut {NoStop}%
\bibitem [{\citenamefont {Langrock}\ \emph {et~al.}(2023)\citenamefont {Langrock}, \citenamefont {Krzywda}, \citenamefont {Focke}, \citenamefont {Seidler}, \citenamefont {Schreiber},\ and\ \citenamefont {Cywi\'{n}ski}}]{Langrock23}%
  \BibitemOpen
  \bibfield  {author} {\bibinfo {author} {\bibfnamefont {V.}~\bibnamefont {Langrock}}, \bibinfo {author} {\bibfnamefont {J.~A.}\ \bibnamefont {Krzywda}}, \bibinfo {author} {\bibfnamefont {N.}~\bibnamefont {Focke}}, \bibinfo {author} {\bibfnamefont {I.}~\bibnamefont {Seidler}}, \bibinfo {author} {\bibfnamefont {L.~R.}\ \bibnamefont {Schreiber}},\ and\ \bibinfo {author} {\bibfnamefont {L.}~\bibnamefont {Cywi\'{n}ski}},\ }\bibfield  {title} {\bibinfo {title} {Blueprint of a scalable spin qubit shuttle device for coherent mid-range qubit transfer in disordered {Si/SiGe/SiO}$_{2}$},\ }\href {https://doi.org/10.1103/PRXQuantum.4.020305} {\bibfield  {journal} {\bibinfo  {journal} {PRX Quantum}\ }\textbf {\bibinfo {volume} {4}},\ \bibinfo {pages} {020305} (\bibinfo {year} {2023})}\BibitemShut {NoStop}%
\bibitem [{\citenamefont {Losert}\ \emph {et~al.}(2023)\citenamefont {Losert}, \citenamefont {Eriksson}, \citenamefont {Joynt}, \citenamefont {Rahman}, \citenamefont {Scappucci}, \citenamefont {Coppersmith},\ and\ \citenamefont {Friesen}}]{Losert23}%
  \BibitemOpen
  \bibfield  {author} {\bibinfo {author} {\bibfnamefont {M.~P.}\ \bibnamefont {Losert}}, \bibinfo {author} {\bibfnamefont {M.~A.}\ \bibnamefont {Eriksson}}, \bibinfo {author} {\bibfnamefont {R.}~\bibnamefont {Joynt}}, \bibinfo {author} {\bibfnamefont {R.}~\bibnamefont {Rahman}}, \bibinfo {author} {\bibfnamefont {G.}~\bibnamefont {Scappucci}}, \bibinfo {author} {\bibfnamefont {S.~N.}\ \bibnamefont {Coppersmith}},\ and\ \bibinfo {author} {\bibfnamefont {M.}~\bibnamefont {Friesen}},\ }\bibfield  {title} {\bibinfo {title} {Practical strategies for enhancing the valley splitting in {Si/SiGe} quantum wells},\ }\href {https://doi.org/https://doi.org/10.1103/PRXQuantum.5.040322} {\bibfield  {journal} {\bibinfo  {journal} {Phys. Rev. B}\ }\textbf {\bibinfo {volume} {108}},\ \bibinfo {pages} {125405} (\bibinfo {year} {2023})}\BibitemShut {NoStop}%
\bibitem [{\citenamefont {Zwanenburg}\ \emph {et~al.}(2013)\citenamefont {Zwanenburg}, \citenamefont {Dzurak}, \citenamefont {Morello}, \citenamefont {Simmons}, \citenamefont {Hollenberg}, \citenamefont {Klimeck}, \citenamefont {Rogge}, \citenamefont {Coppersmith},\ and\ \citenamefont {Eriksson}}]{Zwanenburg13}%
  \BibitemOpen
  \bibfield  {author} {\bibinfo {author} {\bibfnamefont {F.~A.}\ \bibnamefont {Zwanenburg}}, \bibinfo {author} {\bibfnamefont {A.~S.}\ \bibnamefont {Dzurak}}, \bibinfo {author} {\bibfnamefont {A.}~\bibnamefont {Morello}}, \bibinfo {author} {\bibfnamefont {M.~Y.}\ \bibnamefont {Simmons}}, \bibinfo {author} {\bibfnamefont {L.~C.~L.}\ \bibnamefont {Hollenberg}}, \bibinfo {author} {\bibfnamefont {G.}~\bibnamefont {Klimeck}}, \bibinfo {author} {\bibfnamefont {S.}~\bibnamefont {Rogge}}, \bibinfo {author} {\bibfnamefont {S.~N.}\ \bibnamefont {Coppersmith}},\ and\ \bibinfo {author} {\bibfnamefont {M.~A.}\ \bibnamefont {Eriksson}},\ }\bibfield  {title} {\bibinfo {title} {Silicon quantum electronics},\ }\href {https://doi.org/10.1103/RevModPhys.85.961} {\bibfield  {journal} {\bibinfo  {journal} {Rev. Mod. Phys.}\ }\textbf {\bibinfo {volume} {85}},\ \bibinfo {pages} {961} (\bibinfo {year} {2013})}\BibitemShut {NoStop}%
\bibitem [{\citenamefont {Kawakami}\ \emph {et~al.}(2014)\citenamefont {Kawakami}, \citenamefont {Scarlino}, \citenamefont {Ward}, \citenamefont {Braakman}, \citenamefont {Savage}, \citenamefont {Lagally}, \citenamefont {Friesen}, \citenamefont {Coppersmith}, \citenamefont {Eriksson},\ and\ \citenamefont {Vandersypen}}]{Kawakami2014}%
  \BibitemOpen
  \bibfield  {author} {\bibinfo {author} {\bibfnamefont {E.}~\bibnamefont {Kawakami}}, \bibinfo {author} {\bibfnamefont {P.}~\bibnamefont {Scarlino}}, \bibinfo {author} {\bibfnamefont {D.~R.}\ \bibnamefont {Ward}}, \bibinfo {author} {\bibfnamefont {F.~R.}\ \bibnamefont {Braakman}}, \bibinfo {author} {\bibfnamefont {D.~E.}\ \bibnamefont {Savage}}, \bibinfo {author} {\bibfnamefont {M.~G.}\ \bibnamefont {Lagally}}, \bibinfo {author} {\bibfnamefont {M.}~\bibnamefont {Friesen}}, \bibinfo {author} {\bibfnamefont {S.~N.}\ \bibnamefont {Coppersmith}}, \bibinfo {author} {\bibfnamefont {M.~A.}\ \bibnamefont {Eriksson}},\ and\ \bibinfo {author} {\bibfnamefont {L.~M.~K.}\ \bibnamefont {Vandersypen}},\ }\bibfield  {title} {\bibinfo {title} {Electrical control of a long-lived spin qubit in a {Si/SiGe} quantum dot},\ }\href {https://doi.org/10.1038/nnano.2014.153} {\bibfield  {journal} {\bibinfo  {journal} {Nat. Nanotechnol.}\ }\textbf {\bibinfo {volume} {9}},\ \bibinfo {pages} {666} (\bibinfo {year} {2014})}\BibitemShut
  {NoStop}%
\bibitem [{\citenamefont {Veldhorst}\ \emph {et~al.}(2015)\citenamefont {Veldhorst}, \citenamefont {Ruskov}, \citenamefont {Yang}, \citenamefont {Hwang}, \citenamefont {Hudson}, \citenamefont {Flatt\'e}, \citenamefont {Tahan}, \citenamefont {Itoh}, \citenamefont {Morello},\ and\ \citenamefont {Dzurak}}]{Veldhorst_PRB15}%
  \BibitemOpen
  \bibfield  {author} {\bibinfo {author} {\bibfnamefont {M.}~\bibnamefont {Veldhorst}}, \bibinfo {author} {\bibfnamefont {R.}~\bibnamefont {Ruskov}}, \bibinfo {author} {\bibfnamefont {C.~H.}\ \bibnamefont {Yang}}, \bibinfo {author} {\bibfnamefont {J.~C.~C.}\ \bibnamefont {Hwang}}, \bibinfo {author} {\bibfnamefont {F.~E.}\ \bibnamefont {Hudson}}, \bibinfo {author} {\bibfnamefont {M.~E.}\ \bibnamefont {Flatt\'e}}, \bibinfo {author} {\bibfnamefont {C.}~\bibnamefont {Tahan}}, \bibinfo {author} {\bibfnamefont {K.~M.}\ \bibnamefont {Itoh}}, \bibinfo {author} {\bibfnamefont {A.}~\bibnamefont {Morello}},\ and\ \bibinfo {author} {\bibfnamefont {A.~S.}\ \bibnamefont {Dzurak}},\ }\bibfield  {title} {\bibinfo {title} {Spin-orbit coupling and operation of multivalley spin qubits},\ }\href {https://doi.org/10.1103/PhysRevB.92.201401} {\bibfield  {journal} {\bibinfo  {journal} {Phys. Rev. B}\ }\textbf {\bibinfo {volume} {92}},\ \bibinfo {pages} {201401} (\bibinfo {year} {2015})}\BibitemShut {NoStop}%
\bibitem [{\citenamefont {Ferdous}\ \emph {et~al.}(2018)\citenamefont {Ferdous}, \citenamefont {Kawakami}, \citenamefont {Scarlino}, \citenamefont {Nowak}, \citenamefont {Ward}, \citenamefont {Savage}, \citenamefont {Lagally}, \citenamefont {Coppersmith}, \citenamefont {Friesen}, \citenamefont {Eriksson},\ and\ \citenamefont {Vandersypen}}]{Ferdous2018}%
  \BibitemOpen
  \bibfield  {author} {\bibinfo {author} {\bibfnamefont {R.}~\bibnamefont {Ferdous}}, \bibinfo {author} {\bibfnamefont {E.}~\bibnamefont {Kawakami}}, \bibinfo {author} {\bibfnamefont {P.}~\bibnamefont {Scarlino}}, \bibinfo {author} {\bibfnamefont {M.~P.}\ \bibnamefont {Nowak}}, \bibinfo {author} {\bibfnamefont {D.}~\bibnamefont {Ward}}, \bibinfo {author} {\bibfnamefont {D.}~\bibnamefont {Savage}}, \bibinfo {author} {\bibfnamefont {M.}~\bibnamefont {Lagally}}, \bibinfo {author} {\bibfnamefont {S.}~\bibnamefont {Coppersmith}}, \bibinfo {author} {\bibfnamefont {M.}~\bibnamefont {Friesen}}, \bibinfo {author} {\bibfnamefont {M.~A.}\ \bibnamefont {Eriksson}},\ and\ \bibinfo {author} {\bibfnamefont {L.~M.~K.}\ \bibnamefont {Vandersypen}},\ }\bibfield  {title} {\bibinfo {title} {Valley dependent anisotropic spin splitting in silicon quantum dots},\ }\href {https://doi.org/10.1038/s41534-018-0075-1} {\bibfield  {journal} {\bibinfo  {journal} {npj Quantum Inf.}\ }\textbf {\bibinfo {volume} {4}},\ \bibinfo {pages} {26}
  (\bibinfo {year} {2018})}\BibitemShut {NoStop}%
\bibitem [{\citenamefont {Tariq}\ and\ \citenamefont {Hu}(2011)}]{Tariq_NPJQI22}%
  \BibitemOpen
  \bibfield  {author} {\bibinfo {author} {\bibfnamefont {B.}~\bibnamefont {Tariq}}\ and\ \bibinfo {author} {\bibfnamefont {X.}~\bibnamefont {Hu}},\ }\bibfield  {title} {\bibinfo {title} {Impact of the valley orbit coupling on exchange gate for spin qubits in silicon},\ }\href {https://doi.org/10.1038/s41534-022-00554-y} {\bibfield  {journal} {\bibinfo  {journal} {npj Quantum Inf.}\ }\textbf {\bibinfo {volume} {8}},\ \bibinfo {pages} {53} (\bibinfo {year} {2011})}\BibitemShut {NoStop}%
\bibitem [{\citenamefont {Losert}\ \emph {et~al.}(2024)\citenamefont {Losert}, \citenamefont {Oberl\"ander}, \citenamefont {Teske}, \citenamefont {Volmer}, \citenamefont {Schreiber}, \citenamefont {Bluhm}, \citenamefont {Coppersmith},\ and\ \citenamefont {Friesen}}]{Losert24}%
  \BibitemOpen
  \bibfield  {author} {\bibinfo {author} {\bibfnamefont {M.~P.}\ \bibnamefont {Losert}}, \bibinfo {author} {\bibfnamefont {M.}~\bibnamefont {Oberl\"ander}}, \bibinfo {author} {\bibfnamefont {J.~D.}\ \bibnamefont {Teske}}, \bibinfo {author} {\bibfnamefont {M.}~\bibnamefont {Volmer}}, \bibinfo {author} {\bibfnamefont {L.~R.}\ \bibnamefont {Schreiber}}, \bibinfo {author} {\bibfnamefont {H.}~\bibnamefont {Bluhm}}, \bibinfo {author} {\bibfnamefont {S.}~\bibnamefont {Coppersmith}},\ and\ \bibinfo {author} {\bibfnamefont {M.}~\bibnamefont {Friesen}},\ }\bibfield  {title} {\bibinfo {title} {Strategies for enhancing spin-shuttling fidelities in $\mathrm{Si}$/$\mathrm{Si}$$\mathrm{Ge}$ quantum wells with random-alloy disorder},\ }\href {https://doi.org/https://doi.org/10.1103/PRXQuantum.5.040322} {\bibfield  {journal} {\bibinfo  {journal} {PRX Quantum}\ }\textbf {\bibinfo {volume} {5}},\ \bibinfo {pages} {040322} (\bibinfo {year} {2024})}\BibitemShut {NoStop}%
\bibitem [{\citenamefont {Volmer}\ \emph {et~al.}(2026)\citenamefont {Volmer}, \citenamefont {Struck}, \citenamefont {Tu}, \citenamefont {Trellenkamp}, \citenamefont {Esposti}, \citenamefont {Scappucci}, \citenamefont {Cywi{\'n}ski}, \citenamefont {Bluhm},\ and\ \citenamefont {Schreiber}}]{Volmer25}%
  \BibitemOpen
  \bibfield  {author} {\bibinfo {author} {\bibfnamefont {M.}~\bibnamefont {Volmer}}, \bibinfo {author} {\bibfnamefont {T.}~\bibnamefont {Struck}}, \bibinfo {author} {\bibfnamefont {J.-S.}\ \bibnamefont {Tu}}, \bibinfo {author} {\bibfnamefont {S.}~\bibnamefont {Trellenkamp}}, \bibinfo {author} {\bibfnamefont {D.~D.}\ \bibnamefont {Esposti}}, \bibinfo {author} {\bibfnamefont {G.}~\bibnamefont {Scappucci}}, \bibinfo {author} {\bibfnamefont {{\L}.}~\bibnamefont {Cywi{\'n}ski}}, \bibinfo {author} {\bibfnamefont {H.}~\bibnamefont {Bluhm}},\ and\ \bibinfo {author} {\bibfnamefont {L.~R.}\ \bibnamefont {Schreiber}},\ }\bibfield  {title} {\bibinfo {title} {Impact of the local valley splitting on the coherence of conveyor-belt spin shuttling in $^{28}${Si}/{SiGe}},\ }\href {https://doi.org/10.1038/s41467-026-74382-5} {\bibfield  {journal} {\bibinfo  {journal} {Nat. Commun.}\ }\textbf {\bibinfo {volume} {17}},\ \bibinfo {pages} {5448} (\bibinfo {year} {2026})}\BibitemShut {NoStop}%
\bibitem [{\citenamefont {Volmer}\ \emph {et~al.}(2024)\citenamefont {Volmer}, \citenamefont {Struck}, \citenamefont {Sala}, \citenamefont {Chen}, \citenamefont {Oberländer}, \citenamefont {Offermann}, \citenamefont {Xue}, \citenamefont {Visser}, \citenamefont {Tu}, \citenamefont {Trellenkamp}, \citenamefont {Cywi{\'n}ski}, \citenamefont {Bluhm},\ and\ \citenamefont {Schreiber}}]{volmer24}%
  \BibitemOpen
  \bibfield  {author} {\bibinfo {author} {\bibfnamefont {M.}~\bibnamefont {Volmer}}, \bibinfo {author} {\bibfnamefont {T.}~\bibnamefont {Struck}}, \bibinfo {author} {\bibfnamefont {A.}~\bibnamefont {Sala}}, \bibinfo {author} {\bibfnamefont {B.}~\bibnamefont {Chen}}, \bibinfo {author} {\bibfnamefont {M.}~\bibnamefont {Oberländer}}, \bibinfo {author} {\bibfnamefont {T.}~\bibnamefont {Offermann}}, \bibinfo {author} {\bibfnamefont {R.}~\bibnamefont {Xue}}, \bibinfo {author} {\bibfnamefont {L.}~\bibnamefont {Visser}}, \bibinfo {author} {\bibfnamefont {J.-S.}\ \bibnamefont {Tu}}, \bibinfo {author} {\bibfnamefont {S.}~\bibnamefont {Trellenkamp}}, \bibinfo {author} {\bibfnamefont {{\L}.}~\bibnamefont {Cywi{\'n}ski}}, \bibinfo {author} {\bibfnamefont {H.}~\bibnamefont {Bluhm}},\ and\ \bibinfo {author} {\bibfnamefont {L.~R.}\ \bibnamefont {Schreiber}},\ }\bibfield  {title} {\bibinfo {title} {Mapping of valley splitting by conveyor-mode spin-coherent electron shuttling},\ }\href
  {https://doi.org/10.1038/s41534-024-00852-7} {\bibfield  {journal} {\bibinfo  {journal} {npj Quantum Inf.}\ }\textbf {\bibinfo {volume} {10}},\ \bibinfo {pages} {61} (\bibinfo {year} {2024})}\BibitemShut {NoStop}%
\bibitem [{\citenamefont {Lima}\ and\ \citenamefont {Burkard}(2025)}]{delima25_2}%
  \BibitemOpen
  \bibfield  {author} {\bibinfo {author} {\bibfnamefont {J.~R.~F.}\ \bibnamefont {Lima}}\ and\ \bibinfo {author} {\bibfnamefont {G.}~\bibnamefont {Burkard}},\ }\bibfield  {title} {\bibinfo {title} {Partial landau-zener transitions and applications to qubit shuttling},\ }\href {https://doi.org/https://doi.org/10.1103/y4xf-zjjx} {\bibfield  {journal} {\bibinfo  {journal} {Phys. Rev. B}\ }\textbf {\bibinfo {volume} {111}},\ \bibinfo {pages} {235439} (\bibinfo {year} {2025})}\BibitemShut {NoStop}%
\bibitem [{\citenamefont {Veldhorst}\ \emph {et~al.}(2017)\citenamefont {Veldhorst}, \citenamefont {Eenink}, \citenamefont {Yang},\ and\ \citenamefont {Dzurak}}]{Veldhorst2017}%
  \BibitemOpen
  \bibfield  {author} {\bibinfo {author} {\bibfnamefont {M.}~\bibnamefont {Veldhorst}}, \bibinfo {author} {\bibfnamefont {H.~G.~J.}\ \bibnamefont {Eenink}}, \bibinfo {author} {\bibfnamefont {C.~H.}\ \bibnamefont {Yang}},\ and\ \bibinfo {author} {\bibfnamefont {A.~S.}\ \bibnamefont {Dzurak}},\ }\bibfield  {title} {\bibinfo {title} {Silicon {CMOS} architecture for a spin-based quantum computer},\ }\href {https://doi.org/https://doi.org/10.1038/s41467-017-01905-6} {\bibfield  {journal} {\bibinfo  {journal} {Nat. Commun.}\ }\textbf {\bibinfo {volume} {8}},\ \bibinfo {pages} {1766} (\bibinfo {year} {2017})}\BibitemShut {NoStop}%
\bibitem [{\citenamefont {Patom{\"a}ki}\ \emph {et~al.}(2024)\citenamefont {Patom{\"a}ki}, \citenamefont {Gonzalez-Zalba}, \citenamefont {Fogarty}, \citenamefont {Cai}, \citenamefont {Benjamin},\ and\ \citenamefont {Morton}}]{Patomaeki24}%
  \BibitemOpen
  \bibfield  {author} {\bibinfo {author} {\bibfnamefont {S.~M.}\ \bibnamefont {Patom{\"a}ki}}, \bibinfo {author} {\bibfnamefont {M.~F.}\ \bibnamefont {Gonzalez-Zalba}}, \bibinfo {author} {\bibfnamefont {M.~A.}\ \bibnamefont {Fogarty}}, \bibinfo {author} {\bibfnamefont {Z.}~\bibnamefont {Cai}}, \bibinfo {author} {\bibfnamefont {S.~C.}\ \bibnamefont {Benjamin}},\ and\ \bibinfo {author} {\bibfnamefont {J.~J.~L.}\ \bibnamefont {Morton}},\ }\bibfield  {title} {\bibinfo {title} {Pipeline quantum processor architecture for silicon spin qubits},\ }\href {https://doi.org/https://doi.org/10.1038/s41534-024-00823-y} {\bibfield  {journal} {\bibinfo  {journal} {npj Quantum Inf.}\ }\textbf {\bibinfo {volume} {10}},\ \bibinfo {pages} {31} (\bibinfo {year} {2024})}\BibitemShut {NoStop}%
\bibitem [{\citenamefont {Friesen}\ \emph {et~al.}(2007)\citenamefont {Friesen}, \citenamefont {Chutia}, \citenamefont {Tahan},\ and\ \citenamefont {Coppersmith}}]{Friesen2007}%
  \BibitemOpen
  \bibfield  {author} {\bibinfo {author} {\bibfnamefont {M.}~\bibnamefont {Friesen}}, \bibinfo {author} {\bibfnamefont {S.}~\bibnamefont {Chutia}}, \bibinfo {author} {\bibfnamefont {C.}~\bibnamefont {Tahan}},\ and\ \bibinfo {author} {\bibfnamefont {S.~N.}\ \bibnamefont {Coppersmith}},\ }\bibfield  {title} {\bibinfo {title} {Valley splitting theory of $\mathrm{Si}\mathrm{Ge}/\mathrm{Si}/\mathrm{Si}\mathrm{Ge}$ quantum wells},\ }\href {https://doi.org/10.1103/PhysRevB.75.115318} {\bibfield  {journal} {\bibinfo  {journal} {Phys. Rev. B}\ }\textbf {\bibinfo {volume} {75}},\ \bibinfo {pages} {115318} (\bibinfo {year} {2007})}\BibitemShut {NoStop}%
\bibitem [{\citenamefont {Cywi{\'n}ski}\ \emph {et~al.}()\citenamefont {Cywi{\'n}ski}, \citenamefont {Volmer}, \citenamefont {Struck}, \citenamefont {Scappucci},\ and\ \citenamefont {Schreiber}}]{Cywinski26}%
  \BibitemOpen
  \bibfield  {author} {\bibinfo {author} {\bibfnamefont {{\L}.}~\bibnamefont {Cywi{\'n}ski}}, \bibinfo {author} {\bibfnamefont {M.}~\bibnamefont {Volmer}}, \bibinfo {author} {\bibfnamefont {T.}~\bibnamefont {Struck}}, \bibinfo {author} {\bibfnamefont {G.}~\bibnamefont {Scappucci}},\ and\ \bibinfo {author} {\bibfnamefont {L.}~\bibnamefont {Schreiber}},\ }\bibfield  {title} {\bibinfo {title} {Singlet-triplet oscillations in multivalley {Si} quantum dots},\ }\href {10.48550/arXiv.2604.24689} {\ }\Eprint {https://arxiv.org/abs/arXiv:2604.24689} {arXiv:2604.24689} \BibitemShut {NoStop}%
\bibitem [{\citenamefont {King}\ \emph {et~al.}()\citenamefont {King}, \citenamefont {Kim}, \citenamefont {Reily}, \citenamefont {Marcks}, \citenamefont {Friesen}, \citenamefont {Woods},\ and\ \citenamefont {Eriksson}}]{King26}%
  \BibitemOpen
  \bibfield  {author} {\bibinfo {author} {\bibfnamefont {D.~J.}\ \bibnamefont {King}}, \bibinfo {author} {\bibfnamefont {M.}~\bibnamefont {Kim}}, \bibinfo {author} {\bibfnamefont {J.}~\bibnamefont {Reily}}, \bibinfo {author} {\bibfnamefont {J.~C.}\ \bibnamefont {Marcks}}, \bibinfo {author} {\bibfnamefont {M.}~\bibnamefont {Friesen}}, \bibinfo {author} {\bibfnamefont {B.~D.}\ \bibnamefont {Woods}},\ and\ \bibinfo {author} {\bibfnamefont {M.~A.}\ \bibnamefont {Eriksson}},\ }\bibfield  {title} {\bibinfo {title} {Complete measurement of tunnel- and valley-coupling parameters in a silicon double quantum dot},\ }\href {https://arxiv.org/abs/2607.09638} {\ }\Eprint {https://arxiv.org/abs/2607.09638} {arXiv:2607.09638} \BibitemShut {NoStop}%
\bibitem [{\citenamefont {Tagliaferri}\ \emph {et~al.}(2018)\citenamefont {Tagliaferri}, \citenamefont {Bavdaz}, \citenamefont {Huang}, \citenamefont {Dzurak}, \citenamefont {Culcer},\ and\ \citenamefont {Veldhorst}}]{Tagliaferri_PRB18}%
  \BibitemOpen
  \bibfield  {author} {\bibinfo {author} {\bibfnamefont {M.~L.~V.}\ \bibnamefont {Tagliaferri}}, \bibinfo {author} {\bibfnamefont {P.~L.}\ \bibnamefont {Bavdaz}}, \bibinfo {author} {\bibfnamefont {W.}~\bibnamefont {Huang}}, \bibinfo {author} {\bibfnamefont {A.~S.}\ \bibnamefont {Dzurak}}, \bibinfo {author} {\bibfnamefont {D.}~\bibnamefont {Culcer}},\ and\ \bibinfo {author} {\bibfnamefont {M.}~\bibnamefont {Veldhorst}},\ }\bibfield  {title} {\bibinfo {title} {Impact of valley phase and splitting on readout of silicon spin qubits},\ }\href {https://doi.org/10.1103/PhysRevB.97.245412} {\bibfield  {journal} {\bibinfo  {journal} {Phys. Rev. B}\ }\textbf {\bibinfo {volume} {97}},\ \bibinfo {pages} {245412} (\bibinfo {year} {2018})}\BibitemShut {NoStop}%
\bibitem [{\citenamefont {Teske}\ \emph {et~al.}(2023)\citenamefont {Teske}, \citenamefont {Butt}, \citenamefont {Cerfontaine}, \citenamefont {Burkard},\ and\ \citenamefont {Bluhm}}]{Teske_PRB23}%
  \BibitemOpen
  \bibfield  {author} {\bibinfo {author} {\bibfnamefont {J.~D.}\ \bibnamefont {Teske}}, \bibinfo {author} {\bibfnamefont {F.}~\bibnamefont {Butt}}, \bibinfo {author} {\bibfnamefont {P.}~\bibnamefont {Cerfontaine}}, \bibinfo {author} {\bibfnamefont {G.}~\bibnamefont {Burkard}},\ and\ \bibinfo {author} {\bibfnamefont {H.}~\bibnamefont {Bluhm}},\ }\bibfield  {title} {\bibinfo {title} {Flopping-mode electron dipole spin resonance in the strong-driving regime},\ }\href {https://doi.org/10.1103/PhysRevB.107.035302} {\bibfield  {journal} {\bibinfo  {journal} {Phys. Rev. B}\ }\textbf {\bibinfo {volume} {107}},\ \bibinfo {pages} {035302} (\bibinfo {year} {2023})}\BibitemShut {NoStop}%
\bibitem [{\citenamefont {Losert}\ \emph {et~al.}()\citenamefont {Losert}, \citenamefont {G\"ung\"ord\"u}, \citenamefont {Coppersmith}, \citenamefont {Friesen},\ and\ \citenamefont {Tahan}}]{Losert_arXiv25}%
  \BibitemOpen
  \bibfield  {author} {\bibinfo {author} {\bibfnamefont {M.~P.~R.}\ \bibnamefont {Losert}}, \bibinfo {author} {\bibfnamefont {U.}~\bibnamefont {G\"ung\"ord\"u}}, \bibinfo {author} {\bibfnamefont {S.~N.}\ \bibnamefont {Coppersmith}}, \bibinfo {author} {\bibfnamefont {M.}~\bibnamefont {Friesen}},\ and\ \bibinfo {author} {\bibfnamefont {C.}~\bibnamefont {Tahan}},\ }\bibfield  {title} {\bibinfo {title} {The effects of alloy disorder on strongly-driven flopping mode qubits in {Si/SiGe}},\ }\href {10.48550/arXiv.2512.19658} {\ }\Eprint {https://arxiv.org/abs/arXiv:2512.19658} {arXiv:2512.19658} \BibitemShut {NoStop}%
\bibitem [{\citenamefont {Pazhedath}\ \emph {et~al.}(2025)\citenamefont {Pazhedath}, \citenamefont {David}, \citenamefont {Oberl\"ander}, \citenamefont {M\"uller}, \citenamefont {Calarco}, \citenamefont {Bluhm},\ and\ \citenamefont {Motzoi}}]{Pazhedath24}%
  \BibitemOpen
  \bibfield  {author} {\bibinfo {author} {\bibfnamefont {A.~M.}\ \bibnamefont {Pazhedath}}, \bibinfo {author} {\bibfnamefont {A.}~\bibnamefont {David}}, \bibinfo {author} {\bibfnamefont {M.}~\bibnamefont {Oberl\"ander}}, \bibinfo {author} {\bibfnamefont {M.~M.}\ \bibnamefont {M\"uller}}, \bibinfo {author} {\bibfnamefont {T.}~\bibnamefont {Calarco}}, \bibinfo {author} {\bibfnamefont {H.}~\bibnamefont {Bluhm}},\ and\ \bibinfo {author} {\bibfnamefont {F.}~\bibnamefont {Motzoi}},\ }\bibfield  {title} {\bibinfo {title} {Large spin-shuttling oscillations enabling high-fidelity single-qubit gates},\ }\href {https://doi.org/https://doi.org/10.1103/4lky-413f} {\bibfield  {journal} {\bibinfo  {journal} {Phys. Rev. Appl.}\ }\textbf {\bibinfo {volume} {24}},\ \bibinfo {pages} {034029} (\bibinfo {year} {2025})}\BibitemShut {NoStop}%
\bibitem [{\citenamefont {N\'emeth}\ \emph {et~al.}(2026)\citenamefont {N\'emeth}, \citenamefont {Bandaru}, \citenamefont {Alves}, \citenamefont {Brann}, \citenamefont {Eskandari}, \citenamefont {Soomro}, \citenamefont {Vivrekar}, \citenamefont {Eriksson}, \citenamefont {Losert},\ and\ \citenamefont {Friesen}}]{Nemeth24}%
  \BibitemOpen
  \bibfield  {author} {\bibinfo {author} {\bibfnamefont {R.}~\bibnamefont {N\'emeth}}, \bibinfo {author} {\bibfnamefont {V.~K.}\ \bibnamefont {Bandaru}}, \bibinfo {author} {\bibfnamefont {P.}~\bibnamefont {Alves}}, \bibinfo {author} {\bibfnamefont {E.}~\bibnamefont {Brann}}, \bibinfo {author} {\bibfnamefont {O.~M.}\ \bibnamefont {Eskandari}}, \bibinfo {author} {\bibfnamefont {H.}~\bibnamefont {Soomro}}, \bibinfo {author} {\bibfnamefont {A.}~\bibnamefont {Vivrekar}}, \bibinfo {author} {\bibfnamefont {M.}~\bibnamefont {Eriksson}}, \bibinfo {author} {\bibfnamefont {M.~P.}\ \bibnamefont {Losert}},\ and\ \bibinfo {author} {\bibfnamefont {M.}~\bibnamefont {Friesen}},\ }\bibfield  {title} {\bibinfo {title} {Omnidirectional shuttling to avoid valley excitations in $\mathrm{Si}$/$\mathrm{Si}\mathrm{Ge}$ quantum wells},\ }\href {https://doi.org/10.1103/615j-xjyh} {\bibfield  {journal} {\bibinfo  {journal} {PRX Quantum}\ }\textbf {\bibinfo {volume} {7}},\ \bibinfo {pages} {010336} (\bibinfo {year} {2026})}\BibitemShut
  {NoStop}%
\bibitem [{\citenamefont {McJunkin}\ \emph {et~al.}(2021)\citenamefont {McJunkin}, \citenamefont {MacQuarrie}, \citenamefont {Tom}, \citenamefont {Neyens}, \citenamefont {Dodson}, \citenamefont {Thorgrimsson}, \citenamefont {Corrigan}, \citenamefont {Ercan}, \citenamefont {Savage}, \citenamefont {Lagally}, \citenamefont {Joynt}, \citenamefont {Coppersmith}, \citenamefont {Friesen},\ and\ \citenamefont {Eriksson}}]{McJunkin21}%
  \BibitemOpen
  \bibfield  {author} {\bibinfo {author} {\bibfnamefont {T.}~\bibnamefont {McJunkin}}, \bibinfo {author} {\bibfnamefont {E.~R.}\ \bibnamefont {MacQuarrie}}, \bibinfo {author} {\bibfnamefont {L.}~\bibnamefont {Tom}}, \bibinfo {author} {\bibfnamefont {S.~F.}\ \bibnamefont {Neyens}}, \bibinfo {author} {\bibfnamefont {J.~P.}\ \bibnamefont {Dodson}}, \bibinfo {author} {\bibfnamefont {B.}~\bibnamefont {Thorgrimsson}}, \bibinfo {author} {\bibfnamefont {J.}~\bibnamefont {Corrigan}}, \bibinfo {author} {\bibfnamefont {H.~E.}\ \bibnamefont {Ercan}}, \bibinfo {author} {\bibfnamefont {D.~E.}\ \bibnamefont {Savage}}, \bibinfo {author} {\bibfnamefont {M.~G.}\ \bibnamefont {Lagally}}, \bibinfo {author} {\bibfnamefont {R.}~\bibnamefont {Joynt}}, \bibinfo {author} {\bibfnamefont {S.~N.}\ \bibnamefont {Coppersmith}}, \bibinfo {author} {\bibfnamefont {M.}~\bibnamefont {Friesen}},\ and\ \bibinfo {author} {\bibfnamefont {M.~A.}\ \bibnamefont {Eriksson}},\ }\bibfield  {title} {\bibinfo {title} {Valley splittings in {Si}/{SiGe}
  quantum dots with a germanium spike in the silicon well},\ }\href {https://doi.org/https://doi.org/10.1103/PhysRevB.104.085406} {\bibfield  {journal} {\bibinfo  {journal} {Phys. Rev. B}\ }\textbf {\bibinfo {volume} {104}},\ \bibinfo {pages} {085406} (\bibinfo {year} {2021})}\BibitemShut {NoStop}%
\bibitem [{\citenamefont {Paquelet~Wuetz}\ \emph {et~al.}(2022)\citenamefont {Paquelet~Wuetz}, \citenamefont {Losert}, \citenamefont {Koelling}, \citenamefont {Stehouwer}, \citenamefont {Zwerver}, \citenamefont {Philips}, \citenamefont {Mądzik}, \citenamefont {Xue}, \citenamefont {Zheng}, \citenamefont {Lodari}, \citenamefont {Amitonov}, \citenamefont {Samkharadze}, \citenamefont {Sammak}, \citenamefont {Vandersypen}, \citenamefont {Rahman}, \citenamefont {Coppersmith}, \citenamefont {Moutanabbir}, \citenamefont {Friesen},\ and\ \citenamefont {Scappucci}}]{Wuetz2022}%
  \BibitemOpen
  \bibfield  {author} {\bibinfo {author} {\bibfnamefont {B.}~\bibnamefont {Paquelet~Wuetz}}, \bibinfo {author} {\bibfnamefont {M.~P.}\ \bibnamefont {Losert}}, \bibinfo {author} {\bibfnamefont {S.}~\bibnamefont {Koelling}}, \bibinfo {author} {\bibfnamefont {L.~E.~A.}\ \bibnamefont {Stehouwer}}, \bibinfo {author} {\bibfnamefont {A.-M.~J.}\ \bibnamefont {Zwerver}}, \bibinfo {author} {\bibfnamefont {S.~G.~J.}\ \bibnamefont {Philips}}, \bibinfo {author} {\bibfnamefont {M.~T.}\ \bibnamefont {Mądzik}}, \bibinfo {author} {\bibfnamefont {X.}~\bibnamefont {Xue}}, \bibinfo {author} {\bibfnamefont {G.}~\bibnamefont {Zheng}}, \bibinfo {author} {\bibfnamefont {M.}~\bibnamefont {Lodari}}, \bibinfo {author} {\bibfnamefont {S.~V.}\ \bibnamefont {Amitonov}}, \bibinfo {author} {\bibfnamefont {N.}~\bibnamefont {Samkharadze}}, \bibinfo {author} {\bibfnamefont {A.}~\bibnamefont {Sammak}}, \bibinfo {author} {\bibfnamefont {L.~M.~K.}\ \bibnamefont {Vandersypen}}, \bibinfo {author} {\bibfnamefont {R.}~\bibnamefont {Rahman}},
  \bibinfo {author} {\bibfnamefont {S.~N.}\ \bibnamefont {Coppersmith}}, \bibinfo {author} {\bibfnamefont {O.}~\bibnamefont {Moutanabbir}}, \bibinfo {author} {\bibfnamefont {M.}~\bibnamefont {Friesen}},\ and\ \bibinfo {author} {\bibfnamefont {G.}~\bibnamefont {Scappucci}},\ }\bibfield  {title} {\bibinfo {title} {Atomic fluctuations lifting the energy degeneracy in {Si}/{SiGe} quantum dots},\ }\href {https://doi.org/10.1038/s41467-022-35458-0} {\bibfield  {journal} {\bibinfo  {journal} {Nat. Commun.}\ }\textbf {\bibinfo {volume} {13}},\ \bibinfo {pages} {7730} (\bibinfo {year} {2022})}\BibitemShut {NoStop}%
\bibitem [{\citenamefont {{Degli Esposti}}\ \emph {et~al.}(2024)\citenamefont {{Degli Esposti}}, \citenamefont {Stehouwer}, \citenamefont {G{\"u}l}, \citenamefont {Samkharadze}, \citenamefont {D{\'e}prez}, \citenamefont {Meyer}, \citenamefont {Meijer}, \citenamefont {Tryputen}, \citenamefont {Karwal}, \citenamefont {Botifoll}, \citenamefont {Arbiol}, \citenamefont {Amitonov}, \citenamefont {Vandersypen}, \citenamefont {Sammak}, \citenamefont {Veldhorst},\ and\ \citenamefont {Scappucci}}]{Esposti23}%
  \BibitemOpen
  \bibfield  {author} {\bibinfo {author} {\bibfnamefont {D.}~\bibnamefont {{Degli Esposti}}}, \bibinfo {author} {\bibfnamefont {L.~E.~A.}\ \bibnamefont {Stehouwer}}, \bibinfo {author} {\bibfnamefont {{\"O}.}~\bibnamefont {G{\"u}l}}, \bibinfo {author} {\bibfnamefont {N.}~\bibnamefont {Samkharadze}}, \bibinfo {author} {\bibfnamefont {C.}~\bibnamefont {D{\'e}prez}}, \bibinfo {author} {\bibfnamefont {M.}~\bibnamefont {Meyer}}, \bibinfo {author} {\bibfnamefont {I.~N.}\ \bibnamefont {Meijer}}, \bibinfo {author} {\bibfnamefont {L.}~\bibnamefont {Tryputen}}, \bibinfo {author} {\bibfnamefont {S.}~\bibnamefont {Karwal}}, \bibinfo {author} {\bibfnamefont {M.}~\bibnamefont {Botifoll}}, \bibinfo {author} {\bibfnamefont {J.}~\bibnamefont {Arbiol}}, \bibinfo {author} {\bibfnamefont {S.~V.}\ \bibnamefont {Amitonov}}, \bibinfo {author} {\bibfnamefont {L.~M.~K.}\ \bibnamefont {Vandersypen}}, \bibinfo {author} {\bibfnamefont {A.}~\bibnamefont {Sammak}}, \bibinfo {author} {\bibfnamefont {M.}~\bibnamefont {Veldhorst}},\ and\
  \bibinfo {author} {\bibfnamefont {G.}~\bibnamefont {Scappucci}},\ }\bibfield  {title} {\bibinfo {title} {Low disorder and high valley splitting in silicon},\ }\href {https://doi.org/10.1038/s41534-024-00826-9} {\bibfield  {journal} {\bibinfo  {journal} {npj Quantum Inf.}\ }\textbf {\bibinfo {volume} {10}},\ \bibinfo {pages} {32} (\bibinfo {year} {2024})}\BibitemShut {NoStop}%
\bibitem [{\citenamefont {Klos}\ \emph {et~al.}(2024)\citenamefont {Klos}, \citenamefont {Tr{\"o}ger}, \citenamefont {Keutgen}, \citenamefont {Losert}, \citenamefont {Abrosimov}, \citenamefont {Knoch}, \citenamefont {Bracht}, \citenamefont {Coppersmith}, \citenamefont {Friesen}, \citenamefont {Cojocaru-Mir{\'e}din}, \citenamefont {Schreiber},\ and\ \citenamefont {Bougeard}}]{Klos2024}%
  \BibitemOpen
  \bibfield  {author} {\bibinfo {author} {\bibfnamefont {J.}~\bibnamefont {Klos}}, \bibinfo {author} {\bibfnamefont {J.}~\bibnamefont {Tr{\"o}ger}}, \bibinfo {author} {\bibfnamefont {J.}~\bibnamefont {Keutgen}}, \bibinfo {author} {\bibfnamefont {M.~P.}\ \bibnamefont {Losert}}, \bibinfo {author} {\bibfnamefont {N.~V.}\ \bibnamefont {Abrosimov}}, \bibinfo {author} {\bibfnamefont {J.}~\bibnamefont {Knoch}}, \bibinfo {author} {\bibfnamefont {H.}~\bibnamefont {Bracht}}, \bibinfo {author} {\bibfnamefont {S.~N.}\ \bibnamefont {Coppersmith}}, \bibinfo {author} {\bibfnamefont {M.}~\bibnamefont {Friesen}}, \bibinfo {author} {\bibfnamefont {O.}~\bibnamefont {Cojocaru-Mir{\'e}din}}, \bibinfo {author} {\bibfnamefont {L.~R.}\ \bibnamefont {Schreiber}},\ and\ \bibinfo {author} {\bibfnamefont {D.}~\bibnamefont {Bougeard}},\ }\bibfield  {title} {\bibinfo {title} {Atomistic compositional details and their importance for spin qubits in isotope-purified silicon quantum wells},\ }\href {https://doi.org/10.1002/advs.202407442}
  {\bibfield  {journal} {\bibinfo  {journal} {Adv. Sci.}\ }\textbf {\bibinfo {volume} {11}},\ \bibinfo {pages} {e2407442} (\bibinfo {year} {2024})}\BibitemShut {NoStop}%
\bibitem [{\citenamefont {Thayil}\ \emph {et~al.}(2025)\citenamefont {Thayil}, \citenamefont {Ermoneit},\ and\ \citenamefont {Kantner}}]{thayil25}%
  \BibitemOpen
  \bibfield  {author} {\bibinfo {author} {\bibfnamefont {A.}~\bibnamefont {Thayil}}, \bibinfo {author} {\bibfnamefont {L.}~\bibnamefont {Ermoneit}},\ and\ \bibinfo {author} {\bibfnamefont {M.}~\bibnamefont {Kantner}},\ }\bibfield  {title} {\bibinfo {title} {Theory of valley splitting in {Si/SiGe} spin qubits: Interplay of strain, resonances, and random alloy disorder},\ }\href {https://doi.org/10.1103/4sdz-f9cr} {\bibfield  {journal} {\bibinfo  {journal} {Phys. Rev. B}\ }\textbf {\bibinfo {volume} {112}},\ \bibinfo {pages} {115303} (\bibinfo {year} {2025})}\BibitemShut {NoStop}%
\bibitem [{\citenamefont {Thayil}\ \emph {et~al.}()\citenamefont {Thayil}, \citenamefont {Ermoneit}, \citenamefont {Schreiber}, \citenamefont {Koprucki},\ and\ \citenamefont {Kantner}}]{thayil2025arXiv}%
  \BibitemOpen
  \bibfield  {author} {\bibinfo {author} {\bibfnamefont {A.}~\bibnamefont {Thayil}}, \bibinfo {author} {\bibfnamefont {L.}~\bibnamefont {Ermoneit}}, \bibinfo {author} {\bibfnamefont {L.~R.}\ \bibnamefont {Schreiber}}, \bibinfo {author} {\bibfnamefont {T.}~\bibnamefont {Koprucki}},\ and\ \bibinfo {author} {\bibfnamefont {M.}~\bibnamefont {Kantner}},\ }\bibfield  {title} {\bibinfo {title} {Optimization of si/sige heterostructures for large and robust valley splitting in silicon qubits},\ }\href {https://arxiv.org/abs/2512.18064} {\ }\Eprint {https://arxiv.org/abs/2512.18064} {arXiv:2512.18064} \BibitemShut {NoStop}%
\bibitem [{\citenamefont {Cvitkovich}\ \emph {et~al.}()\citenamefont {Cvitkovich}, \citenamefont {Stano}, \citenamefont {Bougeard}, \citenamefont {Niquet},\ and\ \citenamefont {Loss}}]{Cvitkovich_arXiv26}%
  \BibitemOpen
  \bibfield  {author} {\bibinfo {author} {\bibfnamefont {L.}~\bibnamefont {Cvitkovich}}, \bibinfo {author} {\bibfnamefont {P.}~\bibnamefont {Stano}}, \bibinfo {author} {\bibfnamefont {D.}~\bibnamefont {Bougeard}}, \bibinfo {author} {\bibfnamefont {Y.-M.}\ \bibnamefont {Niquet}},\ and\ \bibinfo {author} {\bibfnamefont {D.}~\bibnamefont {Loss}},\ }\bibfield  {title} {\bibinfo {title} {Increasing valley splitting in {Si/SiGe} by practically achievable heterostructure profiles},\ }\href {10.48550/arXiv.2603.19769} {\ }\Eprint {https://arxiv.org/abs/arXiv:2603.19769} {arXiv:2603.19769} \BibitemShut {NoStop}%
\bibitem [{\citenamefont {Rahlff}\ \emph {et~al.}()\citenamefont {Rahlff}, \citenamefont {Richter}, \citenamefont {Schmidbauer}, \citenamefont {Oezkent}, \citenamefont {Remmele}, \citenamefont {Hanke}, \citenamefont {Schreiber}, \citenamefont {Dütz}, \citenamefont {Umezawa}, \citenamefont {Albrecht}, \citenamefont {Niquet}, \citenamefont {Salamone}, \citenamefont {Diaz}, \citenamefont {Schroeder}, \citenamefont {Martin},\ and\ \citenamefont {Gradwohl}}]{rahlff_arXiv2026}%
  \BibitemOpen
  \bibfield  {author} {\bibinfo {author} {\bibfnamefont {I.}~\bibnamefont {Rahlff}}, \bibinfo {author} {\bibfnamefont {C.}~\bibnamefont {Richter}}, \bibinfo {author} {\bibfnamefont {M.}~\bibnamefont {Schmidbauer}}, \bibinfo {author} {\bibfnamefont {M.}~\bibnamefont {Oezkent}}, \bibinfo {author} {\bibfnamefont {T.}~\bibnamefont {Remmele}}, \bibinfo {author} {\bibfnamefont {M.}~\bibnamefont {Hanke}}, \bibinfo {author} {\bibfnamefont {L.~R.}\ \bibnamefont {Schreiber}}, \bibinfo {author} {\bibfnamefont {D.}~\bibnamefont {Dütz}}, \bibinfo {author} {\bibfnamefont {S.}~\bibnamefont {Umezawa}}, \bibinfo {author} {\bibfnamefont {M.}~\bibnamefont {Albrecht}}, \bibinfo {author} {\bibfnamefont {Y.-M.}\ \bibnamefont {Niquet}}, \bibinfo {author} {\bibfnamefont {T.}~\bibnamefont {Salamone}}, \bibinfo {author} {\bibfnamefont {B.~M.}\ \bibnamefont {Diaz}}, \bibinfo {author} {\bibfnamefont {T.}~\bibnamefont {Schroeder}}, \bibinfo {author} {\bibfnamefont {J.}~\bibnamefont {Martin}},\ and\ \bibinfo {author} {\bibfnamefont
  {K.-P.}\ \bibnamefont {Gradwohl}},\ }\bibfield  {title} {\bibinfo {title} {Sharp periodic ge concentration modulations beyond the conduction band valley wavevector $k_0$ in nuclear spin-free si quantum wells},\ }\href {https://arxiv.org/abs/2605.31358} {\ }\Eprint {https://arxiv.org/abs/arXiv:2605.31358} {arXiv:2605.31358} \BibitemShut {NoStop}%
\bibitem [{\citenamefont {Woods}\ \emph {et~al.}(2024)\citenamefont {Woods}, \citenamefont {Soomro}, \citenamefont {Joseph}, \citenamefont {Frink}, \citenamefont {Joynt}, \citenamefont {Eriksson},\ and\ \citenamefont {Friesen}}]{woods24}%
  \BibitemOpen
  \bibfield  {author} {\bibinfo {author} {\bibfnamefont {B.~D.}\ \bibnamefont {Woods}}, \bibinfo {author} {\bibfnamefont {H.}~\bibnamefont {Soomro}}, \bibinfo {author} {\bibfnamefont {E.~S.}\ \bibnamefont {Joseph}}, \bibinfo {author} {\bibfnamefont {C.~C.~D.}\ \bibnamefont {Frink}}, \bibinfo {author} {\bibfnamefont {R.}~\bibnamefont {Joynt}}, \bibinfo {author} {\bibfnamefont {M.~A.}\ \bibnamefont {Eriksson}},\ and\ \bibinfo {author} {\bibfnamefont {M.}~\bibnamefont {Friesen}},\ }\bibfield  {title} {\bibinfo {title} {Coupling conduction-band valleys in {SiGe} heterostructures via shear strain and {Ge} concentration oscillations},\ }\href {https://doi.org/10.1038/s41534-024-00853-6} {\bibfield  {journal} {\bibinfo  {journal} {npj Quantum Inf.}\ }\textbf {\bibinfo {volume} {10}},\ \bibinfo {pages} {54} (\bibinfo {year} {2024})}\BibitemShut {NoStop}%
\bibitem [{\citenamefont {Woods}\ \emph {et~al.}()\citenamefont {Woods}, \citenamefont {Losert}, \citenamefont {Elston}, \citenamefont {Eriksson}, \citenamefont {Coppersmith}, \citenamefont {Joynt},\ and\ \citenamefont {Friesen}}]{Woods25}%
  \BibitemOpen
  \bibfield  {author} {\bibinfo {author} {\bibfnamefont {B.~D.}\ \bibnamefont {Woods}}, \bibinfo {author} {\bibfnamefont {M.~P.}\ \bibnamefont {Losert}}, \bibinfo {author} {\bibfnamefont {N.~R.}\ \bibnamefont {Elston}}, \bibinfo {author} {\bibfnamefont {M.~A.}\ \bibnamefont {Eriksson}}, \bibinfo {author} {\bibfnamefont {S.~N.}\ \bibnamefont {Coppersmith}}, \bibinfo {author} {\bibfnamefont {R.}~\bibnamefont {Joynt}},\ and\ \bibinfo {author} {\bibfnamefont {M.}~\bibnamefont {Friesen}},\ }\bibfield  {title} {\bibinfo {title} {Statistical characterization of valley coupling in {Si/SiGe} quantum dots via $g$-factor measurements near a valley vortex}\ }\href@noop {} {}\Eprint {https://arxiv.org/abs/2507.05160} {arXiv:2507.05160} \BibitemShut {NoStop}%
\bibitem [{\citenamefont {Woods}\ \emph {et~al.}(2026)\citenamefont {Woods}, \citenamefont {Losert}, \citenamefont {Joynt},\ and\ \citenamefont {Friesen}}]{Woods24_2}%
  \BibitemOpen
  \bibfield  {author} {\bibinfo {author} {\bibfnamefont {B.~D.}\ \bibnamefont {Woods}}, \bibinfo {author} {\bibfnamefont {M.~P.}\ \bibnamefont {Losert}}, \bibinfo {author} {\bibfnamefont {R.}~\bibnamefont {Joynt}},\ and\ \bibinfo {author} {\bibfnamefont {M.}~\bibnamefont {Friesen}},\ }\bibfield  {title} {\bibinfo {title} {$g$-factor theory of {Si/SiGe} quantum dots: Spin-valley and giant renormalization effects},\ }\href {https://doi.org/10.1103/rd84-5g2x} {\bibfield  {journal} {\bibinfo  {journal} {Phys. Rev. Lett.}\ }\textbf {\bibinfo {volume} {136}},\ \bibinfo {pages} {206201} (\bibinfo {year} {2026})}\BibitemShut {NoStop}%
\bibitem [{\citenamefont {Marcogliese}\ \emph {et~al.}(2026)\citenamefont {Marcogliese}, \citenamefont {Sabapathy}, \citenamefont {Richter}, \citenamefont {Tu}, \citenamefont {Bougeard},\ and\ \citenamefont {Schreiber}}]{marcogliese25}%
  \BibitemOpen
  \bibfield  {author} {\bibinfo {author} {\bibfnamefont {L.}~\bibnamefont {Marcogliese}}, \bibinfo {author} {\bibfnamefont {O.}~\bibnamefont {Sabapathy}}, \bibinfo {author} {\bibfnamefont {R.}~\bibnamefont {Richter}}, \bibinfo {author} {\bibfnamefont {J.-S.}\ \bibnamefont {Tu}}, \bibinfo {author} {\bibfnamefont {D.}~\bibnamefont {Bougeard}},\ and\ \bibinfo {author} {\bibfnamefont {L.~R.}\ \bibnamefont {Schreiber}},\ }\bibfield  {title} {\bibinfo {title} {Fabrication, characterization, and mechanical loading of {Si}/{SiGe} membranes for spin-qubit devices},\ }\href {https://doi.org/https://doi.org/10.1103/rw7k-pcvw} {\bibfield  {journal} {\bibinfo  {journal} {Phys. Rev. Appl.}\ }\textbf {\bibinfo {volume} {25}},\ \bibinfo {pages} {014054} (\bibinfo {year} {2026})}\BibitemShut {NoStop}%
\bibitem [{\citenamefont {Borselli}\ \emph {et~al.}(2011)\citenamefont {Borselli}, \citenamefont {Ross}, \citenamefont {Kiselev}, \citenamefont {Croke}, \citenamefont {Holabird}, \citenamefont {Deelman}, \citenamefont {Warren}, \citenamefont {Alvarado-Rodriguez}, \citenamefont {Milosavljevic}, \citenamefont {Ku}, \citenamefont {Wong}, \citenamefont {Schmitz}, \citenamefont {Sokolich}, \citenamefont {Gyure},\ and\ \citenamefont {Hunter}}]{Borselli11}%
  \BibitemOpen
  \bibfield  {author} {\bibinfo {author} {\bibfnamefont {M.~G.}\ \bibnamefont {Borselli}}, \bibinfo {author} {\bibfnamefont {R.~S.}\ \bibnamefont {Ross}}, \bibinfo {author} {\bibfnamefont {A.~A.}\ \bibnamefont {Kiselev}}, \bibinfo {author} {\bibfnamefont {E.~T.}\ \bibnamefont {Croke}}, \bibinfo {author} {\bibfnamefont {K.~S.}\ \bibnamefont {Holabird}}, \bibinfo {author} {\bibfnamefont {P.~W.}\ \bibnamefont {Deelman}}, \bibinfo {author} {\bibfnamefont {L.~D.}\ \bibnamefont {Warren}}, \bibinfo {author} {\bibfnamefont {I.}~\bibnamefont {Alvarado-Rodriguez}}, \bibinfo {author} {\bibfnamefont {I.}~\bibnamefont {Milosavljevic}}, \bibinfo {author} {\bibfnamefont {F.~C.}\ \bibnamefont {Ku}}, \bibinfo {author} {\bibfnamefont {W.~S.}\ \bibnamefont {Wong}}, \bibinfo {author} {\bibfnamefont {A.~E.}\ \bibnamefont {Schmitz}}, \bibinfo {author} {\bibfnamefont {M.}~\bibnamefont {Sokolich}}, \bibinfo {author} {\bibfnamefont {M.~F.}\ \bibnamefont {Gyure}},\ and\ \bibinfo {author} {\bibfnamefont {A.~T.}\ \bibnamefont
  {Hunter}},\ }\bibfield  {title} {\bibinfo {title} {Measurement of valley splitting in high-symmetry {Si}/{SiGe} quantum dots},\ }\href {https://doi.org/10.1063/1.3569717} {\bibfield  {journal} {\bibinfo  {journal} {Appl. Phys. Lett.}\ }\textbf {\bibinfo {volume} {98}},\ \bibinfo {pages} {123118} (\bibinfo {year} {2011})}\BibitemShut {NoStop}%
\bibitem [{\citenamefont {Dodson}\ \emph {et~al.}(2022)\citenamefont {Dodson}, \citenamefont {Ercan}, \citenamefont {Corrigan}, \citenamefont {Losert}, \citenamefont {Holman}, \citenamefont {McJunkin}, \citenamefont {Edge}, \citenamefont {Friesen}, \citenamefont {Coppersmith},\ and\ \citenamefont {Eriksson}}]{Dodson22}%
  \BibitemOpen
  \bibfield  {author} {\bibinfo {author} {\bibfnamefont {J.~P.}\ \bibnamefont {Dodson}}, \bibinfo {author} {\bibfnamefont {H.~E.}\ \bibnamefont {Ercan}}, \bibinfo {author} {\bibfnamefont {J.}~\bibnamefont {Corrigan}}, \bibinfo {author} {\bibfnamefont {M.~P.}\ \bibnamefont {Losert}}, \bibinfo {author} {\bibfnamefont {N.}~\bibnamefont {Holman}}, \bibinfo {author} {\bibfnamefont {T.}~\bibnamefont {McJunkin}}, \bibinfo {author} {\bibfnamefont {L.~F.}\ \bibnamefont {Edge}}, \bibinfo {author} {\bibfnamefont {M.}~\bibnamefont {Friesen}}, \bibinfo {author} {\bibfnamefont {S.~N.}\ \bibnamefont {Coppersmith}},\ and\ \bibinfo {author} {\bibfnamefont {M.~A.}\ \bibnamefont {Eriksson}},\ }\bibfield  {title} {\bibinfo {title} {How valley-orbit states in silicon quantum dots probe quantum well interfaces},\ }\href {https://doi.org/10.1103/PhysRevLett.128.146802} {\bibfield  {journal} {\bibinfo  {journal} {Phys. Rev. Lett.}\ }\textbf {\bibinfo {volume} {128}},\ \bibinfo {pages} {146802} (\bibinfo {year} {2022})}\BibitemShut
  {NoStop}%
\bibitem [{\citenamefont {Hollmann}\ \emph {et~al.}(2020)\citenamefont {Hollmann}, \citenamefont {Struck}, \citenamefont {Langrock}, \citenamefont {Schmidbauer}, \citenamefont {Schauer}, \citenamefont {Leonhardt}, \citenamefont {Sawano}, \citenamefont {Riemann}, \citenamefont {Abrosimov}, \citenamefont {Bougeard},\ and\ \citenamefont {Schreiber}}]{Hollmann20}%
  \BibitemOpen
  \bibfield  {author} {\bibinfo {author} {\bibfnamefont {A.}~\bibnamefont {Hollmann}}, \bibinfo {author} {\bibfnamefont {T.}~\bibnamefont {Struck}}, \bibinfo {author} {\bibfnamefont {V.}~\bibnamefont {Langrock}}, \bibinfo {author} {\bibfnamefont {A.}~\bibnamefont {Schmidbauer}}, \bibinfo {author} {\bibfnamefont {F.}~\bibnamefont {Schauer}}, \bibinfo {author} {\bibfnamefont {T.}~\bibnamefont {Leonhardt}}, \bibinfo {author} {\bibfnamefont {K.}~\bibnamefont {Sawano}}, \bibinfo {author} {\bibfnamefont {H.}~\bibnamefont {Riemann}}, \bibinfo {author} {\bibfnamefont {N.~V.}\ \bibnamefont {Abrosimov}}, \bibinfo {author} {\bibfnamefont {D.}~\bibnamefont {Bougeard}},\ and\ \bibinfo {author} {\bibfnamefont {L.~R.}\ \bibnamefont {Schreiber}},\ }\bibfield  {title} {\bibinfo {title} {Large, tunable valley splitting and single-spin relaxation mechanisms in a {Si}/{Si}$_{x}${Ge}$_{1\ensuremath{-}x}$ quantum dot},\ }\href {https://doi.org/https://doi.org/10.1103/PhysRevApplied.13.034068} {\bibfield  {journal} {\bibinfo
  {journal} {Phys. Rev. Appl.}\ }\textbf {\bibinfo {volume} {13}},\ \bibinfo {pages} {034068} (\bibinfo {year} {2020})}\BibitemShut {NoStop}%
\bibitem [{\citenamefont {Chen}\ \emph {et~al.}(2021)\citenamefont {Chen}, \citenamefont {Raach}, \citenamefont {Pan}, \citenamefont {Kiselev}, \citenamefont {Acuna}, \citenamefont {Blumoff}, \citenamefont {Brecht}, \citenamefont {Choi}, \citenamefont {Ha}, \citenamefont {Hulbert}, \citenamefont {Jura}, \citenamefont {Keating}, \citenamefont {Noah}, \citenamefont {Sun}, \citenamefont {Thomas}, \citenamefont {Borselli}, \citenamefont {Jackson}, \citenamefont {Rakher},\ and\ \citenamefont {Ross}}]{Chen21}%
  \BibitemOpen
  \bibfield  {author} {\bibinfo {author} {\bibfnamefont {E.~H.}\ \bibnamefont {Chen}}, \bibinfo {author} {\bibfnamefont {K.}~\bibnamefont {Raach}}, \bibinfo {author} {\bibfnamefont {A.}~\bibnamefont {Pan}}, \bibinfo {author} {\bibfnamefont {A.~A.}\ \bibnamefont {Kiselev}}, \bibinfo {author} {\bibfnamefont {E.}~\bibnamefont {Acuna}}, \bibinfo {author} {\bibfnamefont {J.~Z.}\ \bibnamefont {Blumoff}}, \bibinfo {author} {\bibfnamefont {T.}~\bibnamefont {Brecht}}, \bibinfo {author} {\bibfnamefont {M.~D.}\ \bibnamefont {Choi}}, \bibinfo {author} {\bibfnamefont {W.}~\bibnamefont {Ha}}, \bibinfo {author} {\bibfnamefont {D.~R.}\ \bibnamefont {Hulbert}}, \bibinfo {author} {\bibfnamefont {M.~P.}\ \bibnamefont {Jura}}, \bibinfo {author} {\bibfnamefont {T.~E.}\ \bibnamefont {Keating}}, \bibinfo {author} {\bibfnamefont {R.}~\bibnamefont {Noah}}, \bibinfo {author} {\bibfnamefont {B.}~\bibnamefont {Sun}}, \bibinfo {author} {\bibfnamefont {B.~J.}\ \bibnamefont {Thomas}}, \bibinfo {author} {\bibfnamefont {M.~G.}\ \bibnamefont
  {Borselli}}, \bibinfo {author} {\bibfnamefont {C.}~\bibnamefont {Jackson}}, \bibinfo {author} {\bibfnamefont {M.~T.}\ \bibnamefont {Rakher}},\ and\ \bibinfo {author} {\bibfnamefont {R.~S.}\ \bibnamefont {Ross}},\ }\bibfield  {title} {\bibinfo {title} {Detuning axis pulsed spectroscopy of valley-orbital states in {Si}/{SiGe} quantum dots},\ }\href {https://doi.org/10.1103/PhysRevApplied.15.044033} {\bibfield  {journal} {\bibinfo  {journal} {Phys. Rev. Appl.}\ }\textbf {\bibinfo {volume} {15}},\ \bibinfo {pages} {044033} (\bibinfo {year} {2021})}\BibitemShut {NoStop}%
\bibitem [{\citenamefont {Marcks}\ \emph {et~al.}(2025)\citenamefont {Marcks}, \citenamefont {Eagen}, \citenamefont {Brann}, \citenamefont {Losert}, \citenamefont {Oh}, \citenamefont {Reily}, \citenamefont {Wang}, \citenamefont {Keith}, \citenamefont {Mohiyaddin}, \citenamefont {Luthi}, \citenamefont {Curry}, \citenamefont {Zhang}, \citenamefont {Heremans}, \citenamefont {Friesen},\ and\ \citenamefont {Eriksson}}]{Marcks25}%
  \BibitemOpen
  \bibfield  {author} {\bibinfo {author} {\bibfnamefont {J.~C.}\ \bibnamefont {Marcks}}, \bibinfo {author} {\bibfnamefont {E.}~\bibnamefont {Eagen}}, \bibinfo {author} {\bibfnamefont {E.~C.}\ \bibnamefont {Brann}}, \bibinfo {author} {\bibfnamefont {M.~P.}\ \bibnamefont {Losert}}, \bibinfo {author} {\bibfnamefont {T.}~\bibnamefont {Oh}}, \bibinfo {author} {\bibfnamefont {J.}~\bibnamefont {Reily}}, \bibinfo {author} {\bibfnamefont {C.~S.}\ \bibnamefont {Wang}}, \bibinfo {author} {\bibfnamefont {D.}~\bibnamefont {Keith}}, \bibinfo {author} {\bibfnamefont {F.~A.}\ \bibnamefont {Mohiyaddin}}, \bibinfo {author} {\bibfnamefont {F.}~\bibnamefont {Luthi}}, \bibinfo {author} {\bibfnamefont {M.~J.}\ \bibnamefont {Curry}}, \bibinfo {author} {\bibfnamefont {J.}~\bibnamefont {Zhang}}, \bibinfo {author} {\bibfnamefont {F.~J.}\ \bibnamefont {Heremans}}, \bibinfo {author} {\bibfnamefont {M.}~\bibnamefont {Friesen}},\ and\ \bibinfo {author} {\bibfnamefont {M.~A.}\ \bibnamefont {Eriksson}},\ }\bibfield  {title} {\bibinfo
  {title} {Valley splitting correlations across a silicon quantum well containing germanium},\ }\href {https://doi.org/https://www.nature.com/articles/s41467-025-67325-z} {\bibfield  {journal} {\bibinfo  {journal} {Nat. Commun.}\ }\textbf {\bibinfo {volume} {16}},\ \bibinfo {pages} {11381} (\bibinfo {year} {2025})}\BibitemShut {NoStop}%
\bibitem [{\citenamefont {Ruskov}\ \emph {et~al.}(2018)\citenamefont {Ruskov}, \citenamefont {Veldhorst}, \citenamefont {Dzurak},\ and\ \citenamefont {Tahan}}]{Ruskov_PRB18}%
  \BibitemOpen
  \bibfield  {author} {\bibinfo {author} {\bibfnamefont {R.}~\bibnamefont {Ruskov}}, \bibinfo {author} {\bibfnamefont {M.}~\bibnamefont {Veldhorst}}, \bibinfo {author} {\bibfnamefont {A.~S.}\ \bibnamefont {Dzurak}},\ and\ \bibinfo {author} {\bibfnamefont {C.}~\bibnamefont {Tahan}},\ }\bibfield  {title} {\bibinfo {title} {Electron $g$-factor of valley states in realistic silicon quantum dots},\ }\href {https://doi.org/10.1103/PhysRevB.98.245424} {\bibfield  {journal} {\bibinfo  {journal} {Phys. Rev. B}\ }\textbf {\bibinfo {volume} {98}},\ \bibinfo {pages} {245424} (\bibinfo {year} {2018})}\BibitemShut {NoStop}%
\bibitem [{\citenamefont {Kanaar}\ \emph {et~al.}()\citenamefont {Kanaar}, \citenamefont {Martinez}, \citenamefont {Zhang},\ and\ \citenamefont {Gyure}}]{kanaar26}%
  \BibitemOpen
  \bibfield  {author} {\bibinfo {author} {\bibfnamefont {D.~W.}\ \bibnamefont {Kanaar}}, \bibinfo {author} {\bibfnamefont {E.}~\bibnamefont {Martinez}}, \bibinfo {author} {\bibfnamefont {P.}~\bibnamefont {Zhang}},\ and\ \bibinfo {author} {\bibfnamefont {M.~F.}\ \bibnamefont {Gyure}},\ }\bibfield  {title} {\bibinfo {title} {Silicon-germanium heterostructures with enhanced valley splitting for spin qubits},\ }\href {https://arxiv.org/abs/2607.09652} {\ }\Eprint {https://arxiv.org/abs/2607.09652} {arXiv:2607.09652} \BibitemShut {NoStop}%
\bibitem [{\citenamefont {Albrecht}\ \emph {et~al.}(2017)\citenamefont {Albrecht}, \citenamefont {Moers},\ and\ \citenamefont {Hermanns}}]{Albrecht17}%
  \BibitemOpen
  \bibfield  {author} {\bibinfo {author} {\bibfnamefont {W.}~\bibnamefont {Albrecht}}, \bibinfo {author} {\bibfnamefont {J.}~\bibnamefont {Moers}},\ and\ \bibinfo {author} {\bibfnamefont {B.}~\bibnamefont {Hermanns}},\ }\bibfield  {title} {\bibinfo {title} {{HNF} - {Helmholtz} {Nano} {Facility}},\ }\href {https://doi.org/10.17815/jlsrf-3-158} {\bibfield  {journal} {\bibinfo  {journal} {J. Large Scale Res. Facil. \mbox{(JLSRF)}}\ }\textbf {\bibinfo {volume} {3}},\ \bibinfo {pages} {A112} (\bibinfo {year} {2017})}\BibitemShut {NoStop}%
\bibitem [{\citenamefont {Paquelet~Wuetz}\ \emph {et~al.}(2023)\citenamefont {Paquelet~Wuetz}, \citenamefont {Degli~Esposti}, \citenamefont {Zwerver}, \citenamefont {Amitonov}, \citenamefont {Botifoll}, \citenamefont {Arbiol}, \citenamefont {Sammak}, \citenamefont {Vandersypen}, \citenamefont {Russ},\ and\ \citenamefont {Scappucci}}]{paqueletwuetz23}%
  \BibitemOpen
  \bibfield  {author} {\bibinfo {author} {\bibfnamefont {B.}~\bibnamefont {Paquelet~Wuetz}}, \bibinfo {author} {\bibfnamefont {D.}~\bibnamefont {Degli~Esposti}}, \bibinfo {author} {\bibfnamefont {A.-M.~J.}\ \bibnamefont {Zwerver}}, \bibinfo {author} {\bibfnamefont {S.~V.}\ \bibnamefont {Amitonov}}, \bibinfo {author} {\bibfnamefont {M.}~\bibnamefont {Botifoll}}, \bibinfo {author} {\bibfnamefont {J.}~\bibnamefont {Arbiol}}, \bibinfo {author} {\bibfnamefont {A.}~\bibnamefont {Sammak}}, \bibinfo {author} {\bibfnamefont {L.~M.~K.}\ \bibnamefont {Vandersypen}}, \bibinfo {author} {\bibfnamefont {M.}~\bibnamefont {Russ}},\ and\ \bibinfo {author} {\bibfnamefont {G.}~\bibnamefont {Scappucci}},\ }\bibfield  {title} {\bibinfo {title} {Reducing charge noise in quantum dots by using thin silicon quantum wells},\ }\href {https://doi.org/10.1038/s41467-023-36951-w} {\bibfield  {journal} {\bibinfo  {journal} {Nat. Commun.}\ }\textbf {\bibinfo {volume} {14}},\ \bibinfo {pages} {1385} (\bibinfo {year} {2023})}\BibitemShut
  {NoStop}%
\bibitem [{\citenamefont {Volmer}\ \emph {et~al.}(2021)\citenamefont {Volmer}, \citenamefont {Seidler}, \citenamefont {Bisswanger}, \citenamefont {Tu}, \citenamefont {Schreiber}, \citenamefont {Stampfer},\ and\ \citenamefont {Beschoten}}]{Volmer21}%
  \BibitemOpen
  \bibfield  {author} {\bibinfo {author} {\bibfnamefont {F.}~\bibnamefont {Volmer}}, \bibinfo {author} {\bibfnamefont {I.}~\bibnamefont {Seidler}}, \bibinfo {author} {\bibfnamefont {T.}~\bibnamefont {Bisswanger}}, \bibinfo {author} {\bibfnamefont {J.-S.}\ \bibnamefont {Tu}}, \bibinfo {author} {\bibfnamefont {L.~R.}\ \bibnamefont {Schreiber}}, \bibinfo {author} {\bibfnamefont {C.}~\bibnamefont {Stampfer}},\ and\ \bibinfo {author} {\bibfnamefont {B.}~\bibnamefont {Beschoten}},\ }\bibfield  {title} {\bibinfo {title} {How to solve problems in micro- and nanofabrication caused by the emission of electrons and charged metal atoms during e-beam evaporation},\ }\href {https://doi.org/10.1088/1361-6463/abe89b} {\bibfield  {journal} {\bibinfo  {journal} {J. Phys. D: Appl. Phys.}\ }\textbf {\bibinfo {volume} {54}},\ \bibinfo {pages} {225304} (\bibinfo {year} {2021})}\BibitemShut {NoStop}%
\end{thebibliography}

%

\end{document}